\documentclass[12pt]{elsarticle}                                          
\usepackage{graphicx}                                                         
\usepackage{a41}                                                                   
\usepackage{xcolor}                     
\usepackage{pstricks}              
\usepackage{pstricks-add}
\usepackage[rflt]{floatflt}
\usepackage{float}
\usepackage{hyperref}
\hypersetup{colorlinks=true, linkcolor=blue, 
filecolor=blue, urlcolor=mygreen}
\usepackage{breakurl}  
\usepackage{times}     %
\usepackage{array}
\usepackage[english]{babel}
\usepackage[latin1]{inputenc}
\usepackage[T1]{fontenc}
\usepackage{ae}
\usepackage{url}
\usepackage{amsmath, amsthm, amssymb}

\usepackage{rotating}
\usepackage{graphicx}
\newcounter{mmacnt}
\def\restartmma{\setcounter{mmacnt}{0}}
\restartmma \catcode`|=\active
\def|#1|{\mathrm{#1}}
\catcode`|=12
\newenvironment{mma}{
\par\smallskip
\catcode`|=\active
\parskip=0pt\parindent=0pt 
\small
\def\In##1\\{%
\def\linebreak{\hfill\break\null\qquad}%
\refstepcounter{mmacnt}
\hangindent=2.5em\hangafter=0
\leavevmode
\llap{\tiny\sffamily In[\arabic{mmacnt}]:=\kern.5em}%
\mathversion{bold}\footnotesize$\displaystyle##1$\normalsize
\mathversion{normal}\par
 }%
\def\Print##1\\{%
\def\linebreak{\hfill\break}%
\hangindent=2.5em\hangafter=0
\leavevmode ##1\par}%
\def\Out##1\\{%
\def\linebreak{$\hfill\break\null\hfill$}%
\kern\abovedisplayskip\par
\hangindent=2.5em\hangafter=0
\leavevmode
\llap{\tiny\sffamily Out[\arabic{mmacnt}]=\kern.5em}
\footnotesize$\displaystyle##1$\normalsize\hfill\null\par
\kern\belowdisplayskip
}%
\def\Warning##1##2\\{%
\def\linebreak{\hfill\break}%
\hangindent=2.5em\hangafter=0
\leavevmode
{\scriptsize##1 : ##2}\par}%
}{%
\par\smallskip
}

\usepackage{color}
\newenvironment{fshaded}{%
\MakeFramed {\FrameRestore}
}%
{\endMakeFramed}

\makeatletter
\def\ps@pprintTitle{%
\let\@oddhead\@empty
\let\@evenhead\@empty
\def\@oddfoot{\reset@font\hfil\thepage\hfil}
\let\@evenfoot\@oddfoot
}
\makeatother
\usepackage{tikz}
\usetikzlibrary{matrix}
\allowdisplaybreaks[4]
\newcommand{\Fh}[2]{\,{}_#1F_#2}
\newcommand{\Fs}[3]{\!\!\left[
\begin{array}{c}#1\,;\\#2\,;\end{array}#3\right]}
\newcommand{\Fz}[3]{\Fs{#1}{#2}{#3}}
\begin{document}
\begin{frontmatter}
\title{\huge Scalar one-loop Feynman integrals with complex 
internal masses revisited}
\author{Khiem Hong Phan}
\ead{phkhiem@hcmus.edu.vn}
\author{Thinh Nguyen Hoang Pham}
\address{VNUHCM-University of Science, 
$227$ Nguyen Van Cu, Dist.~$5$, Ho Chi Minh City, Vietnam}
\pagestyle{myheadings}
\markright{}
\begin{abstract}
In this paper, we study systematically scalar one-loop two-, 
three-, and four-point Feynman integrals with complex internal 
masses. Our analytic results presented in this report are valid 
for both real and complex internal masses. The calculations 
are then implemented into a {\tt Mathematica} (version $9$) 
package and {\tt FORTRAN} program. Our program is cross-checked 
numerically with {\tt LoopTools} (version $2.14$) in real 
as well as complex internal masses. We find a perfect agreement 
between our results and {\tt LoopTools} for all cases. 
Additionally, this work is applied for evaluating scalar one-loop 
Feynman integrals developed leading Landau singularities which 
may appear in real scattering processes at colliders.
Last but not least, the method used in this report can
also extend to evaluate tensor one-loop integrals.
Therefore, this may open a new approach which can solve 
the inverse Gram determinant problem analytically.
\end{abstract}
\end{frontmatter}
\section{Introduction} 
\noindent
Future experimental programs at the High-Luminosity 
Large Hadron Collider (HL-LHC)~\cite{ATLAS:2013hta,CMS:2013xfa}
and the International Linear Collider (ILC)~\cite{Baer:2013cma} aim 
to measure precisely the properties of Higgs boson, of top quark and 
vector bosons in order to explore the nature of the Higgs sector as 
well as search indirectly for Physics Beyond the Standard Model (BSM). 
For matching high-precision of experimental data in the near future, 
theoretical predictions including higher-order corrections to 
multi-particle processes at the colliders are required. It means that 
the detailed evaluations of one-loop multi-leg and higher-loop at 
general scale and mass assignments are necessary.

In general, the cross-sections of scattering processes can be obtained 
by integrating over the phase space of squared amplitudes which decomposed 
frequently into the tensor integrals. At the level of one-loop corrections, 
the tensor one-loop integrals are then reduced to scalar one-loop one-, 
two-, three- and four-point functions (they are called master integrals). 
Following the tensor reductions developed 
in~\cite{Passarino:1978jh,Denner:2005nn} we may encounter the problem of 
the small value of the inverse Gram determinants (henceforth it is called 
Gram determinant problem) at several kinematic points in the phase space. 
Consequently, it leads to numerical instabilities. The suitable experimental 
cuts may be applied to get a better stability 
of the numerical results. However, when we consider one-loop multi-leg 
processes, for instance, $2\rightarrow 5,6,$ etc, the resulting 
master integrals have arbitrary configurations of their 
kinematic invariants. As a result, we can not cure the problem 
as former case. It is also noticeable that the master integrals at 
general scale and mass assignments, all space-like kinematic invariants 
as an example, have to be taken into consideration.

In evaluating for multi-particle processes 
at the colliders in which Feynman diagrams involve internal 
unstable particles that can be on-shell, one has to redefine 
their propagators with a complex mass term in the denominator. 
In other words, we have to perform the perturbative 
renormalization in the Complex-Mass Scheme 
(see Ref.~\cite{Denner:2005fg} for more detail). Thus, the 
calculations for scalar and tensor Feynman loop integrals with 
complex internal masses are also of great interest.

There have been available many papers for evaluating  
scalar and tensor one-loop integrals in space-time 
dimension $D =4 -2\varepsilon$ at $\varepsilon^0$-expansion. 
In Refs.~\cite{'tHooft:1978xw,vanOldenborgh:1989wn}, 
the authors have been derived an analytic formula for 
scalar one-loop one-, two-, three-point functions with 
real/complex internal masses and four-point functions 
with real internal masses. Then, a more compact expression 
for scalar four-point functions with real internal masses 
has been presented in Ref.~\cite{Denner:1991qq}. More recently, 
scalar box integrals with complex internal masses have been 
computed in Refs.~\cite{Nhung:2009pm,Denner:2010tr}. 
Other calculations for one-loop integrals have been 
carried out in Refs.~\cite{Ellis:2007qk,vanHameren:2010cp,
Binoth:2008uq,Cullen:2011kv,Guillet:2013msa,Bern:1992em}. 
Furthermore, based on the above calculations, various packages 
have been built for evaluating numerically of one-loop 
integrals such as~{\tt FF}~\cite{vanOldenborgh:1990yc}, 
{\tt LoopTools}~\cite{Hahn:1998yk}, 
{\tt Golem95}~\cite{Binoth:2008uq,Cullen:2011kv} 
and others~\cite{Ellis:2007qk,vanHameren:2010cp,
Berger:2008ag,Ossola:2007ax,Carrazza:2016gav,
Actis:2016mpe, Denner:2016kdg}. Along with the analytic 
calculations, several pure numerical techniques 
have been developed for computing one-loop and 
higher-loop integrals~\cite{Yuasa:2009ym,
Yuasa:2011kt,Heinrich:2008si, Borowka:2012yc,
Borowka:2015mxa,Borowka:2017idc,Gluza:2007rt,Dubovyk:2017cqw}.
As far as our knowledge, all of the above works could not 
solve either analytically or completely the 
Gram determinant problem. Besides, many of 
them have not provided the analytical results 
(or packages) to deal with one-loop integrals with complex 
internal masses as well as with all 
space-like of the kinematic invariants.

It is therefore unquestionable that solving the Gram 
determinant problem analytically and providing an alternative 
method for evaluating one-loop Feynman integrals with including
complex internal masses and at general 
configuration of kinematic invariants are mandatory. There 
already exists several methods which can solve the Gram 
determinant problem analytically. One of these approaches
is using scalar one-loop 
integrals at higher space-time dimensions $D\geq 4$ in tensor 
reduction, as pointed out in Refs.~\cite{Davydychev:1991va,Bern:1993kr,
Campbell:1996zw,Binoth:2005ff,Fleischer:2010sq}. 
Alternatively, one may consider calculating directly the tensor 
one-loop integrals. At the moment, our interest 
focuses on tackling the 
problem by following the second approach. We rely on the method
developed in Refs.~\cite{Kreimer:1991wj, Kreimer:1992ps, FranzPHD}
for evaluating tensor one-loop integrals.
In the early stage of this project, we first study systematically 
scalar one-loop Feynman integrals. We are going to 
extend previous results in~\cite{Kreimer:1991wj, 
Kreimer:1992ps, FranzPHD, khiemD0} to compute the scalar one-loop 
integrals at general external momentum assignments and with 
complex internal masses.Additionally, we 
implement our analytic 
results into a {\tt Mathematica} package and {\tt FORTRAN} program. 
We also cross-checked numerically this work with {\tt LoopTools} 
in both real- and complex-mass cases. This work is also applied 
to evaluate several one-loop Feynman integrals which could develop 
leading Landau singularities in real scattering 
processes at the colliders. 

The layout of the paper is as follows: In section $2$, 
we present in detail the method for evaluating scalar 
one-loop functions. Section $3$ shows the numerical comparison 
of this work with {\tt LoopTools}. Applications 
of our work are also discussed in this section. Conclusions 
and outlooks are devoted in section $4$. Several useful 
formulas used in this calculation can be found in the 
appendixes.
\section{The calculation} 
In this section, based on the method introduced in 
Refs.~\cite{Kreimer:1991wj, Kreimer:1992ps,FranzPHD}, 
we present in detail the calculations 
for scalar one-loop functions with complex internal masses. 
It should be noted that the analytic results for scalar one-loop 
two-, three-point functions with real internal masses have 
been reported in Refs.~\cite{Kreimer:1991wj, Kreimer:1992ps}. 
Also, scalar one-loop four-point functions with real internal 
masses have been discussed in Ref.~\cite{FranzPHD}. Recently, 
the author of Ref.~\cite{khiemD0} have been extended results 
in Ref.~\cite{FranzPHD} to evaluate the four-point functions 
with complex masses. However, in all the above papers, the 
calculations have been restricted to the cases in which the
one-loop integrals having at least one time-like 
external momentum. In this paper, we are going to extend
the previous works to calculate scalar one-loop functions 
with complex internal masses at general configurations 
of external momentum. In the following subsections, we 
use the same notations in 
Refs.~\cite{Kreimer:1991wj, Kreimer:1992ps, FranzPHD, khiemD0}.
\subsection{One-loop two-point functions}
We first mention the simplest case which is scalar one-loop
one-point functions. The Feynman integrals 
are given (see Appendix A for deriving these 
functions in more detail)
\begin{eqnarray}
\label{j1feyn}
J_1(m^2) &=& \int d^Dl\;\dfrac{1}{l^2 - m^2 + i\rho} 
=  -i\pi^{\frac{D}{2}}
\Gamma\left(\frac{2-D}{2}\right)(m^2-i\rho)^{\frac{D-2}{2}}.
\end{eqnarray}
This result has been presented in many papers, such as
\cite{'tHooft:1978xw,FranzPHD}, etc. As
$m^2 \in \mathbb{R^+}$ or $m^2 \in \mathbb{C}$, 
one can take $\rho \rightarrow 0$ in Eq.~(\ref{j1feyn}).

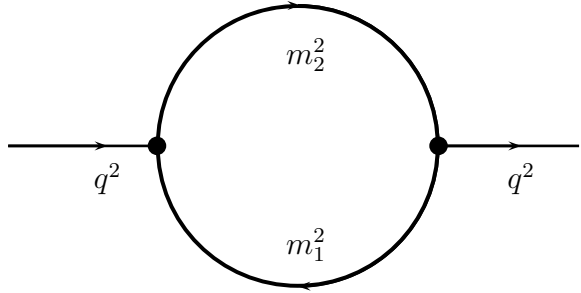
\begin{figure}[ht]
\vspace*{-2cm}
\begin{center}
\hspace{-6cm}
\begin{pspicture}(-2, -1)(6, 5)
\psset{linewidth=1.0pt}
\psset{unit = 1.1}
\psline(1, 1)(2.8,1)
\psline{->}(1, 1)(2.2,1)
\psarc{<-}(4.5, 1){1.7}{89}{90}
\psarc{<-}(4.5, 1){1.7}{-90}{90}
\pscircle(4.5,1){1.7}
\psline(6.2, 1)(7.9,1)
\psline{->}(6.2, 1)(7.2,1)
\rput(2.2,0.6){$q^2$}
\rput(7.2,0.6){$q^2$}
\rput(4.6,-0.2){$m_1^2$}
\rput(4.6,2.1){$m_2^2$}
\psset{dotsize=7pt}
\psdots(2.8,1)(6.2, 1)
\end{pspicture}
\caption{\label{2point} 
One-loop two-point diagrams.}
\end{center}
\end{figure}

We then consider scalar one-loop two-point 
functions. The functions are defined as follows:
\begin{eqnarray}
\label{j2}
J_2(q^2, m_1^2, m_2^2) 
= \int d^Dl\; \dfrac{1}{\mathcal{P}_1\mathcal{P}_2}.
\end{eqnarray}
Where the inverse Feynman propagators are given by
\begin{eqnarray}
 \mathcal{P}_1 &=& (l+q)^2 -m_1^2 +i\rho,\\
 \mathcal{P}_2 &=& l^2 -m_2^2 +i\rho.
\end{eqnarray}
Here, $q$ is the external momentum and 
$m_1, m_2$ are the internal masses, as described in 
Fig.~(\ref{2point}). In this report, the internal 
masses are taken the form of 
\begin{eqnarray}
\label{complexscheme}
 m_k^2 = m_{0k}^2 - i m_{0k}\;\Gamma_k,
\end{eqnarray}
with $k=1,2$. $\Gamma_k$ are decay width of unstable particles. 
$J_2$ is a function of $q^2, m_1^2, m_2^2$ and its symmetric
under the interchange of $m_1^2 \leftrightarrow m_2^2$. 
\subsubsection{$q^2>0$} 
For time-like momentum, e.g. $q^2>0$, we derived 
again the results in Ref.~\cite{Kreimer:1991wj}. 
By working in the rest frame of $q$, one has 
\begin{eqnarray}
 q = q(q_{10},\overrightarrow{0}_{D-1}),
\end{eqnarray}
In the parallel space spanned by external momentum and 
its orthogonal space, as explained in 
Ref.~\cite{Kreimer:1991wj}, $J_2$ takes the form of 
\begin{eqnarray}
\label{J2paraqpos}
J_2(q_{10}, m_1^2, m_2^2) 
=\dfrac{2\pi^{\frac{D-1}{2}} }{ \Gamma\left( \frac{D-1}{2} \right) }
\int\limits_{-\infty}^{\infty}dl_0 \int\limits_{0}^{\infty}dl_{\bot}
\dfrac{l_{\bot}^{D-2} }{ \mathcal{P}_1\mathcal{P}_2 }.
\end{eqnarray}
Where the inverse Feynman propagators are written as
\begin{eqnarray}
\mathcal{P}_1 &=& (l_0+q_{10})^2 - l_{\bot}^2 -m_1^2 +i\rho,\\
\mathcal{P}_2 &=& l_0^2 - l_{\bot}^2  -m_2^2 +i\rho.
\end{eqnarray}

Partitioning the integrand is as follows
\begin{eqnarray}
\label{partitionj2}
\dfrac{1}{\mathcal{P}_1\mathcal{P}_2 }=
\dfrac{1}{\mathcal{P}_1(\mathcal{P}_2-\mathcal{P}_1)}
+ 
\dfrac{1}{\mathcal{P}_2(\mathcal{P}_1-\mathcal{P}_2)}.
\end{eqnarray}
We then make a shift $l_0' = l_0 + q_{10}$. The 
result reads
\begin{eqnarray}
\hspace{-0.6cm}
J_2 &=& \dfrac{2\pi^{\frac{D-1}{2}} }
{ \Gamma\left( \frac{D-1}{2} \right) }
\frac{1}{2 q_{10}}
\int\limits_{-\infty}^{\infty}dl_0 
\int\limits_{0}^{\infty}dl_{\bot} 
\times \\
&&\hspace{0cm}\times \left\{ 
\dfrac{l_{\bot}^{D-2}}
{[l_0^2 -l_{\bot}^2 -m_2^2 + i\rho]
\left[ l_0 + \left(\frac{q_{10}}{2}  - M_d\right) \right]}
-\dfrac{l_{\bot}^{D-2}}
{[l_0^2 -l_{\bot}^2 -m_1^2 + i\rho]
\left[ l_0 - \left(\frac{q_{10}}{2} + M_d\right) \right]}
\right\}, \nonumber
\end{eqnarray}
with $M_d =\dfrac{m_1^2-m_2^2}{2q_{10}}$.
The $l_{\bot}$-integration could be performed 
easily by applying Eq.~(\ref{lbot}) in appendix B. 
We arrive at 
\begin{eqnarray}
\label{l0intqpos}
\dfrac{J_2}
{\Gamma\left( \frac{3-D}{2} \right) } 
=-\frac{\pi^{\frac{D-1}{2}} \; e^{i\pi(3-D)/2}}{2 q_{10}}
\int\limits_{-\infty}^{\infty}dl_0 \left\{ 
\dfrac{(l_0^2 - m_2^2 + i\rho)^{\frac{D-3}{2}}}
{l_0 + \left(\frac{q_{10}}{2}  - M_d\right)}
-\dfrac{(l_0^2 - m_1^2 + i\rho)^{\frac{D-3}{2}} }
{l_0 - \left(\frac{q_{10}}{2} + M_d\right)}
\right\}.
\end{eqnarray}
For case of $m_1,m_2 \in \mathbb{R}$, or $\Gamma_1 = \Gamma_2$, 
the integrands in Eq.~(\ref{l0intqpos}) have singularity poles
in real axis. However, we check that the numerators of 
(\ref{l0intqpos}) at these points are 
\begin{eqnarray}
 (l_0^2 - m_2^2)|_{l_0=-\left(\frac{q_{10}}{2}  - M_d\right)^2}
 &=&\left(\frac{q_{10}}{2}  - M_d\right)^2 -m_2^2 
 = \dfrac{\lambda(q_{10}^2, m_1^2, m_2^2)}{4q_{10}^2}, \\
 (l_0^2 - m_1^2)|_{l_0=+\left(\frac{q_{10}}{2}  + M_d\right)^2}
 &=& \left(\frac{q_{10}}{2} + M_d\right)^2 -m_1^2
 = \dfrac{\lambda(q_{10}^2, m_1^2, m_2^2)}{4q_{10}^2}.
\end{eqnarray}
Where 
\begin{eqnarray}
\lambda(x,y,z) = x^2 + y^2 + z^2 - 2xy - 2yz -2 xz
\end{eqnarray}
is the K\"allen function.
As a result, Hadamard's finite part~\cite{Hadamard} 
of these integrals will be cancelled each other. We now can present 
these integrals in terms of $\mathcal{R}$-functions as follows.
By multiplying $l_0 - \left(\dfrac{q_{10}}{2} - M_{d} \right)$,
$l_0 + \left(\dfrac{q_{10}}{2} + M_{d} \right)$ to the first 
and the second term of the integrand in (\ref{l0intqpos}) respectively,
the resulting integrand now depends on $l_0^2$. Hence, the integration 
region can be split into $[0, +\infty]$. Finally, the analytic result
for $J_2$ can be expressed by means of 
$\mathcal{R}$-functions~\cite{B.C.carlson} as follows
\begin{eqnarray}
\label{j2finalqpos}
\dfrac{J_2}
{\Gamma\left(3-\frac{D}{2}\right) }
&=&\dfrac{\pi^{(D-1)/2} e^{i\pi(3-D)/2} }{2\;q_{10}} \times
\mathcal{B}\left(\frac{4-D}{2}, \frac{1}{2} \right) \times \\
&&\hspace{0.0cm} \times
\Bigg\{ \left(\frac{q_{10}}{2} + M_d\right) 
\mathcal{R}_{\frac{D-4}{2}}
\left(\frac{3-D}{2},1;-m_1^2 +i\rho, 
- \left(\frac{q_{10}}{2}+M_d \right)^2 
\right)  
\nonumber\\
&&\hspace{0.2cm} + 
\left(\frac{q_{10}}{2} - M_d\right) 
\mathcal{R}_{\frac{D-4}{2}}
\left(\frac{3-D}{2}, 1;-m_2^2 +i\rho,
-\left(\frac{q_{10}}{2}-M_d \right)^2 \right)
\Bigg\}, \nonumber
\end{eqnarray}
It can be seen that the right hand side 
of Eq.~(\ref{j2finalqpos}) is symmetric under 
the interchange of $m_1^2 \leftrightarrow m_2^2$. 
This reflects the symmetry of the $J_2$ as mentioned 
previously. We have just derived again Eq.~($7$) in 
Ref.~\cite{Kreimer:1991wj}.

In $D =4 -2 \varepsilon$, 
by applying the expansion formula for 
$\mathcal{R}_{-\varepsilon}$ in appendix C, 
(see Eq.~(\ref{expR1}) for more detail), 
we get 
\begin{eqnarray}
\label{expansionJ2qpos}
\dfrac{J_2}{i\pi^2} &=& \frac{1}{\varepsilon} 
 -\gamma_E -\ln\pi +
\left\{ 1+ \left(1 + \frac{M_d}{q_{10} }\right) 
\left[ 
Z\ln \left( \dfrac{Z-1}{Z+1} \right) -\ln m^2_1 
\right]  \right\}
\nonumber\\
&& 
\nonumber\\
&& 
\hspace{2.6cm}
+ 
\left\{ m_1^2 \leftrightarrow m_2^2 \right\}-\text{term},
\end{eqnarray}
with 
\begin{eqnarray}
Z = \sqrt{1- \dfrac{4(m_1^2-i\rho)q_{10}^2}{(q_{10}^2 +m_1^2 - m_2^2)^2}} 
= \sqrt{ \dfrac{\lambda\left(q_{10}^2, m_1^2, m_2^2\right) }
{(q_{10}^2 +m_1^2 - m_2^2)^2 } }. 
\end{eqnarray}
The $1/\varepsilon$ pole corresponds 
to an ultraviolet divergence. Here, in the function 
$\lambda\left(q_{10}^2, m_1^2, m_2^2\right)$, $m_i^2$ are 
understood as $m_i^2 -i\rho$ for $i=1,2$. As a result,
in the case of real internal masses, the arguments of 
logarithm functions in (\ref{expansionJ2qpos}) are 
never in the negative real axis. This is totally in 
agreement with Eq.~($8$) in Ref.~\cite{Kreimer:1991wj}.

From Eq.~(\ref{j2finalqpos}), by putting $q_{10}^2 = p^2$, 
we arrive at 
\begin{eqnarray}
 \label{j2finalppos}
\dfrac{J_2}
{\Gamma\left(3-\frac{D}{2}\right) }
&=&\dfrac{\pi^{(D-1)/2} e^{i\pi(3-D)/2} }{2} \times
\mathcal{B}\left(\frac{4-D}{2}, \frac{1}{2} \right) \times \\
&&\hspace{0.0cm} \times
\Bigg\{ \left(\dfrac{p^2 +m_1^2 -m_2^2}{2 p^2}\right) 
\mathcal{R}_{\frac{D-4}{2}}
\left(\frac{3-D}{2},1;-m_1^2 +i\rho, 
-\dfrac{(p^2 +m_1^2-m_2^2)^2}{4p^2} 
\right)  
\nonumber\\
&&\hspace{0.2cm} + 
\left(\dfrac{p^2 -m_1^2 +m_2^2}{2 p^2}\right) 
\mathcal{R}_{\frac{D-4}{2}}
\left(\frac{3-D}{2}, 1;-m_2^2 +i\rho,
-\dfrac{(p^2 -m_1^2+m_2^2)^2}{4p^2}\right) 
\Bigg\}. \nonumber
\end{eqnarray}
The representation assumes to be valid also for
the case of $p^2<0$. We will confirm this conclusion
in next subsection. 
\subsubsection{$q^2<0$}  
In this article, we are also interested in 
calculating two-point functions with $q^2<0$. 
For this case, we work in the configuration of 
external momentum as follows
\begin{eqnarray}
 q = q(0, q_{11}, \overrightarrow{0}_{D-2}).
\end{eqnarray}

In the parallel and orthogonal space, 
the integral $J_2$ in Eq.~(\ref{j2}) takes 
the form of 
\begin{eqnarray}
\label{j2po}
J_2(q_{11}, m_1^2, m_2^2) 
= \dfrac{\pi^{\frac{D-2}{2}}}{\Gamma\left( \frac{D-2}{2}\right)}
\int\limits_{-\infty}^{\infty}dl_0 \int\limits_{-\infty}^{\infty}dl_1
\int\limits_{-\infty}^{\infty}dl_{\bot}\dfrac{l_{\bot}^{D-3}}
{\mathcal{P}_1\mathcal{P}_2}.
\end{eqnarray}
The inverse Feynman propagators are given  
\begin{eqnarray}
\mathcal{P}_1 &=& l_0^2-(l_1+q_{11})^2 -l_{\bot}^2 -m_1^2 +i\rho,\\
\mathcal{P}_2 &=& l_0^2 -l_1^2-l_{\bot}^2-m_2^2 +i\rho.
\end{eqnarray}

The integration written in terms of $l_0$ 
have singularity poles which are
\begin{eqnarray}
l_0 &=& \pm \sqrt{(l_1+q_{11})^2 + l_{\bot}^2 + m_1^2 -i\rho}, \\
l_0 &=& \pm \sqrt{l_1^2 + l_{\bot}^2 + m_2^2 -i\rho}. 
\end{eqnarray}
Because $m_1^2$ and $m_2^2$ have negative imaginary parts, 
as shown in Eq.~(\ref{complexscheme}), we can verify that 
these poles locate in the second and the fourth quarters 
of $l_0$-complex plane. As a result, if we close the integration 
contour in the first and the third quarters of the $l_0$-complex 
plane, there will be no residue contributions from these poles. 
Subsequently, we can derive the following relation
\begin{eqnarray}
 \int\limits_{-\infty}^{\infty}dl_0 
 = \int\limits_{-i\infty}^{+i\infty}dl_0. 
\end{eqnarray}
Using this relation, we next applied 
a Wick rotation $l_0 \rightarrow i l_0$.
After all, converting 
the Cartesian coordinates of $l_0$ and $l_{\bot}$ 
to polar coordinates $l'_{\bot}$, $\theta$
by implying following transformation: 
\begin{eqnarray}
 &&l_0 \rightarrow l'_{\bot} \cos\theta, 
\quad l_{\bot} \rightarrow l'_{\bot} \sin\theta.
\end{eqnarray}
Integrating over $\theta$, labeling 
$l'_{\bot}$ as $l_{\bot}$, the $J_2$ then reads
\begin{eqnarray}
\label{j2negative}
J_2 = \dfrac{2 i\; \pi^{\frac{D-1}{2}}}
{\Gamma\left( \frac{D-1}{2}\right)}
\int\limits_{-\infty}^{\infty}dl_1
\int\limits_{0}^{\infty}dl_{\bot}\dfrac{l_{\bot}^{D-2}}
{\mathcal{P}_1 \mathcal{P}_2}.
\end{eqnarray}
Where the Feynman propagators are now taken
\begin{eqnarray}
\mathcal{P}_1 &=& (l_1+q_{11})^2 + l_{\bot}^2 + m_1^2 - i\rho, \\
\mathcal{P}_2 &=& l_1^2 + l_{\bot}^2 + m_2^2 - i\rho. 
\end{eqnarray}

Applying the same previous procedure, we 
first use the partition for the integrand
as Eq.~(\ref{partitionj2}). We then make 
a shift $l_1 \rightarrow l_1+q_{11}$, to 
get
\begin{eqnarray}
J_2 &=& \dfrac{2 i\; \pi^{\frac{D-1}{2}}}
{\Gamma\left( \frac{D-1}{2}\right)}
\dfrac{1}{2 q_{11}}\int\limits_{-\infty}^{\infty}dl_1
\int\limits_{0}^{\infty}dl_{\bot} \times \\
&&\hspace{0cm}
\times \left\{
\dfrac{l_{\bot}^{D-2}}{[l_1^2 + l_{\bot}^2 + m_1^2 - i\rho]
[l_1 -\left(\frac{q_{11}}{2}-M_d\right)]}
-\dfrac{l_{\bot}^{D-2}}{[l_1^2 + l_{\bot}^2 + m_2^2 - i\rho]
[l_1 + \left(\frac{q_{11}}{2}+M_d \right)]}
\right\}
\nonumber
\end{eqnarray}
with $M_d =\dfrac{m_1^2-m_2^2}{2q_{11}}$. 
The $l_{\bot}$-integration can be carried out 
by using Eq.~(\ref{lbot}) in appendix B.
The result reads
\begin{eqnarray}
\label{l0intqneg}
\dfrac{J_2}
{\Gamma\left( \frac{3-D}{2} \right)} =
\frac{\pi^{\frac{D-1}{2}} \; i}{2 q_{11}}
\int\limits_{-\infty}^{\infty}dl_1\left\{
\dfrac{(l_1^2 + m_1^2 - i\rho)^{\frac{D-3}{2}}}
{l_1 - \left(\frac{q_{11}}{2} - M_d\right)}
-\dfrac{(l_1^2 + m_2^2 - i\rho)^{\frac{D-3}{2}} }
{l_1 + \left(\frac{q_{11}}{2} + M_d\right)}
\right\}. 
\end{eqnarray}
As previous case, each integrand 
in this equation has singularity pole in real axis when 
masses are real or complex having equal imaginary parts. We 
also verify that the Hadamard's finite part~\cite{Hadamard} 
of these integrals will be cancelled each other.

By comparing Eq.~(\ref{l0intqneg}) with 
Eq.~(\ref{l0intqpos}), one can realize that this 
integral has the same form as $J_2$ in 
(\ref{l0intqpos}). Thus, we can also present $J_2$ 
in terms of $\mathcal{R}$-functions as follows
\begin{eqnarray}
\label{j2rnegative}
\dfrac{J_2}
{\Gamma\left(3-\frac{D}{2}\right)}
&=&\dfrac{i\pi^{(D-1)/2} e^{i\pi(D-3)/2} }{2q_{11}}
\mathcal{B}\left(\frac{4-D}{2}, \frac{1}{2}\right) \times \\
&&\hspace{0.2cm} \times
\Bigg\{ \left(\frac{q_{11}}{2} - M_d\right)
\mathcal{R}_{\frac{D-4}{2}}
\left(\frac{3-D}{2}, 1; -m_1^2 + i\rho, 
\left(\frac{q_{11}}{2} - M_d \right)^2\right)
\nonumber\\
&&\hspace{0.6cm} +
\left(\frac{q_{11}}{2} + M_d\right)
\mathcal{R}_{\frac{D-4}{2}}
\left(\frac{3-D}{2}, 1; -m_2^2 +i\rho, 
\left(\frac{q_{11}}{2} + M_d \right)^2 \right)
\Bigg\}. \nonumber
\end{eqnarray}
As expected, Eq.~(\ref{j2rnegative}) shows the symmetry 
of $J_2$ under the interchange of $m_1^2 \leftrightarrow m_2^2$. 

In the space-time $D = 4 -2 \varepsilon$, with the help of 
$\varepsilon$-expansion for $\mathcal{R}_{-\varepsilon}$ 
in Eq.~(\ref{expR1}) in appendix C, the result reads
\begin{eqnarray}
 \dfrac{J_2}{i\pi^2} &=& \frac{1}{\varepsilon} 
 -\gamma_E -\ln\pi +
 \left\{ 1+ \left( 1- \frac{M_d}{q_{11}} \right) 
 \left[ 
 Z\ln \left( \dfrac{Z-1}{Z+1} \right) -\ln m^2_1 
 \right]  \right\}\nonumber\\
 &&
\nonumber\\
&&
\hspace{2.6cm}
+ \left\{ m_1^2 \leftrightarrow m_2^2 \right\}-\text{term}.
\end{eqnarray}
Where $Z$ now has the form of
\begin{eqnarray}
Z = \sqrt{1+ \dfrac{4(m_1^2-i\rho)q_{11}^2 }
{(q_{11}^2-m_1^2+ m_2^2)^2 )^2}}
= \sqrt{ \dfrac{\lambda\left(-q_{11}^2, m_1^2, m_2^2\right) }
{(q_{11}^2-m_1^2+m_2^2)^2 } }. 
\end{eqnarray}
It should be reminded that $m_i^2 \rightarrow m_i^2 -i\rho$ 
for $i=1,2$ in the arguments of the K\"allen function. 
Consequently, the arguments of the logarithmic functions 
are never in negative real axis. 

By substituting $q_{11}^2 = -p^2$ into 
Eq.~(\ref{j2rnegative}), we confirm again the result for 
$J_2$ in (\ref{j2finalppos}).
We can derive other representations
for $J_2$ by imploying the transformations in appendix C
for $\mathcal{R}$-functions from (\ref{relation1}) 
to (\ref{relation6}).
For example, using Euler's transformation (\ref{relation5})
for $\mathcal{R}$-functions (\ref{relation5}), 
Eq.~(\ref{j2finalppos}) becomes
\begin{eqnarray}
 \label{j2qtend0}
\dfrac{J_2}
{\Gamma\left(3-\frac{D}{2}\right) }
&=&-\pi^{(D-1)/2} e^{i\pi(3-D)/2}  \times
\mathcal{B}\left(\frac{4-D}{2}, \frac{1}{2} \right) \times \\
&&\hspace{0.0cm} \times
\Bigg\{\dfrac{(-m_1^2 + i\rho)^{\frac{D-3}{2} } }{p^2 +m_1^2 -m_2^2}
\mathcal{R}_{-\frac{1}{2}}
\left(\frac{5-D}{2},2;\frac{-1}{m_1^2 -i\rho}, 
\dfrac{-4p^2}{(p^2 +m_1^2-m_2^2)^2} 
\right)  
\nonumber\\
&&\hspace{0.2cm} + 
\dfrac{ (-m_2^2 + i\rho)^{\frac{D-3}{2} } }{p^2 -m_1^2 +m_2^2}
\mathcal{R}_{-\frac{1}{2}}
\left(\frac{5-D}{2}, 2;\frac{-1}{m_2^2 -i\rho},
\dfrac{-4p^2}{(p^2 -m_1^2+m_2^2)^2}\right) 
\Bigg\}. \nonumber
\end{eqnarray}
At threshold $p^2 = (m_1+m_2)^2$ or
pseudo threshold $p^2 = (m_1-m_2)^2$, the second
argument of these $\mathcal{R}$-functions will be
equal to the first one. Using (\ref{Rzz}), the result reads 
\begin{eqnarray}
\label{j2finalqpos2F1Bthreshold}
\dfrac{J_2}{\Gamma\left(2-\frac{D}{2}\right) } 
&=&\pi^{(D-1)/2} e^{i\pi(3-D)/2}  \Bigg\{ 
\left(\frac{p^2 + m_1^2 -m_2^2}{4 p^2}\right) 
(-m_1^2+i\rho)^{\frac{D-4}{2}} 
+ (m_1^2 \leftrightarrow m_2^2)
\Bigg\}. 
\nonumber\\
\end{eqnarray}
Eq.~(\ref{j2finalqpos2F1Bthreshold}) shows that 
$J_2$ can be reduced to two $J_1$ functions
with the space-time dimension 
shifted $D \rightarrow D-2$.
\subsubsection{$q^2=0$}
It is well-known that if internal masses are different from zero 
then $J_2$ is finite at $q^2=0$. As a result, 
the transition of $J_2$ from $q^2>0$ to the case of $q^2<0$ must be 
smooth. This property of $J_2$ has been addressed in 
Ref.~\cite{Berends:1996gs}. It means that we can perform the 
analytic continuation for $J_2$ in the limit of $q^2 \rightarrow 0$.
It can be done by taking $q^2 \rightarrow 0$ in Eq.~(\ref{j2qtend0}),
the resulting equation reads
\begin{eqnarray}
 \label{j2qtend00}
\dfrac{J_2}
{\Gamma\left(3-\frac{D}{2}\right) }
&=&-\pi^{(D-1)/2} e^{i\pi(3-D)/2}  \times
\mathcal{B}\left(\frac{4-D}{2}, \frac{1}{2} \right) \times \\
&&\hspace{0cm} \times
\Bigg\{\dfrac{(-m_1^2 + i\rho)^{\frac{D-3}{2} } }{m_1^2 -m_2^2}
\mathcal{R}_{-\frac{1}{2}}
\left(\frac{5-D}{2},2;\frac{-1}{m_1^2 -i\rho}, 
0\right) \nonumber\\
&&\hspace{0.5cm}+
\dfrac{ (-m_2^2 + i\rho)^{\frac{D-3}{2} } }{m_2^2 -m_1^2}
\mathcal{R}_{-\frac{1}{2}}
\left(\frac{5-D}{2}, 2;\frac{-1}{m_2^2 -i\rho},
0\right) 
\Bigg\}. \nonumber
\end{eqnarray}
Applying the relation (\ref{Rz1}), we will get Eq.~(\ref{j2q000}).

Alternatively, we can take the limit of 
$q_{10}\rightarrow 0$ in Eq.~(\ref{J2paraqpos}), 
or $q_{11} \rightarrow 0$ in (\ref{j2negative}).
$J_2$ is then reduced to two of $J_1$.
In particular, $J_2$ is expressed as
\begin{eqnarray}
\label{j2q000}
J_2 &=& i\pi^{\frac{D}{2} }\; 
\Gamma\left( 1-\frac{D}{2} \right)
\dfrac{ (m_2^2)^{\frac{D}{2} -1} 
- (m_1^2)^{\frac{D}{2}-1} }{m_2^2-m_1^2}.
\end{eqnarray}
If $m_1 = m_2=m$, $J_2$ can be presented as
\begin{eqnarray}
J_2 &=& i\pi^{\frac{D}{2}}\;
\Gamma\left(2-\frac{D}{2} \right) 
(m^2)^{\frac{D}{2} -2}.
\end{eqnarray}
The result in right hand side of this equation 
is proportional to $J_1$ with $D \rightarrow D-2$.

In the space-time $D = 4-2\varepsilon$, 
a series expansion of $J_2$ up to $\varepsilon^0$
yields
\begin{eqnarray}
J_2&=& \frac{1}{\varepsilon} - \gamma_E 
+ 1 - \sum\limits_{i=1}^2 
(-1)^i \dfrac{ 
m_i^2 \ln(m^2_i)}{m_1^2 -m_2^2}.
\end{eqnarray}
\subsubsection{$m_1^2= m_2^2=m^2$}
For $q^2>0$, we can take the limit of 
$m_1^2 \rightarrow m_2^2$ in Eq.~(\ref{j2finalqpos}), 
the resulting equation reads
\begin{eqnarray}
\label{j2m1m2m}
\dfrac{J_2}{\Gamma\left(3-\frac{D}{2}\right)}
&=& \dfrac{\pi^{(D-1)/2} e^{i\pi(3-D)/2} }{2}
\mathcal{B}\left(\frac{4-D}{2}, \frac{1}{2}\right)
\mathcal{R}_{\frac{D-4}{2}}\left(\frac{3-D}{2},1; 
-m^2 +i\rho, -\frac{q^2_{10}}{4} \right).
\end{eqnarray}
When $m^2 \rightarrow 0$, Eq.~(\ref{j2m1m2m})
becomes 
\begin{eqnarray}
\label{j2m1m20}
\dfrac{J_2}{\Gamma\left(3-\frac{D}{2}\right)}
&=& \dfrac{\pi^{(D-1)/2} e^{i\pi(3-D)/2} }{2}
\mathcal{B}\left(\frac{4-D}{2}, \frac{1}{2}\right)
\mathcal{R}_{\frac{D-4}{2}}\left(\frac{3-D}{2},1; 
0, -\frac{q^2_{10}}{4} \right).
\end{eqnarray}
With the help of (\ref{Rz0}), we obtain
\begin{eqnarray}
\label{j2m1m201}
\dfrac{J_2}{\Gamma\left(3-\frac{D}{2}\right)}
&=& \dfrac{\pi^{\frac{D-1}{2} } e^{i\pi(3-D)/2} }{2}
\mathcal{R}_{\frac{D-4}{2}}\left(1;-\frac{q^2_{10}}{4} \right)
=\dfrac{\pi^{\frac{D-1}{2} } e^{i\pi(3-D)/2} }{2}
\left(-\frac{q^2_{10}}{4} \right)^{\frac{D-4}{2}}.
\end{eqnarray}
In this expression, we note that 
$q_{10}^2 \rightarrow q_{10}^2 + i\rho$. 
In the limit of $q^2 = 4m^2$ and applying 
(\ref{Rzz}), Eq.~(\ref{j2m1m2m}) becomes
\begin{eqnarray}
\label{j2m1m2q4m}
\dfrac{J_2}{\Gamma\left(3-\frac{D}{2}\right)}
&=& \dfrac{\pi^{(D-1)/2} e^{i\pi(3-D)/2} }{2}
\mathcal{B}\left(\frac{4-D}{2}, \frac{1}{2}\right)
\mathcal{R}_{\frac{D-4}{2}}\left(\frac{3-D}{2},1; 
-m^2 +i\rho, -m^2 +i\rho  \right)  \nonumber\\
&=& \dfrac{\pi^{(D-1)/2} e^{i\pi(3-D)/2}}{2}
(-m^2 +i\rho)^{\frac{D-4}{2}}.
\end{eqnarray}
\subsubsection{Gauss hypergeometric presentation for $J_2$} 
Recently, the hypergeometric representations for one-loop 
integrals (like Gauss, Appell and Lauricella representations) 
have been paid attention, and asymptotic expansions of these 
functions have been studied by many authors. These are 
useful to obtain higher-order $\varepsilon$-expansions 
for Feynman loop integrals. Thus, we are also interested 
in expressing our analytic results for two and three-point 
functions in terms of the Gauss and Appell $F_1$ series in 
this literature.

The relations between Carlson's functions 
(or $\mathcal{R}$-functions) and the generalized 
hypergeometric series have been derived in 
Ref.~\cite{carlsonFD}. 
From Eq.~(\ref{R2F}), $J_2$ in (\ref{j2finalqpos}) can 
be presented in terms of the Appell $F_1$ hypergeometric 
series as follows
\begin{eqnarray}
\label{j2finalqpos2F1}
&& \hspace{-1cm}\dfrac{J_2} {\Gamma\left(3-\frac{D}{2}\right) }
=\pi^{(D-1)/2} e^{i\pi(3-D)/2}
\mathcal{B}\left(\frac{4-D}{2}, \frac{1}{2}\right)
\Bigg\{ 
\left(\frac{q_{10}^2 + m_1^2 -m_2^2}{4 q_{10}^2}\right)  
\times \\
&&\hspace{-0.2cm} \times F_1\left[\frac{4-D}{2}; \frac{3-D}{2},1; 
\frac{5-D}{2}; 1 + m_1^2 - i\rho, 
1+ \left(\dfrac{q_{10}^2 +m_1^2 -m_2^2 }{2q_{10}}\right)^2  
\right] 
+ (m_1^2 \leftrightarrow m_2^2)
\Bigg\}. \nonumber
\end{eqnarray}
Using (\ref{FN2FN1}), we can reduce the Appell $F_1$ function to
the Gauss $\Fh21$ hypergeometric functions.
The result reads
\begin{eqnarray}
\label{j2finalqpos2F1A}
&&\hspace{-1.2cm}
\dfrac{J_2} {\Gamma\left(3-\frac{D}{2}\right) }
=\pi^{(D-1)/2} e^{i\pi(3-D)/2} 
\mathcal{B}\left(\frac{4-D}{2}, \frac{1}{2}\right)
\times \\
&&
\hspace{-0.9cm} \times
\Bigg\{ 
\left(\frac{q_{10}^2 + m_1^2 -m_2^2}{4q_{10}^2}\right) 
(-m_1^2+i\rho)^{\frac{D-4}{2}} 
\Fh21\Fz{\frac{4-D}{2},1}{\frac{5-D}{2} }
{-\dfrac{\lambda(q_{10}^2,m_1^2, m_2^2) }
{4q_{10}^2m_1^2} }
+ (m_1^2 \leftrightarrow m_2^2) 
\Bigg\}. 
\nonumber
\end{eqnarray}
By applying the analytic continuation formula for 
the Gauss hypergeometric function (\ref{Fz21z}), we
have
\begin{eqnarray}
\label{2F1forJ2}
\hspace{-0.5cm}
\dfrac{J_2}{\Gamma(2-\frac{D}{2})}    
&=& \dfrac{\sqrt{\pi} 
\;\Gamma(\frac{D}{2}-1) }{\Gamma(\frac{D-1}{2} ) } 
\frac{~~~\left( \overline{m}_{2}^2 \right)^{\frac{D-2}{2}} }
{2\lambda_{12} } 
\left[ \dfrac{\partial_2 \lambda_{12}}
{\sqrt{1- m_1^2/\overline{m}_{2}^2 }} 
+ (1\leftrightarrow 2) \right]   \\
&&\nonumber\\
&&\hspace{-1cm}
- \dfrac{\Gamma( \frac{D}{2} -1 )}{\Gamma(\frac{D}{2}) }
\left \{
\left(
\dfrac{\partial_2 \lambda_{12}}{ 2 \lambda_{12}  }
\right)
\dfrac{(m_1^2)^{\frac{D-2}{2}} }
{\sqrt{1 - m_1^2/\overline{m}_{2}^2} }  
\;\Fh21\Fz{\frac{D-2}{2}, \frac{1}{2} } {\frac{D}{2}}
{\frac{m_1^2}{ \overline{m}_{2}^2 }}  
+  (1\leftrightarrow 2)      \right \},  \nonumber
\end{eqnarray}
with $\lambda_{12} = \lambda(q^2_{10}, m_1^2, m_2^2)$, 
$\overline{m}_{2}^2 = -\dfrac{\lambda_{12} }{4q_{10}^2}$ and 
$\partial_i = \dfrac{\partial}{\partial m_i^2}$
for $i=1,2$. This is totally in agreement 
with Eq.~($53$) of Ref.~\cite{Fleischer:2003rm}.

From (\ref{2F1forJ2}), we can take the limit
of $m_{1,2} \rightarrow 0$. The terms in second line
of (\ref{2F1forJ2}) will vanish under the condition of
$\mathcal{R}$e$\left(\frac{D-2}{2}\right)>0$. In the limit, 
$\overline{m}_{2}^2 \rightarrow -\dfrac{q_{10}^2}{4}$, we then 
confirm again the result in Eq.~(\ref{j2m1m20}). Furthermore, 
at the threshold $q^2 = (m_1+m_2)^2$ or pseudo threshold 
$q^2= (m_1-m_2)^2$, the arguments of hypergeometric functions 
become $0$, the resulting equation reads as 
(\ref{j2finalqpos2F1Bthreshold}).
\subsection{One-loop three-point functions} 
\begin{figure}[ht]
\vspace{1.5cm}
\begin{center}
\begin{pspicture}(-3, -3)(4, 3)
\psset{linewidth=1.0pt}
\psset{unit = 0.9}
\psline(-2.5,-2)(0,2)
\psline{->}(-2.5, -2)(-1.25, 0)
\psline(0,2)(2.5,-2)
\psline{->}(0, 2)(1.25, 0)
\psline(-2.5,-2)(2.5,-2)
\psline{->}(2.5, -2)(0, -2)
\psline(-2.5,-2)(-5,-4)
\psline{->}(-5, -4)(-3.8, -3)
\psline(2.5,-2)(5,-4)
\psline{->}(5,-4)(3.8,-3)
\psline(0,2)(0,5)
\psline{->}(0,5)(0,3)
\rput(0.5, 3.5){$p_2^2$}
\rput(-3.7,-2.5){$ p_1^2$}
\rput(3.9, -2.5){$ p_3^2$}
\rput(0, -1.5){$m_3^2$}
\rput(-2,0){$m_1^2$}
\rput(1.9, 0){$ m_2^2$}
\psset{dotsize=7pt}
\psdots(-2.5, -2)(2.5, -2)(0, 2)
\end{pspicture}
\end{center}
\caption{\label{j3diagram} 
One-loop three-point Feynman diagrams.}
\end{figure}
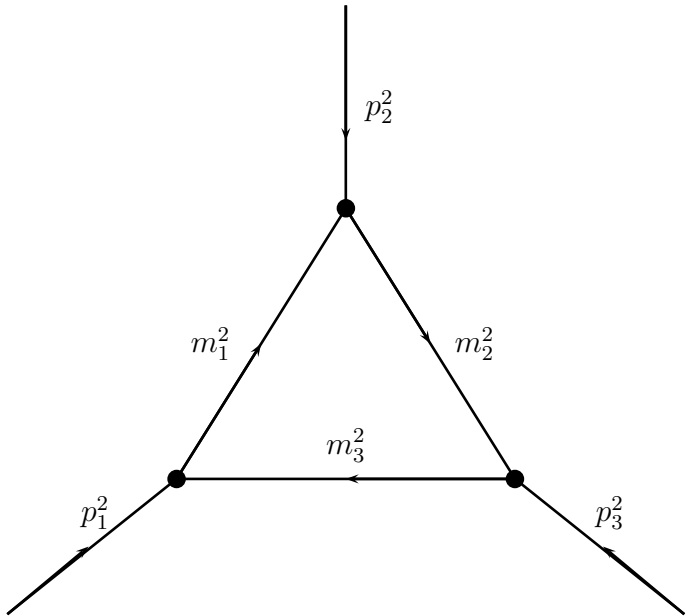

This subsection will be dealing with the scalar 
one-loop three-point functions with complex internal 
masses. The Feynman integrals for these functions 
are defined as follows
\begin{eqnarray}
J_3(p_1^2, p_2^2, p_3^2, m_1^2, m_2^2, m_3^2)
= \int d^Dl \; 
\dfrac{1}{\mathcal{P}_1\mathcal{P}_2\mathcal{P}_3}.
\end{eqnarray}
Here, the inverse Feynman propagators are given by
\begin{eqnarray}
\mathcal{P}_1 &=& (l+p_1)^2     - m_1^2 + i\rho,\\
\mathcal{P}_2 &=& (l+p_1+p_2)^2 - m_2^2 + i\rho,\\
\mathcal{P}_3 &=& l^2           - m_3^2 + i\rho.
\end{eqnarray}
We keep the convention that all external momenta flow 
inward as described in Fig.~\ref{j3diagram}. The complex 
internal masses take the same form as in 
Eq.~(\ref{complexscheme}). 
The $J_3$ is a function of six independent parameters:
$p_1^2, p_2^2, p_3^2, m_1^2, m_2^2, m_3^2$.
The symmetry of $J_3$ under the interchange of these
parameters is indicated in the Table~\ref{symJ3}.
\begin{table}[H]
\begin{center}
\begin{tabular}{c@{\hspace{1cm}}c@{\hspace{1cm}}c@{\hspace{1cm}}
c@{\hspace{1cm}}c@{\hspace{1cm}}c} \hline
$p_1^2$& $p_2^2$ &$p_3^2$& $m_1^2$& $m_2^2$ & $m_3^2$\\ \hline
$p_1^2$& $p_3^2$ &$p_2^2$& $m_3^2$& $m_2^2$ & $m_1^2$\\ \hline
$p_2^2$& $p_3^2$ &$p_1^2$& $m_2^2$& $m_3^2$ & $m_1^2$\\ \hline
$p_2^2$& $p_1^2$ &$p_3^2$& $m_1^2$& $m_3^2$ & $m_2^2$\\ \hline
$p_3^2$& $p_1^2$ &$p_2^2$& $m_3^2$& $m_1^2$ & $m_2^2$\\ \hline
$p_3^2$& $p_2^2$ &$p_1^2$& $m_2^2$& $m_1^2$ & $m_3^2$\\ \hline
\end{tabular}
\caption{\label{symJ3} Symmetry of $J_3$.}
\end{center}
\end{table}
\subsubsection{At least one time-like 
external momentum, e.g. $p_1^2>0$}    
In this case, working in rest frame of 
$p_1$, the external momenta take the 
following configuration
\begin{eqnarray}
q_1 &=& p_1 = q_1(q_{10}, 0,          
\overrightarrow{0}_{D-2}),\\
q_2 &=& p_1+p_2 = q_2(q_{20}, q_{21},  
\overrightarrow{0}_{D-2}).
\end{eqnarray}

The integral $J_3$ in the parallel and orthogonal space 
gets the form of
\begin{eqnarray}
J_3(q_{10}, q_{20}, q_{21}, m_1^2,m_2^2,m_3^2) 
&=&  \dfrac{\pi^{\frac{D-2}{2}}}{\Gamma\left( \frac{D-2}{2}\right)}
\int\limits_{-\infty}^{\infty}dl_0 \int\limits_{-\infty}^{\infty}dl_1
\int\limits_{-\infty}^{\infty}dl_{\bot}\dfrac{l_{\bot}^{D-3}}
{ \mathcal{P}_1\mathcal{P}_2\mathcal{P}_3},
\end{eqnarray}
in which the Feynman propagators become 
explicitly as
\begin{eqnarray}
 \mathcal{P}_1 &=& (l_0+q_{10})^2 - l_1^2 -l_{\bot}^2 - m_1^2 +i\rho,\\
 \mathcal{P}_2 &=& (l_0+q_{20})^2 - (l_1+q_{21})^2-l_{\bot}^2 - m_2^2 +i\rho, \\
 \mathcal{P}_3 &=& l_0^2 -l_1^2   - l_{\bot}^2 - m_3^2 +i\rho.
\end{eqnarray}

Applying the same procedure as for $J_2$, we first perform the 
partition for the integrand of $J_3$ as follows
\begin{eqnarray}
\dfrac{1}
{ \mathcal{P}_1\mathcal{P}_2\mathcal{P}_3} 
= \sum\limits_{k=1}^{3} \dfrac{1}
{\mathcal{P}_k\prod\limits_{\substack{l=1\\k\neq l}}^3
(\mathcal{P}_l- \mathcal{P}_k)}.
\end{eqnarray}
One then makes a shift $l_i = l_i + q_{ki}$, 
for $i=0,1$ and $k=1,2,3$, which gives
\begin{eqnarray}
J_3= \dfrac{2\pi^{\frac{D-2}{2}}}{\Gamma\left( \frac{D-2}{2}\right)}
 \int\limits_{-\infty}^{\infty}dl_0 \int\limits_{-\infty}^{\infty}dl_1
 \int\limits_{0}^{\infty}dl_{\bot} \sum\limits_{k=1}^{3}
 \dfrac{l_{\bot}^{D-3}}{[l_0^2-l_1^2-l_{\bot}^2-m_k^2+i\rho]
 \prod\limits_{\substack{l=1\\k\neq l}}^3(a_{lk}l_0 + b_{lk}l_1 +c_{lk})}.
\end{eqnarray}
In the above formula, we have introduced new 
kinematic variables 
\begin{eqnarray}
a_{lk} &=& 2(q_{l0}-q_{k0}), \quad 
b_{lk} = -2(q_{l1}-q_{k1}),  \quad 
c_{lk} = (q_k-q_l)^2 +m_k^2 -m_l^2. 
\end{eqnarray}
It is important to note that 
$a_{lk}, b_{lk}\in \mathbb{R}$ and $c_{lk}\in \mathbb{C}$.

Using Eq.~(\ref{lbot}) in appendix B, 
the $l_{\bot}$-integration yields
\begin{eqnarray}
\int\limits_0^{\infty} dl_{\bot}  
\dfrac{l_{\bot}^{D-3}}{[l_0^2-l_1^2-l_{\bot}^2-m_k^2+i\rho]}
= -\dfrac{\Gamma\left( \frac{D-2}{2}\right) 
\Gamma\left(2-\frac{D}{2}\right)}{2}
\left( -l_0^2 + l_1^2+m_k^2-i\rho\right)^{\frac{D}{2}-2}.
\end{eqnarray}

After integrating over $l_{\bot}$, the $J_3$ becomes
\begin{eqnarray}
\label{J3twofold}
\dfrac{J_3}{\Gamma\left(2-\frac{D}{2}\right)} 
&=&-\pi^{\frac{D-2}{2}}
\int\limits_{-\infty}^{\infty}dl_0 
\int\limits_{-\infty}^{\infty}dl_1
\sum\limits_{k=1}^{3}
\dfrac{\left(-l_0^2 + l_1^2+m_k^2-i\rho\right)^{\frac{D}{2}-2} }
{\prod\limits_{\substack{l=1\\k\neq l}}^3(a_{lk}l_0 + b_{lk}l_1 +c_{lk})}. 
\end{eqnarray}

We have just arrived at the two-fold integrals (\ref{J3twofold}), 
which can be evaluated by applying the residue theorem. 
The $l_0$-integration can be carried out first by linearizing 
the $l_0$, .i.e $\tilde{l}_1 = l_1 + l_0$. Under this 
transformation, the Jacobian is $1$ and the integration region 
is unchanged. By relabeling $\tilde{l}_1$ to $l_1$, $J_3$ is 
casted into the form of
\begin{eqnarray}
\dfrac{J_3}{ \Gamma\left(2-\frac{D}{2}\right)} = 
-\pi^{\frac{D-2}{2} }\;
\sum\limits_{k=1}^3\int\limits_{-\infty}^{\infty} dl_0 \
\int\limits_{-\infty}^{\infty} dl_1
\dfrac{[l_1^2 - 2l_1 l_0 +m_k^2 - i\rho ]^{\frac{D}{2}-2} }
{\prod\limits_{\substack{l=1\\k\neq l}}^3 
\left[AB_{lk} l_0 + b_{lk} l_1 +c_{lk} \right] },
\end{eqnarray}
where 
\begin{eqnarray}
AB_{lk} = a_{lk} - b_{lk} \in \mathbb{R}. 
\end{eqnarray} 

The integrand now depends linearly on $l_0$. Next,
we analyze the $l_0$-poles of $J_3$'s 
integrand. Its first pole is   
\begin{eqnarray}
\label{rel0J3}
 l_0 = \dfrac{l_1^2 + m_k^2 -i\rho}{2l_1},
\end{eqnarray}
with the imaginary part
\begin{eqnarray}
\label{iml0J3}
\text{Im}(l_0)= - \dfrac{m_{0k}\Gamma_k + \rho}{2l_1}.
\end{eqnarray}
From Eqs.~(\ref{rel0J3},~\ref{iml0J3}), we can confirm 
that the location of this pole on the $l_0$-complex plane 
is determined by the sign of $l_1$. For example, it is located 
in upper (lower) half-plane of $l_0$ when $l_1>0$ ($l_1<0$)
(see Fig.~(\ref{l0contour}) for more detail).

Two other poles of the $l_0$-integrand 
are the roots of the following equation: 
\begin{eqnarray}
(AB_{lk}) l_0 + b_{lk} l_1 +c_{lk} = 0,
\end{eqnarray}
which are written explicitly as
\begin{eqnarray}
\label{residueJ3l0}
l_0 = -\dfrac{ b_{lk}\; l_1 +c_{lk} }{AB_{lk}},\quad 
\text{with}\quad \text{Im}(l_0) = - 
\text{Im}\left(\dfrac{c_{lk}}{AB_{lk} }\right).
\end{eqnarray}
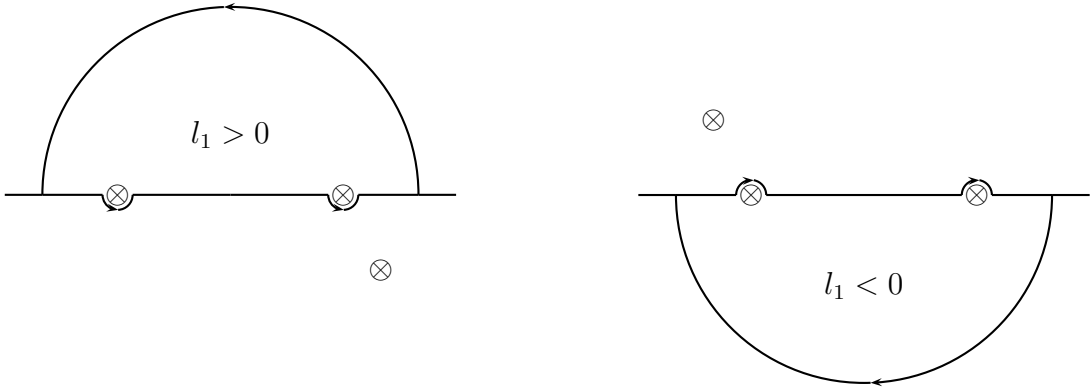
\begin{figure}[H]
\vspace*{-2cm}
\begin{tabular}{cc} 
\begin{pspicture}(-1, -2)(6, 5)
\label{zneg}
\psarc{->}(3, 1){2.5}{0}{92}
\psarc(3, 1){2.5}{92}{180}
\psline(5.5,1)(4.7, 1)
\psline(4.3, 1)(3, 1)
\psline(3, 1)(1.7, 1)
\psline(1.3, 1)(0.5,1)
\psarc{->}(4.5, 1){0.2}{180}{272}
\psarc(4.5, 1){0.2}{272}{360}
\psarc{->}(1.5, 1){0.2}{180}{272}
\psarc(1.5, 1){0.2}{272}{360}
\psline(6,1)(5.5, 1)
\psline(0.5,1)(0, 1)
\rput(4.5,1){$\otimes$}
\rput(1.5,1){$\otimes$}
\rput(5,0){$\otimes$}
\rput(3, 1.8){$ l_1 > 0$}
\end{pspicture}
&
\begin{pspicture}(-8,0)(6, 5)
\label{zpos}
\psarc(-3, 3){2.5}{180}{272}
\psarc{<-}(-3, 3){2.5}{272}{360}
\psline(-5.5,3)(-4.7, 3)
\psline(-4.3, 3)(-3, 3)
\psline(-3, 3)(-1.7, 3)
\psline(-1.3, 3)(-0.5,3)
\psarc{<-}(-4.5, 3){0.2}{80}{180}
\psarc(-4.5, 3){0.2}{0}{80}
 \psarc{<-}(-1.5, 3){0.2}{80}{180}
\psarc(-1.5, 3){0.2}{0}{80}
\psline(-6,3)(-5.5, 3)
\psline(-0.5,3)(0, 3)
\rput(-4.5,3){$\otimes$}
\rput(-1.5,3){$\otimes$}
\rput(-5,4){$\otimes$}
\rput(-3, 1.8){$ l_1 < 0$}
\end{pspicture}
\end{tabular}
\caption{\label{l0contour} 
The contour for the $l_0$-integration.}
\end{figure}
We should split the $l_1$-integration into two domains 
which are $[-\infty, 0]$ and $[0, \infty]$ as shown 
in (\ref{l0contour}). By closing 
the integration contour in the lower (upper) 
half-plane of $l_0$ for $l_1<0$ ($l_1>0$) respectively, 
and taking into account the residue contributions 
from the poles in Eq.~(\ref{residueJ3l0}), the 
result reads 
\begin{eqnarray}
\label{J3last}
\dfrac{J_3}{\Gamma\left(2-\frac{D}{2}\right) } &=& 
-\pi^{\frac{D}{2} }i\;
\sum\limits_{k=1}^3\sum\limits_{\substack{l=1\\k\neq l}}^3\; 
\dfrac{[1-\delta(AB_{lk})]}{A_{mlk}}\times \\
&&\hspace{0cm}\times \left\lbrace
f^{+}_{lk}\; \int\limits_{0}^{\infty} dz
\dfrac{\left[\left(1+\dfrac{2b_{lk}}{AB_{lk}} \right)z^2 
+ 2\dfrac{c_{lk}}{AB_{lk} }z 
+m_k^2 - i\rho \right]^{\frac{D}{2}-2} }
{\left(z +F_{mlk} \right) } \right.
\nonumber\\
&&
\hspace{0.5cm} \left.
+ f^{-}_{lk}\; \int\limits_{0}^{\infty} d z
\dfrac{\left[\left(1+\dfrac{2b_{lk}}{AB_{lk}} \right) 
z^2 - 2\dfrac{c_{lk}}{AB_{lk} }z +m_k^2 
- i\rho \right]^{\frac{D}{2}-2} }
{\left(z -F_{mlk} \right) } 
\right\rbrace, \nonumber
\end{eqnarray}
for $m\neq l$. In this equation, we have 
relabeled $z = l_1$ and introduced the 
following notations
\begin{eqnarray}
A_{mlk} &=& -AB_{km}\;b_{lk} + AB_{lk}\;b_{km}
\in \mathbb{R},  
\\
C_{mlk} &=& -AB_{km}\; c_{lk} + AB_{lk}\; c_{km} 
\in \mathbb{C},
\\
F_{mlk} &=& \dfrac{C_{mlk}}{A_{mlk}} \pm i\rho' \in \mathbb{C}, 
\end{eqnarray}
where $l,k=1,2,3, l \neq k$ and 
$m\neq l$ and $\rho' \rightarrow 0^+$.

The $ f^+_{lk}$ and $f_{lk}^{-}$ functions 
determine the location of $l_0$-poles in Eq.~(\ref{residueJ3l0}), 
which contribute to the $l_0$-integration. These functions
are defined as
\begin{eqnarray}
\label{hkl}
f^+_{lk}=
\begin{cases}
2,     & \text{if}   \;\; \text{Im}\left(-\dfrac{c_{lk}}{AB_{lk}}\right)>0,
\vspace{0.2cm} \\ 
1,     & \text{if}   \;\;\text{Im}\left(-\dfrac{c_{lk}}{AB_{lk}}\right)=0,
\vspace{0.2cm} \\
0,     & \text{if}   \;\; \text{Im}\left(-\dfrac{c_{lk}}{AB_{lk}}\right)<0.
\end{cases}
\quad \text{and} \quad 
f_{lk}^{-}=
 \begin{cases}
2,   &\text{if}    
\;\;  \text{Im}\left(-\dfrac{c_{lk}}{AB_{lk}} \right)<0,
\vspace{0.2cm}  \\
1,   & \text{if}    
\;\; \text{Im}\left(-\dfrac{c_{lk}}{AB_{lk}} \right)=0,
\vspace{0.2cm}  \\
0,   &  \text{if}   \;\; \text{Im}\left(-\dfrac{c_{lk}}{AB_{lk}} \right)>0.
\end{cases}
\end{eqnarray}

If $AB_{lk} =0$, there are no residue contributions from $l_0$ 
to the integration. Taking this point into account in the general 
formula for $J_3$, we have introduced the $\delta$-function 
as follows
\begin{eqnarray}
\label{delta}
\delta(x)=
 \begin{cases}
0,   &\text{if}  \quad x \neq 0 ;\\
1,   &\text{if}  \quad x = 0.\\
\end{cases}
\end{eqnarray}

In Eq.~(\ref{J3last}), it is also important to verify that 
\begin{eqnarray}
\text{Im}\left[\left(1+\dfrac{2b_{lk}}{AB_{lk}} \right) z^2
\pm \dfrac{2c_{lk}}{AB_{lk} } z
+m_k^2 - i\rho \right] 
=\mp \;2\;\text{Im}\left(-\dfrac{c_{lk}}{AB_{lk} } \right)z
-m_{0k}\Gamma_k -\rho \leq 0.
\end{eqnarray}
We have just arrived at the one-fold integral as shown in 
Eq.~(\ref{J3last}). It can also be expressed in terms of 
$\mathcal{R}$-functions. We first examine the general case 
in which
\begin{eqnarray}
\alpha_{lk}=1+\dfrac{2b_{lk}}{AB_{lk}} 
= \dfrac{a_{lk}+b_{lk} }{ a_{lk} - b_{lk} } \neq 0.
\end{eqnarray}
In this case, the roots $Z_{lk}^{(1)}$ and $Z_{lk}^{(2)}$  
of the equation
\begin{eqnarray}
z^2 -  2 \dfrac{c_{lk}}{ a_{lk}+ b_{lk} }z 
+ \dfrac{m_k^2 - i\rho}{\alpha_{lk}} =0
\end{eqnarray}
 are given
\begin{eqnarray}
 Z^{(1,2)}_{lk} &=& \dfrac{c_{lk}}{a_{lk} + b_{lk}} \pm 
\sqrt{ \left( \dfrac{c_{lk}}{a_{lk} + b_{lk}} \right)^2 
- \frac{m_k^2 -i\rho}{\alpha_{lk} } }.
\end{eqnarray}
In the real-mass case, 
we can decompose the numerator of $J_3$'s integrand in 
Eq.~(\ref{J3last}) as follows
\begin{eqnarray}
&&\left[ \left(1+\dfrac{2b_{lk}}{AB_{lk}} \right)z^2
\mp 2\dfrac{c_{lk}}{AB_{lk} }z + m_k^2 - i\rho  \right]^{\frac{D-4}{2}}
= \left[ \alpha_{lk} 
\left( z^2 \mp 2 \dfrac{c_{lk}}{ a_{lk}+ b_{lk} }z 
+ \dfrac{m_k^2}{\alpha_{lk}} \right) 
-i\rho \right]^{\frac{D-4}{2}}. \nonumber\\
\end{eqnarray}
Following Eq.~(\ref{abrho}) in appendix B, 
the resulting equation reads
\begin{eqnarray}
\label{decomposeJ3real1}
\left[ \alpha_{lk} 
\left( z^2 \mp 2 \dfrac{c_{lk}}{ a_{lk}+ b_{lk} }z 
+ \dfrac{m_k^2}{\alpha_{lk}} \right) 
-i\rho \right]^{\frac{D-4}{2}}
&=& \left(\alpha_{lk} - i\rho\right)^{\frac{D-4}{2}}
\left[(z\mp Z^{(1)}_{lk})(z\mp Z^{(2)}_{lk}) 
\right]^{\frac{D-4}{2}}. \nonumber\\
\end{eqnarray}
Because of $\text{Im}(Z^{(1)}_{lk})\cdot\text{Im}(Z^{(2)}_{lk}) <0$, 
Eq.~(\ref{decomposeJ3real1}) becomes
\begin{eqnarray}
\label{decomposeJ3real}
\left[ \alpha_{lk} 
\left( z^2 \mp 2 \dfrac{c_{lk}}{ a_{lk}+ b_{lk} }z 
+ \dfrac{m_k^2}{\alpha_{lk}} \right) 
-i\rho \right]^{\frac{D-4}{2}}
&=& \left(\alpha_{lk} - i\rho \right)^{\frac{D-4}{2}}
\left( z\mp Z^{(1)}_{lk} \right)^{\frac{D-4}{2}}
\left( z\mp Z^{(2)}_{lk} \right)^{\frac{D-4}{2}}. 
\nonumber\\
\end{eqnarray}
In the complex-mass case, applying Eq.~(\ref{abrhogeneral}), 
we arrive at another relation
\begin{eqnarray}
\label{decomposeJ3complex1}
\left[ \alpha_{lk} 
\left( z^2 \mp 2 \dfrac{c_{lk}}{ a_{lk}+ b_{lk} }z 
+ \dfrac{m_k^2}{\alpha_{lk}} \right) 
-i\rho \right]^{\frac{D-4}{2}}
&=& 
e^{-2\pi i\; (\frac{D-4}{2}) \theta(-\alpha_{lk})}
\left(\alpha_{lk} \right)^{\frac{D-4}{2}}
\left[(z\mp Z^{(1)}_{lk})(z\mp Z^{(2)}_{lk}) 
\right]^{\frac{D-4}{2}}  
\nonumber\\
&=& 
\left(\alpha_{lk} - i\rho\right)^{\frac{D-4}{2}}
\left[(z\mp Z^{(1)}_{lk})(z\mp Z^{(2)}_{lk}) 
\right]^{\frac{D-4}{2}}.
\end{eqnarray}
Because of the arbitrary value of $\text{Im}(Z^{(1)}_{lk})$
and $\text{Im}(Z^{(2)}_{lk})$, applying Eq.~(\ref{abrhogeneral})
for the second term in right hand side of 
Eq.~(\ref{decomposeJ3complex1}), the result reads
\begin{eqnarray}
\label{decomposeJ3complex}
\left[ \alpha_{lk} 
\left( z^2 \mp 2 \dfrac{c_{lk}}{ a_{lk}+ b_{lk} }z 
+ \dfrac{m_k^2}{\alpha_{lk}} \right) 
-i\rho \right]^{\frac{D-4}{2}}
&=& \mathcal{S}^{\pm}_{lk}
\left( z\pm Z^{(1)}_{lk} \right)^{\frac{D-4}{2}}
\left( z\pm Z^{(2)}_{lk} \right)^{\frac{D-4}{2}}. 
\end{eqnarray}
Here 
\begin{eqnarray}
\mathcal{S}^{\pm}_{lk} &=&
\left(\alpha_{lk} - i\rho\right)^{\frac{D-4}{2}} \times \\
&&\times \text{Exp}
\left[\pi i \theta\left(-\alpha_{lk}\right)
\theta[\mp \text{Im}(Z^{(1)}_{lk})]
\theta[\mp \text{Im}(Z^{(2)}_{lk})]
\left(D-4\right)\right] \times \nonumber\\
&&\times \text{Exp}
\left[-\pi i \theta\left(\alpha_{lk}\right)
\theta[\pm \text{Im}(Z^{(1)}_{lk})]
\theta[\pm \text{Im}(Z^{(2)}_{lk})]
\left(D-4\right)\right]. \nonumber 
\end{eqnarray}

We can show that the decomposition of the numerator of $J_3$-integrand
in Eq.~(\ref{decomposeJ3complex}) also covers the 
relation in Eq.~(\ref{decomposeJ3real}) for the real-mass case. 
In fact, when all masses are real, one can see that 
$\text{Im}(Z^{(1)}_{lk}) = - \text{Im}(Z^{(2)}_{lk})$, 
and Eq.~(\ref{decomposeJ3complex}) gets back to 
Eq.~(\ref{decomposeJ3real}).

The integral in Eq.~(\ref{J3last}) becomes
\begin{eqnarray}
\label{J3last1}
&&\dfrac{J_3}{\Gamma\left(2-\frac{D}{2}\right) } 
= -\pi^{\frac{D}{2} }i\;
\sum\limits_{k=1}^3\sum\limits_{\substack{l=1\\k\neq l}}^3\;
\dfrac{[1-\delta(AB_{lk})]}{A_{mlk}}\; \times                 \\
&& \times \left\lbrace 
\mathcal{S}^{+}_{lk}\;
f^{+}_{lk}\; \int\limits_{0}^{\infty} dz
\dfrac{ (z + Z^{(1)}_{lk})^{\frac{D}{2}-2}
(z + Z^{(2)}_{lk})^{\frac{D}{2}-2} } {z + F_{mlk} }
+ \mathcal{S}^{-}_{lk}\;f^{-}_{lk}\;
\int\limits_{0}^{\infty} dz
\frac{ (z - Z^{(1)}_{lk})^{\frac{D}{2}-2}
(z - Z^{(2)}_{lk})^{\frac{D}{2}-2} }{ z - F_{mlk} }
\right\rbrace, \nonumber
\end{eqnarray}
for $m\neq l$. We note that $ F_{mlk} 
\rightarrow F_{mlk} \pm i\rho'$ with 
$\rho' \rightarrow 0^+$.

The integral in Eq.~(\ref{J3last1}) can be presented 
in terms of $\mathcal{R}$-functions~\cite{B.C.carlson} 
as follows
\begin{eqnarray}
\label{timelikeJ3}
\dfrac{J_3}{\Gamma\left(2-\frac{D}{2}\right) } 
&=& -\pi^{\frac{D}{2} }i
\;\mathcal{B}(4-D,1)\;
\sum\limits_{k=1}^3\sum\limits_{\substack{l=1\\k\neq l}}^3\;
\dfrac{[1-\delta(AB_{lk})]}{A_{mlk}} \times 
\nonumber\\
&&\hspace{0cm}\times
\left\lbrace 
\mathcal{S}^{+}_{lk}\;f^{+}_{lk}\;
\mathcal{R}_{D-4}\left( 2-\frac{D}{2},  2-\frac{D}{2},1; 
Z^{(1)}_{lk}, Z^{(2)}_{lk}, F_{mlk} \right) \right.
 \\
&&\hspace{0.5cm} \left.
+ \mathcal{S}^{-}_{lk}\;f^{-}_{lk}\;
\mathcal{R}_{D-4}\left( 2-\frac{D}{2},  2-\frac{D}{2},1; 
-Z^{(1)}_{lk}, -Z^{(2)}_{lk}, -F_{mlk} \right)
\right\rbrace, \nonumber
\end{eqnarray}
for $m\neq l$. When all internal masses are real, 
$f_{lk}^{+}=f_{lk}^- = 1$ and 
$\mathcal{S}^{\pm}_{lk} = (\alpha_{lk} -i\rho)^{\frac{D-4}{2}}$, 
Eq.~(\ref{timelikeJ3}) confirms the results of, for instance, $J_3$ in the 
Eq.~($11$) of \cite{Kreimer:1992ps}. Therefore, Eq.~(\ref{timelikeJ3}) can be 
considered as a generalization of $J_3$ with real masses obtained in  
Refs.~\cite{Kreimer:1991wj,Kreimer:1992ps,FranzPHD}. 
When all internal masses are real or 
complex having equal imaginary parts, 
the integrands in (\ref{J3last}) may have singularity poles 
$z = \pm F_{mlk}$ which locate in real axis. To present
these integrals in terms of $\mathcal{R}$ functions, we first 
make $F_{mlk} \rightarrow F_{mlk} \pm i\rho'$ 
with $\rho' \rightarrow 0^{+}$. It turn out the integral 
representations for $\mathcal{R}$ functions~(\ref{rintegral})
are applicable. Final result is independent of the chosen
$\pm i\rho'$~\cite{Kreimer:1992ps}.

We can derive other represents for $J_3$ by applying several 
transformations for $\mathcal{R}$-functions, as shown in Appendix C. 
For an example, when $A_{mlk}$ become small, one should use 
Euler's transformation for $\mathcal{R}$-functions. We then get 
another formula for $J_3$ in which the over factor $A_{mlk}$ in 
denominator of Eq.~(\ref{timelikeJ3}) will be cancelled. This 
supports to be a stable numerical representation for $J_3$ 
when $A_{mlk}$ become small, 
\begin{eqnarray}
\label{timelikeJ3Amlksmall}
\dfrac{J_3}{\Gamma\left(2-\frac{D}{2}\right) } 
&=& -\pi^{\frac{D}{2} }i
\;\mathcal{B}(4-D,1)\;
\sum\limits_{k=1}^3\sum\limits_{\substack{l=1\\k\neq l}}^3\;
\dfrac{[1-\delta(AB_{lk})]}{C_{mlk}} \times 
\nonumber\\
&&\hspace{-3cm}\times
\left\lbrace 
\mathcal{S}^{+}_{lk}\;f^{+}_{lk}\; 
(Z^{(1)}_{lk})^{(D-4)/2}(Z^{(2)}_{lk})^{(D-4)/2}
\mathcal{R}_{-1}\left( 3-\frac{D}{2},  3-\frac{D}{2},2; 
1/Z^{(1)}_{lk}, 1/Z^{(2)}_{lk}, 1/F_{mlk} \right) \right.
 \\
&&\hspace{-3cm} \left.
- \mathcal{S}^{-}_{lk}\;f^{-}_{lk}\;
(-Z^{(1)}_{lk})^{(D-4)/2}(-Z^{(2)}_{lk})^{(D-4)/2}
\mathcal{R}_{-1}\left( 3-\frac{D}{2},  3-\frac{D}{2},2; 
-1/Z^{(1)}_{lk}, -1/Z^{(2)}_{lk}, -1/F_{mlk} \right)
\right\rbrace, \nonumber
\end{eqnarray}
for $m\neq l$. When $A_{mlk} \rightarrow 0$, the last argument of
$\mathcal{R}$-functions in (\ref{timelikeJ3Amlksmall}) will be zero. One 
then uses (\ref{Rz0}) and  (\ref{relation4}), the result will
give Eq.~(\ref{J3last1Aklm0}) which will be calculated in next 
paragraphs. 

In the limit of $m_k^2 \rightarrow 0$ for $k=1,2,3$, 
Eq.~(\ref{timelikeJ3}) becomes
\begin{eqnarray}
\label{timelikeJ3Mk0}
\dfrac{J_3}{\Gamma\left(2-\frac{D}{2}\right) } 
&=& -\pi^{\frac{D}{2} }i
\;\mathcal{B}\left(4-D,\frac{D-2}{2}\right)\;
\sum\limits_{k=1}^3\sum\limits_{\substack{l=1\\k\neq l}}^3\;
\dfrac{[1-\delta(AB_{lk})]}{A_{mlk}} 
(\alpha_{lk} -i\rho)^{(D-4)/2}
\times
\\
&&\hspace{0cm}\times
\left\lbrace 
\mathcal{R}_{D-4}\left( 2-\frac{D}{2},1; 
Z^{(2)}_{lk}, F_{mlk}  \right) 
+ \mathcal{R}_{D-4}\left( 2-\frac{D}{2}
,1;-Z^{(2)}_{lk}, -F_{mlk}  \right)
\right\rbrace, \nonumber
\end{eqnarray}
for $m\neq l$. 

It should be noticed that Eq.~(\ref{J3last1}) is only
valid if $A_{mlk}\neq 0$.  For the case of $A_{mlk}= 0$, 
the $J_3$ integral becomes
\begin{eqnarray}
\dfrac{J_3}{\Gamma\left(2-\frac{D}{2}\right) } 
&=&
-\pi^{\frac{D}{2} }i
\sum\limits_{k=1}^3\sum\limits_{\substack{l=1\\k\neq l}}^3 
\dfrac{[1-\delta(AB_{lk})]}{C_{mlk}}\times \\
&&\hspace{0cm}\times \left\lbrace 
f^{+}_{lk}\; \int\limits_{0}^{\infty} dz
\left[ \left( \dfrac{a_{lk}+ b_{lk}}{a_{lk} -b_{lk}} \right)z^2
+ 2\dfrac{c_{lk}}{AB_{lk} }z + m_k^2 - i\rho  
\right]^{\frac{D-4}{2}} \right. 
\nonumber\\
&&\hspace{0.5cm}+ \left.
f^{-}_{lk}\; \int\limits_{0}^{\infty} dz
\left[ \left( \dfrac{a_{lk}+ b_{lk}}{a_{lk} -b_{lk}} \right)z^2
- 2\dfrac{c_{lk}}{AB_{lk} }z + m_k^2 - i\rho  
\right]^{\frac{D-4}{2}}
\right\rbrace, \nonumber
\end{eqnarray}
for $m\neq l$. This integral can be also written
in terms of $\mathcal{R}$-function as follows
\begin{eqnarray}
\label{J3last1Aklm0}
\dfrac{J_3}{\Gamma\left(2-\frac{D}{2}\right) } 
&=&-\pi^{\frac{D}{2} }i\;\mathcal{B}(3-D,1)\;
\sum\limits_{k=1}^3\sum\limits_{\substack{l=1\\k\neq l}}^3
\dfrac{[1-\delta(AB_{lk})]} {C_{mlk}}\times \\ 
&& \hspace*{-3cm}\times \left\lbrace 
\mathcal{S}^{+}_{lk}\;
f^{+}_{lk}\; \mathcal{R}_{D-3}\left( 2-\frac{D}{2},  2-\frac{D}{2}; 
Z^{(1)}_{lk}, Z^{(2)}_{lk} \right)   
+\mathcal{S}^{-}_{lk}\;f^{-}_{lk}\;
\mathcal{R}_{D-3}\left( 2-\frac{D}{2},  2-\frac{D}{2}; 
-Z^{(1)}_{lk}, -Z^{(2)}_{lk}\right)
\right\rbrace, \nonumber
\end{eqnarray}
for $m\neq l$. If all internal masses are zero, 
Eq.~(\ref{J3last1Aklm0}) becomes 
\begin{eqnarray}
\label{J3last1Aklm0massless}
\dfrac{J_3}{\Gamma\left(2-\frac{D}{2}\right) } 
&=&-\pi^{\frac{D}{2} }i\;
\mathcal{B}\left(3-D,\frac{D-2}{2}\right)
\times \\
&& \times 
\sum\limits_{k=1}^3\sum\limits_{\substack{l=1\\k\neq l}}^3
\dfrac{[1-\delta(AB_{lk})]} {C_{mlk}}
(\alpha_{lk}-i\rho)^{\frac{D-4}{2}} \left[ (Z_{lk}^{(2)})^{D-3}
+ (-Z_{lk}^{(2)})^{D-3} \right] 
\nonumber
\end{eqnarray}
for $m\neq l$.

We turn our attention to the special case in which
$\alpha_{lk}=1 + \dfrac{2b_{lk}}{AB_{lk}} =  0$.
In this case, $a_{lk} = -b_{lk}$, $AB_{lk}=2a_{lk} = -2b_{lk}$
and $F_{mlk} = \dfrac{c_{lk}}{b_{lk}}$.
The $J_3$ in Eq.~(\ref{J3last}) can be reduced to the 
following integral
\begin{eqnarray}
\label{J3lastalpha0}
\dfrac{J_3}{\Gamma\left(2-\frac{D}{2}\right) } &=& 
-\pi^{\frac{D}{2} }i
\sum\limits_{k=1}^3\sum\limits_{\substack{l=1\\k\neq l}}^3\; 
\dfrac{[1-\delta(AB_{lk})]}{A_{mlk}}\times \\
&&\hspace{-1cm}\times \left\lbrace 
f^{+}_{lk}\; \int\limits_{0}^{\infty} d z
\dfrac{\left( \dfrac{2c_{lk}}{AB_{lk} }z +m_k^2 
- i\rho \right)^{\frac{D}{2}-2} }
{\left(z + F_{mlk} \right) }  
+ f^{-}_{lk}\; \int\limits_{0}^{\infty} dz
\dfrac{\left(-\dfrac{2c_{lk}}{AB_{lk} }z 
+m_k^2 - i\rho \right)^{\frac{D}{2}-2} }
{\left(z - F_{mlk} \right) } 
\right\rbrace.
\nonumber
\end{eqnarray}
This will be formulated with the help of the 
$\mathcal{R}$-function as
\begin{eqnarray}
\label{alphalk0}
\dfrac{J_3}{\Gamma\left(2-\frac{D}{2}\right) }  &=& 
\pi^{\frac{D}{2} }i\;
\sum\limits_{k=1}^3\sum\limits_{\substack{l=1\\k\neq l}}^3\;
\dfrac{[1-\delta(AB_{lk})]}{A_{mlk}}
\mathcal{B}\left( 2-\frac{D}{2},1\right)
\times \nonumber\\
&&\hspace{0cm}\times \left\lbrace
f^{+}_{lk}\; \left(\dfrac{c_{lk}}{a_{lk}}\right)^{\frac{D-4}{2}}
\;\mathcal{R}_{\frac{D}{2}-2}\left(1, 2-\frac{D}{2};
F_{mlk}, \frac{a_{lk}(m_k^2 -i\rho)}{c_{lk}}\right)  
\right. \\
&&\hspace{0.5cm}  \left.
+ f^{-}_{lk}\; \left(-\dfrac{c_{lk}}{a_{lk}}\right)^{\frac{D-4}{2}}\;
\mathcal{R}_{\frac{D}{2}-2}\left(1, 2-\frac{D}{2};
-F_{mlk}, -\frac{a_{lk}(m_k^2 -i\rho)}{c_{lk}}\right)
\right\rbrace, \nonumber
\end{eqnarray}
for $m\neq l$. When all masses $m_k^2 =0$ for $k =1,2,3$, 
Eq.~(\ref{alphalk0}) gets
\begin{eqnarray}
\label{alphaLK0MK0}
\dfrac{J_3}{\Gamma\left(2-\frac{D}{2}\right) }  &=& 
2\;\pi^{\frac{D}{2} }i\;
\mathcal{B}\left(\frac{4-D}{2},\frac{D-2}{2}\right) 
\sum\limits_{k=1}^3
\sum\limits_{\substack{l=1\\k\neq l}}^3\;
\dfrac{[1-\delta(AB_{lk})]}{A_{mlk}} 
\left(\dfrac{c_{lk}^2}{a_{lk}b_{lk}}\right)^{\frac{D-4}{2}} 
\nonumber\\
\end{eqnarray}
for $m\neq l$.

The $\varepsilon$-expansions for the 
$\mathcal{R}$-functions used in 
this subsection can be found in appendix B 
(see Eq.~(\ref{expR2}) to Eq.~(\ref{expR4}) for more detail). 
There are two dilogarithm 
functions in (\ref{expR2}). Hence, it is verified 
that $J_3$ in (\ref{timelikeJ3}) is presented in terms of $12$ 
dilogarithm functions at $\varepsilon^0-$expansion. 
\subsubsection{All $p_i^2<0$ for $i=1,2,3$}
In this case, external momenta can take 
the form of 
\begin{eqnarray}
q_1 &=& p_1 = q_1(0, q_{11},      
\overrightarrow{0}_{D-2}),\\
q_2 &=& p_1+p_2 = q_2(q_{20}, q_{21},  
\overrightarrow{0}_{D-2}).
\end{eqnarray}

The $J_3$ in the parallel and orthogonal space 
then becomes
\begin{eqnarray}
J_3(q_{11}, q_{20}, q_{21}, m_1^2,m_2^2,m_3^2)
&=& -\dfrac{2\pi^{\frac{D-2}{2}}}{\Gamma\left( \frac{D-2}{2}\right)}
\int\limits_{-\infty}^{\infty}dl_0 \int\limits_{-\infty}^{\infty}dl_1
\int\limits_{0}^{\infty}dl_{\bot}\dfrac{l_{\bot}^{D-3}}
{ \mathcal{P}_1\mathcal{P}_2\mathcal{P}_3},
\end{eqnarray}
with
\begin{eqnarray}
 \mathcal{P}_1 &=& (l_1+q_{11})^2 - l_0^2  + l_{\bot}^2 + m_1^2 - i\rho,\\
 \mathcal{P}_2 &=& (l_1+q_{21})^2 - (l_0+q_{20})^2 + l_{\bot}^2+m_2^2 -i\rho, \\
 \mathcal{P}_3 &=& l_1^2 - l_0^2  + l_{\bot}^2 + m_3^2 +i\rho.
\end{eqnarray}

By repeating the previous calculation as in 2.3.1, 
we obtain the following relation
\begin{eqnarray}
\label{allspacetimeJ3}
J_3(q_{11}, q_{20}, q_{21}, m_1^2,m_2^2,m_3^2) = -
J_3^{(p_1^2>0)} 
(q_{11}, q_{21}, q_{20}, -m_1^2,-m_2^2,-m_3^2).
\end{eqnarray}
Where, the right hand side of Eq.~(\ref{allspacetimeJ3}) 
is the analytic result for $J_3$ in Eq.~(\ref{timelikeJ3}) 
with $q_{10} \rightarrow q_{11}$, $q_{20} \leftrightarrow q_{21}$
and $m_i^2 -i\rho \rightarrow -m_i^2 +i\rho$ for 
$i=1,2,3$.   

We note that Eqs.~(\ref{timelikeJ3}, \ref{allspacetimeJ3}) 
also holds for cases where there is one or two light-like 
momenta. For example, if $p_1^2 =0$, we then can rotate $J_3$ 
(see Table~(\ref{symJ3})) towards 
$q_1 = p_i^2 \neq 0$ for $i=2$ or $i=3$.

For all light-like momenta, .i.e $p_1^2=0, p_2^2=0$, 
and $p_3^2 = (p_1+p_2)^2=0$, the resulting integral 
obtained after the Feynman parametrization 
for $J_3$ has 
\begin{eqnarray}
\label{pi0}
J_3 &=& -\Gamma\left(3-\frac{D}{2}\right)
\pi^{\frac{D}{2}} i
\int\limits_{0}^1 dx \int\limits_{0}^{1-x}dy
\dfrac{1}{\left[ (m_2^2-m_1^2) x + 
(m_3^2 -m_1^2) y + m_1^2 -i\rho\right]^{3-\frac{D}{2}}}
\end{eqnarray}
which is a trivial integral and can be integrated 
easily. The result reads
\begin{eqnarray}
J_3 &=&\dfrac{-4\;\pi^{\frac{D}{2}} i\;
\Gamma\left(3-\frac{D}{2}\right)}  
{ (D-4)(D-2)} 
\left\{
\dfrac{(m_1^2)^{\frac{D-2}{2} } }
{(m_2^2-m_1^2)(m_3^2-m_1^2)} 
+ \dfrac{(m_2^2)^{\frac{D-2}{2} } }
{(m_1^2-m_2^2)(m_3^2-m_2^2)} 
\right. \nonumber\\
&&\left.
\hspace{3.8cm} + \dfrac{(m_3^2)^{\frac{D-2}{2} } }
{(m_2^2-m_3^2)(m_1^2-m_3^2)} 
\right\}. 
\end{eqnarray}
It can be seen that the right hand side of $J_3$ in this case is 
a sum of three $J_1$. If $m_1^2=m_2^2=m_3^2=m^2$, Eq.~(\ref{pi0}) 
becomes
\begin{eqnarray}
J_3 &=&
-\pi^{\frac{D}{2}} i
\dfrac{\Gamma\left(3-\frac{D}{2}\right)}{2} 
(m^2)^{\frac{D}{2}-3}.
\end{eqnarray} 
The term in right hand side of this equation is propotional
to $J_1$ with shifting space-time dimension as 
$D \rightarrow D-4$. 
\subsubsection{Appell $F_1$ hypergeometric    
presentation of $J_3$}                       
As same previous reasons in $2.2.5$, 
we are also interested in expressing the $J_3$ in 
terms of the Appell $F_1$ series. Using Eq.~(\ref{R2F}), 
$J_3$ can be written as follows
\begin{eqnarray}
\label{timelikeJ3Lauri}
\dfrac{J_3}{\Gamma\left(2-\frac{D}{2}\right) } 
&=& -\pi^{\frac{D}{2} }i
\;\mathcal{B}(4-D,1)\;
\sum\limits_{k=1}^3
\sum\limits_{\substack{l=1\\k\neq l}}^3\;
\dfrac{[1-\delta(AB_{lk})]}{A_{mlk}} \times    \\
&&\hspace{-2cm}\times
\left\lbrace 
\mathcal{S}^{+}_{lk}\;f^{+}_{lk}\;
F_D\left(4-D;2-\frac{D}{2},  2-\frac{D}{2},1; 5-D;
1-Z^{(1)}_{lk}, 1-Z^{(2)}_{lk}, 1-F_{mlk} \right) 
\right. \nonumber\\
&&\hspace{-1.5cm} \left.
+ \mathcal{S}^{-}_{lk}\;f^{-}_{lk}\;
F_D\left(4-D; 2-\frac{D}{2},  2-\frac{D}{2},1; 5-D;
1+Z^{(1)}_{lk}, 1+Z^{(2)}_{lk}, 1+F_{mlk}\right)
\right\rbrace, \nonumber
\end{eqnarray}
for $m\neq l$. Where $F_D$ is 
Lauricella hypergeometric series. 
With the help of (\ref{FN2FN1}), 
Eq.~(\ref{timelikeJ3Lauri}) becomes
\begin{eqnarray}
\label{timelikeJ3Appell}
\dfrac{J_3}{\Gamma\left(2-\frac{D}{2}\right) } 
&=& -\pi^{\frac{D}{2} }i
\;\mathcal{B}(4-D,1)\;
\sum\limits_{k=1}^3
\sum\limits_{\substack{l=1\\k\neq l}}^3\;
\dfrac{[1-\delta(AB_{lk})]}{A_{mlk}} 
\times \\
&& \times 
\Big[
\mathcal{S}^{+}_{lk}\;f^{+}_{lk}\; 
(F_{mlk})^{D-4}
+ \mathcal{S}^{-}_{lk}\;f^{-}_{lk}\;
(-F_{mlk})^{D-4}
\Big] \times \nonumber\\
&& \hspace{0cm}
\times 
F_1\left(4-D;2-\frac{D}{2},  2-\frac{D}{2}; 5-D;
1-\frac{Z^{(1)}_{lk}}{F_{mlk} }, 
1-\frac{Z^{(2)}_{lk}}{F_{mlk}}\right) 
\nonumber
\end{eqnarray}
for $m\neq l$. The result is presented in
in terms of six Appell $F_1$ functions. 
We have known analogy representation for $J_3$ 
which has expressed in six Appell $F_1$ functions 
in Refs.~\cite{Fleischer:2003rm,Bluemlein:2017rbi}.

In the limits of 
$m_k^2 \rightarrow 0$ for $k=1,2,3$, 
it can be confirmed that $f_{lk}^{\pm} = 1$
and $Z_{lk}^{(1)} = 0$. 
The $J_3$ now becomes 
\begin{eqnarray}
\label{timelikeJ3Appell}
\dfrac{J_3}{\Gamma\left(2-\frac{D}{2}\right) } 
&=& -\pi^{\frac{D}{2} }i
\dfrac{\Gamma(4-D)\Gamma\left(\frac{D-2}{2} \right) 
}{\Gamma\left(3-\frac{D}{2}\right)}
\sum\limits_{k=1}^3
\sum\limits_{\substack{l=1\\k\neq l}}^3\;
\dfrac{[1-\delta(AB_{lk})]}{A_{mlk}} 
(\alpha_{lk} -i\rho)^{\frac{D-4}{2}}
\times \\
&&\hspace{0cm} 
\times 
\Big[ (F_{mlk})^{D-4}
+ (-F_{mlk})^{D-4}             \Big] 
\Fh21\Fz{4-D,2-\frac{D}{2}}{3-\frac{D}{2}}
{1-\frac{Z^{(2)}_{lk}}{F_{mlk}} }
\nonumber
\end{eqnarray}
for $m\neq l$. We have observed six Gauss hypergeometric functions
for $J_3$ in this case. In Ref.~\cite{Davydychev:1999mq}, 
$J_3$ with massless internal lines has been expressed in terms
of three $\Fh21$. Our result is numerical checked 
with Ref.~\cite{Davydychev:1999mq}. We have found perfect 
agreement to all valid digits of our results and 
Ref.~\cite{Davydychev:1999mq}.

In the case of $A_{mlk} =0$, $J_3$ can be presented 
in terms of Gauss hypergeometric series as
\begin{eqnarray}
\label{J3last1Aklm0F21}
\dfrac{J_3}{\Gamma\left(2-\frac{D}{2}\right) } 
&=&-\pi^{\frac{D}{2} }i\;\mathcal{B}(3-D,1)\;
\sum\limits_{k=1}^3\sum\limits_{\substack{l=1\\k\neq l}}^3
\dfrac{[1-\delta(AB_{lk})]} {C_{mlk}}\times \nonumber \\ 
&& \hspace*{0cm}\times \left\lbrace 
\mathcal{S}^{+}_{lk}\;
f^{+}_{lk}\;(Z_{lk}^{(1)})^{D-3} 
\Fh21\Fz{3-D,2-\frac{D}{2}}{4-D} 
{1-\frac{Z^{(2)}_{lk}}{Z^{(1)}_{lk} }}
\right.  \\
&&\hspace{0.5cm}\left. 
+ \mathcal{S}^{-}_{lk}\;f^{-}_{lk}\;
(-Z_{lk}^{(2)})^{D-3} 
\Fh21\Fz{3-D; 2-\frac{D}{2}}{4-D} 
{1-\frac{Z^{(1)}_{lk}}{Z^{(2)}_{lk}} }
\right\rbrace, \nonumber
\end{eqnarray}
for $m\neq l$. 

In the case of $\alpha_{lk}=1 + \dfrac{2b_{lk}}{AB_{lk}} = 0$,
we represent $J_3$ as
\begin{eqnarray}
\dfrac{J_3}{\Gamma\left(2-\frac{D}{2}\right) }  &=& 
\pi^{\frac{D}{2} }i\;
\sum\limits_{k=1}^3\sum\limits_{\substack{l=1\\k\neq l}}^3\;
\dfrac{[1-\delta(AB_{lk})]}{A_{mlk}}
\mathcal{B}\left( 2-\frac{D}{2},1\right)
\times \\
&&\hspace{-2cm}\times 
(f^{+}_{lk} + f^{-}_{lk}\;) 
\left(\frac{c_{lk}^2}{a_{lk}b_{lk}}\right)^{\frac{D-4}{2}}
\Fh21\Fz{\frac{4-D}{2}, \frac{4-D}{2} }{\frac{6-D}{2}} 
{1 -\frac{a_{lk}b_{lk}}{c_{lk}^2 }(m_k^2 -i\rho) }  
\nonumber
\end{eqnarray}
for $m\neq l$. When all internal masses are zero,
this equation will return to the Eq.~(\ref{alphaLK0MK0}).  
\subsection{One-loop four-point functions}
We apply the same method for evaluating scalar 
one-loop four-point functions with complex internal 
masses. Particularly, we extend the analytic 
results for these functions in~\cite{FranzPHD, khiemD0} 
which have been available for at least one time-like 
external momentum to the case of all space-like of 
kinematic invariants. 

Scalar one-loop four-point functions are defined:
\begin{eqnarray}
\label{d0}
J_4(p_1^2, p_2^2, p_3^2, p_4^2, s,
t, m_1^2, m_2^2, m_3^2, m_4^2) = 
\int\dfrac{d^Dl}{
\mathcal{P}_1 \mathcal{P}_2 
\mathcal{P}_3 \mathcal{P}_4}.
\end{eqnarray}
Where the inverse Feynman propagators are  
given
\begin{eqnarray}
\mathcal{P}_1 &=& (l+p_1)^2 - m_1^2+i\rho,          \\
\mathcal{P}_2 &=& (l+p_1+p_2)^2-m_2^2+i\rho,        \\
\mathcal{P}_3 &=& (l+p_1+p_2+p_3)^2-m_3^2 +i\rho,   \\
\mathcal{P}_4 &=& l^2- m_4^2+i\rho.                             
\end{eqnarray}
Here, $p_i~(m_i)$ for $i=1,2, \cdots, 4$ 
are the external momenta (internal masses) 
respectively. The internal masses have the 
same form of Eq.~(\ref{complexscheme}) in 
the Complex-Mass Scheme.
The $J_4$ is a function of 
$p_1^2, p_2^2, p_3^2, p_4^2, s,
t, m_1^2, m_2^2, m_3^2, m_4^2$ with
$s= (p_1+p_2)^2$, $t= (p_2+p_3)^2$.

In this calculation, we are not going to deal with 
infrared divergence using dimensional regularization 
in $D = 4 \pm 2\varepsilon$. Instead, we introduce 
fictitious mass for photon \cite{Kinoshita:1962ur}. 
Thus, we can work directly in space-time dimension 
$D=4$.  
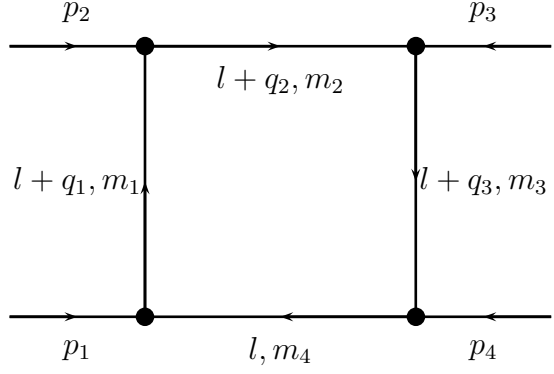
\begin{figure}[ht]
\begin{center}
\begin{pspicture}(-6, -3)(6, 3)
\psset{linewidth=1.0pt}
\psset{unit = 0.9}
\psline(-2,-2)(-2,2)
\psline{->}(-2, -2)(-2, 0)
\psline(2,-2)(2,2)
\psline{->}(2, 2)(2, 0)
\psline(-2,2)(2,2)
\psline{->}(-2, 2)(0, 2)
\psline(-2,-2)(2,-2)
\psline{->}(2, -2)(0, -2)
\psline(-4,-2)(-2,-2)
\psline{->}(-4,-2)(-3,-2)
\psline(-4,2)(-2,2)
\psline{->}(-4,2)(-3,2)
\psline(2,-2)(4,-2)
\psline{->}(4,-2)(3,-2)
\psline(2,2)(4,2)
\psline{->}(4,2)(3,2)
\rput(-3,2.5){$ p_2$}
\rput(-3,-2.5){$ p_1$}
\rput(3,2.5){$ p_3$}
\rput(3,-2.5){$ p_4$}
\rput(0,-2.5){$l, m_4$}
\rput(0,1.5){$l+q_2, m_2$}
\rput(-3,0){$l+q_1, m_1$}
\rput(3,0){$l+q_3, m_3$}
\psset{dotsize=7pt}
\psdots(-2, -2)(2, 2)(-2, 2)(2,- 2)
\end{pspicture}
\caption{\label{d0feyn} The box diagrams.}
\end{center}
\end{figure}

Let us define the momenta (see Fig.~(\ref{d0feyn}) )
\begin{eqnarray}
q_i =\sum\limits_{j=1}^i p_j \quad 
\text{for}\quad i, j=1,2, \cdots,4. 
\end{eqnarray}
If $p_i^2>0$ for $i=1,\cdots, 4$ and $s>0$, $t>0$, 
we should work in the configuration for the external 
momenta as in Eqs.~($7-10$) in Ref.~\cite{khiemD0}, or
\begin{eqnarray}
\label{LorentzOLD}
q_{1}&=&(q_{10},0,0,0), \quad q_{2}= (q_{20},q_{21},0,0),
\quad q_{3}=(q_{30},q_{31},q_{32},0), \quad q_{4} = 0.     
\end{eqnarray}
If there is one of $p_i^2$ (for $i=1,2,\cdots, 4$), $s$ and $t$
that is greater $0$, we can evaluate $J_4$ by two following ways.
First, we can rotate the $J_4$ (as shown in Table $1$ 
of Ref.~\cite{Nhung:2009pm}) towards 
the condition that $q_1^2$ will become one of 
$q_i^2 =\Big(\sum\limits_{j=1}^i p_j\Big)^2 >0 \quad 
\text{for}\quad i, j=1,2, \cdots,4$.
 We back to the case 
(\ref{LorentzOLD}). 
Alternatively, we can work in the configuration of 
the external momenta as in Eq.~(\ref{Lorentz}). We 
also check the consistency of both ways using 
numerical computation.

In the case of all kinematic invariants $p_i^2<0$ 
for $i=1,\cdots, 4$ and $s<0, t<0$ (we call hereby this 
is unphysical configuration), we set configuration of 
external momenta as follows
\begin{eqnarray}
\label{Lorentz}
q_{1}&=&(0,q_{11},0,0), \quad q_{2}= (q_{20},q_{21},0,0),
\quad q_{3}=(q_{30},q_{31},q_{32},0), \quad q_{4} = 0.     
\end{eqnarray}

It is important to mention that $J_4$ in unphysical configuration 
may appear in tensor reduction for higher-point functions. 
They may occur, especially when we consider one-loop 
corrections to multi-particle processes 
like $2\rightarrow 5,6,$ etc.

In the parallel and orthogonal space, $J_4$ takes
the form of
\begin{eqnarray}
J_4 &=&2\int\limits_{-\infty}^{\infty} \;
d l_{0}dl_{1} dl_{2}\int\limits_{0}^{\infty}d l_{\bot}
\times \\
&& \times 
\dfrac{1}{[(l_0+q_{10})^2- (l_1+q_{11})^2 -l_2^2-l_{\bot}^2 - m_1^2+i\rho] 
[(l_0+q_{20})^2 - (l_1+q_{21})^2 -l_2^2-l_{\bot}^2 -m_2^2+i\rho]} 
\nonumber \\
&&\times 
\dfrac{1}{[(l_0+q_{30})^2 - (l_1+q_{31})^2 
-(l_2+q_{32})^2-l_{\bot}^2-m_3^2 +i\rho ] 
[l_0^2 -  l_1^2 -l_2^2-l_{\bot}^2 - m_4^2+i\rho]}.
\nonumber
\end{eqnarray}
After taking $l_0, l_2$-integrations, see Ref.~\cite{khiemD0}
for more detail, and noting $z = l_1, t = l_{\bot}$, 
we have arrived at the following integrals
\begin{eqnarray}
\label{j4integrandyz}
J_4&=& J_4^{++}+ J_4^{+-} + J_4^{-+} + J_4^{--},
\end{eqnarray}
with
\begin{eqnarray}
\label{yint1}
\dfrac{J_4^{++}}{i\pi^2}&=&+
\sum_{k=1}^{4}\sum_{\substack{l=1\\k\neq l}}^{4}
\sum_{\substack{m=1\\m\neq l\\m\neq k}}^{4}
\frac{\Big(1-\delta(AC_{lk})\Big)
\Big(1-\delta(B_{mlk})\Big)}
{AC_{lk}(B_{mlk}A_{nlk}-B_{nlk}A_{mlk})}
\int\limits_{0}^{\infty}dz
\int\limits_{0}^{\infty}dt\;\;\; 
f_{lk}^+g_{mlk}^+ \;\; \mathcal{I}_{nmlk}(z,t),
\nonumber\\
&& \\
\label{yint2}
\dfrac{J_4^{+-}}{i\pi^2}&=&-
\sum_{k=1}^{4}\sum_{\substack{l=1\\k\neq l}}^{4}
\sum_{\substack{m=1\\m\neq l\\m\neq k}}^{4}
\frac{\Big(1-\delta(AC_{lk})\Big)
\Big(1-\delta(B_{mlk})\Big)}
{AC_{lk}(B_{mlk}A_{nlk}-B_{nlk}A_{mlk})}
\int\limits_{0}^{\infty}dz
\int\limits^{0}_{-\infty}dt\;\;\; 
f_{lk}^+g_{mlk}^{-} \;\;  \mathcal{I}_{nmlk}(z,t),
\nonumber\\
&& \\
\label{yint3}
\dfrac{J_4^{-+}}{i\pi^2}&=&-
\sum_{k=1}^{4}\sum_{\substack{l=1\\k\neq l}}^{4}
\sum_{\substack{m=1\\m\neq l\\m\neq k}}^{4}
\frac{\Big(1-\delta(AC_{lk})\Big)
\Big(1-\delta(B_{mlk})\Big)}
{AC_{lk}(B_{mlk}A_{nlk}-B_{nlk}A_{mlk})}
\int\limits^{0}_{-\infty}dz
\int\limits_{0}^{\infty}dt\;\;\; 
f_{lk}^{-}g_{mlk}^+ \;\;  \mathcal{I}_{nmlk}(z,t),
\nonumber \\
&& \\
\label{yint4}
\dfrac{J_4^{--}}{i\pi^2}&=&+
\sum_{k=1}^{4}\sum_{\substack{l=1\\k\neq l}}^{4}
\sum_{\substack{m=1\\m\neq l\\m\neq k}}^{4}
\frac{\Big(1-\delta(AC_{lk})\Big)
\Big(1-\delta(B_{mlk})\Big)}
{AC_{lk}(B_{mlk}A_{nlk}-B_{nlk}A_{mlk})}
\int\limits^{0}_{-\infty}dz
\int\limits^{0}_{-\infty}dt\;\;\; 
f_{lk}^{-}g_{mlk}^{-}\;\;  \mathcal{I}_{nmlk}(z,t).
\nonumber\\
\end{eqnarray}
Where the integrand $\mathcal{I}_{nmlk}$ is given by
\begin{eqnarray}
\label{integrandz}
\mathcal{I}_{nmlk}(z,t)&=&\dfrac{1}{\Big[z+F_{nmlk}\Big]\Big[D_{mlk}z^2 
- 2\Big(\frac{A_{mlk}}{B_{mlk}}-\alpha_{lk}\Big)zt-2\frac{C_{mlk}}{B_{mlk}}t 
-2\frac{d_{lk}}{AC_{lk}}z+t^2-m_k^2+i\rho \Big]}, 
\nonumber\\
\end{eqnarray}
which the related kinematic variables are given
\begin{eqnarray}
a_{lk}  &=& 2(q_{l0} -q_{k0}), \quad 
\quad  \quad \quad b_{lk} = - 2(q_{l1} -q_{k1}),         \\
c_{lk}  &=& - 2(q_{l2} -q_{k2}), 
\quad\quad \quad d_{lk}= (q_l -q_k)^2 -(m_l^2 -m_k^2),  \\
AC_{lk} &=& a_{lk}+c_{lk} \in \mathbb{R},         
\quad \quad \quad  \alpha_{lk} =\frac{b_{lk}}{AC_{lk}},           
\end{eqnarray}
and 
\begin{eqnarray}
A_{mlk} &=& a_{mk}-\frac{a_{lk}}{AC_{lk}}AC_{mk}, 
\quad \quad  B_{mlk} = b_{mk}-\frac{b_{lk}}{AC_{lk}}AC_{mk}, \\
C_{mlk} &=& d_{mk}-\frac{d_{lk}}{AC_{lk}}AC_{mk}, 
\quad \quad 
D_{mlk} = -4\frac{(q_l-q_k)^2}{AC_{lk}^2}, \label{Dmlk},   \\
F_{nmlk}&=& \frac{C_{nlk}B_{mlk}-B_{nlk}C_{mlk}}
{A_{nlk}B_{mlk}-B_{nlk}A_{mlk}} \pm i\rho', \quad 
\text{with} \quad \rho' \rightarrow 0^+.
\end{eqnarray}
We find that the integrands of $J_4$
in (\ref{j4integrandyz}) may have
singularity at $z = \pm F_{nmlk}$ locating in real axis if 
all internal masses are real or complex with possessing the same
imaginary parts. This can again be justified by giving 
$F_{nmlk} \pm i\rho'$ with $\rho' \rightarrow 0^+$. Final result 
is independent of the chosen of $\pm i\rho'$~\cite{khiemD0}.

The functions $ f_{lk}^{\pm}$ are defined as
\begin{eqnarray}
\label{flk}
f_{lk}^+ =
\begin{cases}
0,     & \text{if}   \;\; \text{Im}\left(-\dfrac{d_{lk}}{AC_{lk}}\right)<0;
\vspace{0.2cm}  \\
1,     & \text{if}   \;\;\text{Im}\left(-\dfrac{d_{lk}}{AC_{lk}}\right)=0;
\vspace{0.2cm} \\
2,     & \text{if}   \;\; \text{Im}\left(-\dfrac{d_{lk}}{AC_{lk}}\right)>0.
 \end{cases}
\quad \text{and} \quad 
f_{lk}^{-}=
 \begin{cases}
0,   &\text{if}    \;\;  \text{Im}\left(-\dfrac{d_{lk}}{AC_{lk}} \right)>0;
\vspace{0.2cm} \\
1,   & \text{if}    \;\; \text{Im}\left(-\dfrac{d_{lk}}{AC_{lk}} \right)=0;
\vspace{0.2cm} \\
2,   &  \text{if}   \;\; \text{Im}\left(-\dfrac{d_{lk}}{AC_{lk}} \right)<0.
\end{cases}
\end{eqnarray}
Furthermore, the $g_{mlk}^+$ and $g_{mlk}^-$ are given by
\begin{eqnarray}
\label{gmlk}
g_{mlk}^+= 
\begin{cases}
 0,& \text{if}   \;\; \mathrm{Im} \left(-\dfrac{C_{mlk}}{B_{mlk}} \right)<0;
 \vspace{0.2cm}  \\
 1,& \text{if}   \;\; \mathrm{Im}\left(-\dfrac{C_{mlk}}{B_{mlk}} \right)=0; 
 \vspace{0.2cm} \\
 2,& \text{if}   \;\; \mathrm{Im}\left(-\dfrac{C_{mlk}}{B_{mlk}}\right)>0;
 \end{cases} 
\qquad  \text{and}  \qquad
g_{mlk}^{-}=
\begin{cases}
 0,& \text{if}   \;\; \mathrm{Im}\left (-\dfrac{C_{mlk}}{B_{mlk}} \right)>0;
 \vspace{0.2cm}  \\
 1,& \text{if}   \;\; \mathrm{Im}\left(-\dfrac{C_{mlk}}{B_{mlk}} \right)=0;
 \vspace{0.2cm}  \\
 2,& \text{if}   \;\; \mathrm{Im}\left (-\dfrac{C_{mlk}}{B_{mlk}} \right)<0.
 \end{cases}
\end{eqnarray}

With the definitions of $f^+_{lk}, f^-_{lk}$ in Eq.~(\ref{flk}) and
$g^+_{mlk},g^-_{mlk}$ in Eq.~(\ref{gmlk}), we confirm that 
\begin{eqnarray} 
\label{Imgklm}
\mathrm{Im} \Big[D_{mlk}z^2-2\Big(\frac{A_{mlk}}{B_{mlk}}-\alpha_{lk}\Big)zt 
- 2\frac{C_{mlk}}{B_{mlk}}t -2\frac{d_{lk}}{AC_{lk}}z+t^2-m_k^2+i\rho \Big] > 0.  
\end{eqnarray}

If $D_{mlk} = 0$, it means that some kinematic invariants are 
light-like (see Eq.(\ref{Dmlk})). The integrations written in terms 
of $z$ in (\ref{yint1}, \ref{yint2}, \ref{yint3},\ref{yint4}) are 
trivial and they can be taken first. The $t$-integrations are then 
presented in terms of three basic integrals (\ref{basicsINTZ}). 
The results for this case have been reported in~\cite{khiemD0}.

For $D_{mlk}\neq 0$, we first linearize $t$-integration. 
In \cite{khiemD0}, we have performed the following 
rotation
\begin{equation}
\label{rotation}
\begin{aligned}
z = z' + \beta_{mlk} t' & & z' = \frac{z-\beta_{mlk}
t}{1-\beta_{mlk}\varphi_{mlk}}, \\
&\qquad \Longrightarrow \qquad& \\
t = t' + \varphi_{mlk} z' & & t' = \frac{t-\varphi_{mlk},
z}{1-\beta_{mlk}\varphi_{mlk}}.
\end{aligned}
\end{equation}

To linearize in $t$, $\beta_{mlk}$ and $\varphi_{mlk}$ are
taken 
\begin{eqnarray}
\label{betasol}
\beta_{mlk}^{(1,2)}&=&\dfrac{\Big(\frac{A_{mlk}}{B_{mlk}}-\alpha_{lk}\Big) \pm 
\sqrt{\Big(\frac{A_{mlk}}{B_{mlk}}-\alpha_{lk}\Big)^2-D_{mlk}}}{D_{mlk}}\\
\label{phisol}
\varphi_{mlk}^{(1,2)} &=&\Big(\frac{A_{mlk}}{B_{mlk}} - \alpha_{lk}\Big) \pm 
\sqrt{\Big(\frac{A_{mlk}}{B_{mlk}}-\alpha_{lk}\Big)^2-D_{mlk}}. 
\end{eqnarray}  
As a consequence, in the case of 
$D_{mlk} = \Big(\frac{A_{mlk}}{B_{mlk}}-\alpha_{lk}\Big)^2>0$, 
the Jacobian of this shift is zero. The previous results in \cite{khiemD0} 
have not covered this case. 

In order to cover this case as well, 
instead of using (\ref{rotation}), we performed a shift as 
\begin{eqnarray}
\label{shiftbeta}
z = z' + \beta_{mlk} t.
\end{eqnarray}
The Jacobian of this shift is $1$. Taking $\beta_{mlk}$ as 
one of the solution (\ref{betasol}), 
the integrand of $J_4$ now reads
\begin{eqnarray}
\label{integrandz}
\mathcal{I}_{nmlk}(z,t)&=&\dfrac{1}{\Big[z+\beta_{mlk}t +F_{nmlk}\Big]
\Big[P_{mlk}zt + Q_{mlk}t+ D_{mlk}z^2 + E_{mlk}z -m_k^2+i\rho \Big]},
\nonumber\\
\end{eqnarray}
with new kinematic variables which have been introduced as
\begin{eqnarray}
Q_{mlk} &=& -2\left(\frac{C_{mlk}}{B_{mlk}} \right) 
-2\left(\dfrac{d_{lk}}{AC_{lk}} \right)\beta_{mlk}, \\
P_{mlk} &=& -2\left(\frac{A_{mlk}}{B_{mlk}}-\alpha_{lk}
-\beta_{mlk}D_{mlk}\right), \\
E_{mlk} &=& -2\left(\dfrac{d_{lk}}{AC_{lk}} \right).
\end{eqnarray}
\subsubsection{$D_{mlk}<0$ or 
$0< D_{mlk}< \left(\frac{A_{mlk}}{B_{mlk}}-\alpha_{lk}\right)^2$
and $\frac{A_{mlk}}{B_{mlk}}-\alpha_{lk} \leq 0$}
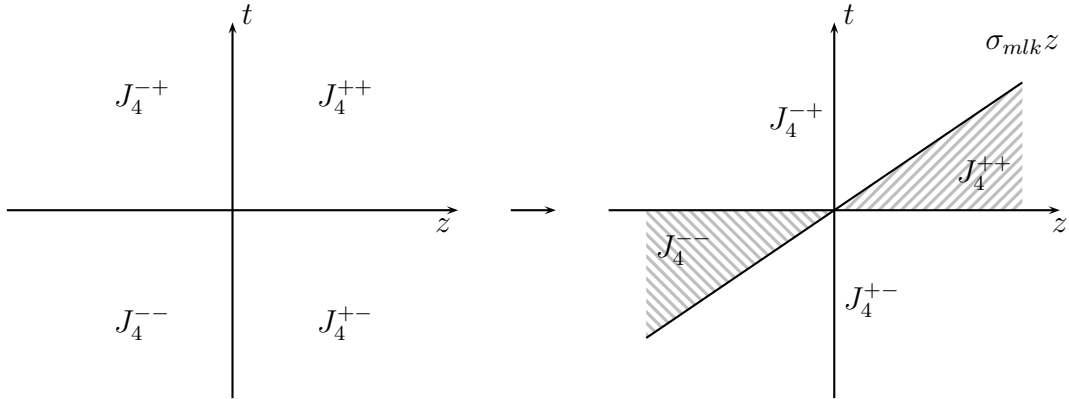
\begin{figure}[ht]
\begin{center}
\begin{pspicture}(-7, -2.5)(7, 2.5)
\psline{->}(-4,-2.5)(-4,2.5)
\psline{->}(4,-2.5)(4,2.5)
\pspolygon[fillstyle=hlines,
linestyle=none, linecolor=lightgray, hatchcolor=lightgray, hatchwidth=1.2pt,
hatchsep=1.8pt](6.5,0)(4, 0)(6.5,1.7)
\pspolygon[fillstyle=hlines,
linestyle=none, linecolor=lightgray, hatchcolor=lightgray, hatchwidth=1.2pt,
hatchsep=1.8pt,hatchangle=135](1.5,-1.7)(4, 0) (1.5,0)
\psline{->}(-7,0)(-1,0)
\psline{->}(1,0)(7,0)
\psline(1.5,-1.7)(6.5,1.7)
\rput(-3.8, 2.6){$t$}
\rput(4.2, 2.6){$t$}
\rput(-1.2,-0.2){$z$}
\rput(7,-0.2){$z$}
\rput(6.5, 2.2){$\sigma_{mlk}z$}
\rput(6., 0.5){$J_4^{++}$}
\rput(4.5,-1.2){$J_4^{+-}$}
\rput(3.5,1.2 ){$J_4^{-+}$}
\rput(2.,-0.5)   {$J_4^{--}$}
\rput(-2.5, 1.5){$J_4^{++}$}
\rput(-2.5,-1.5){$J_4^{+-}$}
\rput(-5.2,1.5 ){$J_4^{-+}$}
\rput(-5.2, -1.5){$J_4^{--}$}
\psline{->}(-0.3,0)(0.3,0)
\end{pspicture}
\caption{\label{beta1} The integration region.}
\end{center}
\end{figure}

In this case, $\beta_{mlk} \leqslant 0$. The integration region 
now looks as Fig.~\ref{beta1}. 
To integrate over $t$, we split the integrations 
written in terms of $t$ as follows
\begin{eqnarray}
J_4^{++} & \longrightarrow & \int\limits_{0}^{\infty} dz 
\int\limits_{0}^{\sigma_{mlk}\; z}dt                         \\
J_4^{+-} &\longrightarrow & \int\limits_{0}^{\infty} dz 
\int\limits^{0}_{-\infty}dt
+ \int\limits^{0}_{-\infty}  dz 
\int\limits^{\sigma_{mlk}\; z}_{-\infty} dt, \\
J_4^{-+} &\longrightarrow &\int\limits^{0}_{-\infty} dz 
\int\limits^{\infty}_{0} dt
+\int\limits^{\infty}_0 dz 
\int\limits^{\infty}_{\sigma_{mlk}\;z} dt,  \\
J_4^{--} &\longrightarrow & \int\limits_{-\infty}^0 dz 
\int\limits_{\sigma_{mlk}}^{0} dt,
\end{eqnarray}
with $\sigma_{mlk} = -\dfrac{1}{\beta_{mlk}}$. Besides that
we also rewrite $\mathcal{I}_{nmlk}(z,t)$ as follows
\begin{eqnarray}
&&\mathcal{I}_{nmlk}(z,t) = 
G(z)\left[
\dfrac{1}{t+ \frac{z+F_{nmlk}}{\beta_{mlk}}} - 
\dfrac{1}{t+ \frac{ D_{mlk} z^2 + 
E_{mlk}z -m_k^2+i\rho}{ P_{mlk}z + Q_{mlk} } } 
\right],
\end{eqnarray}
in which the $G(z)$ is given by 
\begin{eqnarray}
\label{GZ}
G(z) &=& \dfrac{1}
{ Z_{mlk}\;z^2 + 
(E_{mlk} \beta_{mlk} -Q_{mlk} -P_{mlk}F_{nmlk})z 
-\beta_{mlk}(m_k^2 -i\rho) - F_{nmlk}Q_{mlk}  }, \nonumber \\
&=& \dfrac{1}{Z_{mlk}(z-T_{mlk}^{(1)})(z-T_{mlk}^{(2)})},
\quad 
\text{with}\quad Z_{mlk} = D_{mlk}\beta_{mlk} -P_{mlk}.
\end{eqnarray}

After carrying out the $t$-integrations, 
the $J_4$ reads 
\begin{eqnarray}
\label{zint-ln1}
&& \dfrac{J_4}{i\pi^2} =
\sum_{k=1}^{4}\sum_{\substack{l=1\\k\neq l}}^{4}
\sum_{\substack{m=1\\m\neq l\\m\neq k}}^{4}
\frac{\Big(1-\delta(AC_{lk})\Big)\Big(1-\delta(B_{mlk})\Big)}
{AC_{lk}(B_{mlk}A_{nlk}-B_{nlk}A_{mlk})}  \times \\
&&\times \left[ \hspace{0.4cm}\int\limits_{0}^{\infty} dz \; G(z) 
\left\{ 
(f_{lk}^+ g_{mlk}^+ + f_{lk}^- g_{mlk}^+) 
\ln\left(\frac{F_{nmlk}}{\beta_{mlk}}\right)
- f_{lk}^+ g_{mlk}^+ 
\ln\left(\frac{z+F_{nmlk}}{\beta_{mlk}}\right) 
\right.\right.\nonumber\\
&&\hspace{1.5cm}
- f_{lk}^+ g_{mlk}^- 
\ln\left(-\frac{z+F_{nmlk}}{\beta_{mlk}}\right)
- (f_{lk}^- g_{mlk}^+ + f_{lk}^+ g_{mlk}^+) 
\ln\left(\frac{ S(\sigma_{mlk}, z)}{ P_{mlk}z+ Q_{mlk} }\right)    
\nonumber\\
&&\hspace{1.5cm} \left.
+
f_{lk}^+ g_{mlk}^+ \ln\left(\frac{ S(\sigma_{mlk}=0, z)}
{ P_{mlk}z+ Q_{mlk} }\right)
+
f_{lk}^+ g_{mlk}^- \ln\left(-\frac{ S(\sigma_{mlk}=0, z)}
{ P_{mlk}z+ Q_{mlk} }\right)
\hspace{0.5cm} \right\} 
\nonumber\\
&& \hspace{0.7cm}
+\left. \int\limits^{0}_{-\infty} dz \; G(z)
\left\{ 
-f_{lk}^+ g_{mlk}^- 
\ln\left(-\frac{F_{nmlk}}{\beta_{mlk}}\right)
+ (f_{lk}^- g_{mlk}^- + f_{lk}^- g_{mlk}^+)
\ln\left(\frac{z+F_{nmlk}}{\beta_{mlk}}\right)
\right.\right.\nonumber\\
&&\hspace{1.5cm}
-f_{lk}^- g_{mlk}^- \ln\left(\frac{F_{nmlk}}{\beta_{mlk}}\right)
-(f_{lk}^- g_{mlk}^+ + f_{lk}^- g_{mlk}^- )
\ln\left(\frac{ S(\sigma_{mlk}=0, z)}{ P_{mlk}z+ Q_{mlk} }\right)
\nonumber\\
&&\hspace{1.5cm}
\left. \left.
+f_{lk}^- g_{mlk}^-  
\ln\left(\frac{ S(\sigma_{mlk}, z)}{ P_{mlk}z+ Q_{mlk} }\right)
+ 
f_{lk}^+ g_{mlk}^- \ln\left(-\frac{ S(\sigma_{mlk}, z)}
{ P_{mlk}z+ Q_{mlk} }\right)
\hspace{0.3cm}
\right\}
\hspace{0.2cm}
\right. 
\Bigg].
\nonumber
\end{eqnarray}
Where the function $S(\sigma_{mlk},z)$ is defined as
\begin{eqnarray}
\label{SZ}
S(\sigma_{mlk},z) &=& (D_{mlk} + P_{mlk}\sigma_{mlk})z^2
+ (E_{mlk} + Q_{mlk}\sigma_{mlk})z -m_k^2 + i \rho \nonumber\\ 
&=& S_{mlk}^{(\sigma)} (z-Z_{mlk}^{(1\sigma)})(z-Z_{mlk}^{(2\sigma)}), \\
S(\sigma_{mlk}=0,z) &=& D_{mlk}z^2 + E_{mlk}z -m_k^2 + i \rho 
= S_{mlk}^{(0)} (z-Z_{mlk}^{(10)})(z-Z_{mlk}^{(20)}), 
\end{eqnarray}
with $S_{mlk}^{(\sigma)} = D_{mlk} + P_{mlk}\sigma_{mlk}$ and 
$S_{mlk}^{(0)} =  D_{mlk}$.

We realize that
\begin{eqnarray}
\label{Im(e-q)}
\mathrm{Im}\Big(E_{mlk}\Big)=
\begin{cases}
        \geq 0,   &\text{with}    \;\;  f_{lk}^+\;\; g_{mlk}^{\pm}, \\\\
        \leq 0    &\text{with}    \;\;  f_{lk}^-\;\; g_{mlk}^{\pm},
\end{cases}
\end{eqnarray}
and
\begin{eqnarray}
\label{Im(e-qb)}
\mathrm{Im}
\Big(E_{mlk}-\frac{Q_{mlk}}{\beta_{mlk}}\Big)=
\begin{cases}
        \geq 0,   &\text{with}    \;\;  f_{lk}^{\pm} \;\; g_{mlk}^{+}, \\\\
        \leq  0    &\text{with}     \;\; f_{lk}^{\pm} \;\; g_{mlk}^{-}. 
\end{cases}
\end{eqnarray}
Therefore, from Eqs.~(\ref{Im(e-q)}, \ref{Im(e-qb)}), we confirm that
\begin{eqnarray}
\label{Im}
\text{Im}\left( S(\sigma, z)\right) \geq 0,
\end{eqnarray}
with  $\sigma_{mlk}=0, -1/\beta_{mlk}$.

In order to perform the $z$-integrals, we decompose logarithmic functions 
in the $z$-integrands (\ref{zint-ln1}) by using Eq.~(\ref{logdecompose})
\begin{eqnarray}
\label{decomposea}
\ln\Big(\frac{ S(\sigma,z)}{P_{mlk}z+Q_{mlk}}\Big)&=&
\ln\left(S^{(\sigma)}_{mlk} z- S^{(\sigma)}_{mlk} Z^{(1\sigma)}_{mlk}\right)
+ \ln\left(z-Z^{(2\sigma)}_{mlk}\right) - \ln(P_{mlk}z+Q_{mlk})\\
&&\hspace{-3cm}
+2\pi i\theta[\text{Im}\left(S^{(\sigma)}_{mlk} Z^{(1\sigma)}_{mlk}\right)]
\theta[\text{Im}(Z^{(2\sigma)}_{mlk})] - 
2\pi i \theta[-\text{Im}(Q_{mlk})]
\theta\Big[-\text{Im} 
\left(\frac{S(\sigma,z)}{P_{mlk}z+Q_{mlk}} \right) \Big], 
\nonumber
\end{eqnarray}
and 
\begin{eqnarray}
\label{decomposeb}
\ln\Big(-\frac{ S(\sigma,z)}{P_{mlk}z+Q_{mlk}}\Big)&=&
\ln\left(-S^{(\sigma)}_{mlk} z + S^{(\sigma)}_{mlk} Z^{(1\sigma)}_{mlk}\right)
+ \ln\left(z-Z^{(2\sigma)}_{mlk}\right) - \ln(P_{mlk}z+Q_{mlk}) \nonumber\\
&&\hspace{-3.5cm}
-2\pi i\theta[\text{Im}\left(S^{(\sigma)}_{mlk} Z^{(1\sigma)}_{mlk}\right)]
\theta[-\text{Im}(Z^{(2\sigma)}_{mlk})] + 2\pi i \theta[\text{Im}(Q_{mlk})]
\theta\Big[-\text{Im} 
\left(\frac{S(\sigma,z)}{P_{mlk}z+Q_{mlk}} \right) \Big].
\nonumber\\
\end{eqnarray}
We verify that 
\begin{eqnarray}
\text{Im}
\left(\frac{S(\sigma,z)}{P_{mlk}z+Q_{mlk}} \right)
&=& \text{Im}
\left(\frac{S(\sigma=0,z)}{P_{mlk}z+Q_{mlk}} \right)
= A_{mlk}^0 z^2 + B_{mlk}^0 z + C_{mlk}^0, 
\end{eqnarray}
with
\begin{eqnarray}
A_{mlk}^0 & =& \text{Im}\{ D_{mlk}Q_{mlk}^* + P_{mlk}E_{mlk}\}, \\
B_{mlk}^0 & =& \text{Im}\{ E_{mlk}Q^*_{mlk} -P_{mlk}(m_k^2 -i\rho)\}, \\
C_{mlk}^0 & =& -\text{Im}\{ (m_k^2 -i\rho)Q^*_{mlk}\}. 
\end{eqnarray}
With the help of the 
formulas~(\ref{decomposea}, \ref{decomposeb}), $J_4$ can 
be presented in the form 
\begin{eqnarray}
\label{zint-ln1-decom}
&&\hspace{-0.5cm}\dfrac{J_4}{i\pi^2} =
\sum_{k=1}^{4}\sum_{\substack{l=1\\k\neq l}}^{4}
\sum_{\substack{m=1\\m\neq l\\m\neq k}}^{4}
\frac{\Big(1-\delta(AC_{lk})\Big)\Big(1-\delta(B_{mlk})\Big)}
{AC_{lk}(B_{mlk}A_{nlk}-B_{nlk}A_{mlk})}  \times \\
&&\hspace{-0.5cm} 
\times \left[ \hspace{0.4cm}\int\limits_{0}^{\infty} dz \; G(z)
\Bigg\{
\lambda_{nmlk}^+ 
- f_{lk}^+g_{mlk}^+ \ln\left(\dfrac{z+ F_{nmlk}}{\beta_{mlk}}\right) 
- f_{lk}^+g_{mlk}^- \ln\left(-\dfrac{z+ F_{nmlk}}{\beta_{mlk}}\right) 
\right.
\nonumber\\
&& - (f_{lk}^-g_{mlk}^+ + f_{lk}^+g_{mlk}^+)
\ln\left( S^{(\sigma)}_{mlk} z- S^{(\sigma)}_{mlk} Z^{(1\sigma)}_{mlk} \right)
- (f_{lk}^-g_{mlk}^+ + f_{lk}^+g_{mlk}^+)
\ln\left(z-Z^{(2\sigma)}_{mlk} \right) \nonumber\\
&& + f_{lk}^+g_{mlk}^+ 
\ln\left( S^{(0)}_{mlk} z- S^{(0)}_{mlk} Z^{(10)}_{mlk} \right)
+ f_{lk}^+g_{mlk}^- 
\ln\left(-S^{(0)}_{mlk} z+ S^{(0)}_{mlk} Z^{(10)}_{mlk} \right) 
\nonumber\\
&& 
+ (f_{lk}^+g_{mlk}^+ + f_{lk}^+g_{mlk}^-)
\ln\left(z-Z^{(20)}_{mlk} \right) 
- (f_{lk}^+g_{mlk}^- - f_{lk}^-g_{mlk}^+)
\ln\left(P_{mlk}z + Q_{mlk}\right) 
\hspace{0.3cm}
\Bigg\}
\nonumber\\
&& \int\limits^{0}_{-\infty} dz \; G(z)
\Bigg\{
\lambda_{nmlk}^- 
+ f_{lk}^-g_{mlk}^- 
\ln\left( S^{(\sigma)}_{mlk} z- S^{(\sigma)}_{mlk} Z^{(1\sigma)}_{mlk} \right)
\nonumber\\
&& 
- (f_{lk}^-g_{mlk}^+ + f_{lk}^-g_{mlk}^-) 
\ln\left( S^{(0)}_{mlk} z- S^{(0)}_{mlk} Z^{(10)}_{mlk} \right)
+ f_{lk}^+g_{mlk}^- 
\ln\left(-S^{(\sigma)}_{mlk} z + S^{(\sigma)}_{mlk} Z^{(1\sigma)}_{mlk} \right)
\nonumber\\
&& 
- (f_{lk}^-g_{mlk}^+ + f_{lk}^-g_{mlk}^-) 
\ln\left(z-Z^{(20)}_{mlk} \right)
+ (f_{lk}^-g_{mlk}^- + f_{lk}^+g_{mlk}^-) 
\ln\left(z-Z^{(2\sigma)}_{mlk} \right)
\nonumber\\
&&
+ (f_{lk}^-g_{mlk}^- + f_{lk}^-g_{mlk}^+)
\ln\left(\dfrac{z+ F_{nmlk}}{\beta_{mlk}}\right) 
- (f_{lk}^+g_{mlk}^- - f_{lk}^-g_{mlk}^+)
\ln\left(P_{mlk}z + Q_{mlk}\right) 
\hspace{0.3cm}
\Bigg\}
\nonumber\\
&& + 2\pi\; i\;
\Big(f_{lk}^+g^-_{mlk} 
\theta[\text{Im}(Q_{mlk})]
+f^-_{lk}g_{mlk}^+\theta[-\text{Im}(Q_{mlk})]\Big)
\times \nonumber\\
&&\hspace{6cm}
\left. 
\times \int\limits^{\infty}_{-\infty} dz\; G(z)\theta 
\left(-A_{mlk}^0 z^2 - B_{mlk}^0 z 
- C_{mlk}^0 \right)\;\; \right].
\nonumber
\end{eqnarray}
Where $\lambda_{nmlk}^{\pm}$ are given by
\begin{eqnarray}
\lambda_{nmlk}^+ &=& 
(f^+_{lk}g_{mlk}^+ + f^-_{lk}g_{mlk}^+)
\ln\left(\dfrac{F_{nmlk}}{\beta_{mlk}}\right)    \\
&& \nonumber\\
&& + 2\pi \; i \; f^+_{lk}g_{mlk}^+
\theta[\text{Im}(S^{(0)}_{mlk}Z^{(10)}_{mlk})]
\theta[\text{Im}(Z^{(20)}_{mlk})]
\nonumber\\
&& \nonumber\\
&&- 2\pi \; i \; 
(f^-_{lk}g_{mlk}^+ + f^+_{lk}g_{mlk}^+) 
\theta[\text{Im}(S^{(\sigma)}_{mlk}Z^{(1\sigma)}_{mlk})]
\theta[\text{Im}(Z^{(2\sigma)}_{mlk})] \nonumber\\
&& \nonumber\\
&&+2\pi \; i \; f^+_{lk}g_{mlk}^-
\theta[\text{Im}(S^{(0)}_{mlk}Z^{(10)}_{mlk})]
\theta[-\text{Im}(Z^{(20)}_{mlk})], \nonumber \\
&& \nonumber\\
\lambda_{nmlk}^- &=& -f^+_{lk}g_{mlk}^-
\ln\left(-\dfrac{F_{nmlk}}{\beta_{mlk}}\right)  
- f^-_{lk}g_{mlk}^- 
\ln\left(\dfrac{F_{nmlk}}{\beta_{mlk}}\right)   \\
&& \nonumber\\
&&- 2\pi \; i \; 
(f^-_{lk}g_{mlk}^+ + f^-_{lk}g_{mlk}^-) 
\theta[\text{Im}(S^{(0)}_{mlk}Z^{(10)}_{mlk})]
\theta[\text{Im}(Z^{(20)}_{mlk})]
\nonumber\\
&& \nonumber\\
&& + 2\pi \; i \; f^-_{lk}g_{mlk}^-
\theta[\text{Im}(S^{(\sigma)}_{mlk}Z^{(1\sigma)}_{mlk})]
\theta[\text{Im}(Z^{(2\sigma)}_{mlk})]
\nonumber\\
&& \nonumber\\
&& - 2\pi \; i \; f^+_{lk}g_{mlk}^-
\theta[\text{Im}(S^{(\sigma)}_{mlk}Z^{(1\sigma)}_{mlk})]
\theta[-\text{Im}(Z^{(2\sigma)}_{mlk})]. \nonumber
\end{eqnarray}
We first emphasize that $\ln(P_{mlk}z+Q_{mlk})$ 
might have poles in negative real-axes in the real-mass case. 
However, in this case we have 
$f_{lk}^+=f_{lk}^-=1$ and $g_{mlk}^+ = g_{mlk}^- =1$. As a result, 
one checks that $(f^{+}_{lk}g_{mlk}^{-} -f_{lk}^-g_{mlk}^+) 
\ln(P_{mlk}z+Q_{mlk})=0$.  
Thus, we do not need to make $Q_{mlk}\longrightarrow 
Q_{mlk}+i\rho'$ as $F_{nmlk}$ case.  
Secondly, the $z$-integrals now are split into three basic integrals 
(\ref{basicsINTZ}). These integrals can be calculated in 
appendix C. 
\subsubsection{
$0< D_{mlk} \leq \left(\frac{A_{mlk}}{B_{mlk}}-\alpha_{lk}\right)^2$
and $\frac{A_{mlk}}{B_{mlk}}-\alpha_{lk} > 0$}   
In this case, $\beta_{mlk}>0$. The $t$-integrations are split
as follows
\begin{eqnarray}
J_4^{++} & \longrightarrow & \int\limits_{0}^{\infty} dz 
\int\limits_{0}^{\infty}dt + 
\int\limits^{0}_{-\infty} dz
\int\limits_{\sigma_{mlk}z}^{\infty}dt  \\
J_4^{+-} &\longrightarrow & \int\limits_{0}^{\infty} dz 
\int\limits_{\sigma_{mlk}\; z}^{0} dt, \\
J_4^{-+} &\longrightarrow &\int\limits^{0}_{-\infty} dz 
\int\limits_{0}^{\sigma_{mlk}\;z} dt,  \\
J_4^{--} &\longrightarrow & \int\limits_{-\infty}^0 dz 
\int\limits_{-\infty}^{0} dt + 
\int\limits^{\infty}_0 dz
\int\limits_{-\infty}^{\sigma_{mlk}z} dt. 
\end{eqnarray}
Taking the $t$-integrations, the result reads
\begin{eqnarray}
\label{zint-ln2}
&& \dfrac{J_4}{i\pi^2} =
\sum_{k=1}^{4}\sum_{\substack{l=1\\k\neq l}}^{4}
\sum_{\substack{m=1\\m\neq l\\m\neq k}}^{4}
\frac{\Big(1-\delta(AC_{lk})\Big)\Big(1-\delta(B_{mlk})\Big)}
{AC_{lk}(B_{mlk}A_{nlk}-B_{nlk}A_{mlk})}  \times \\
&&\times \left[ \hspace{0.4cm}\int\limits_{0}^{\infty} dz \; G(z) 
\left\{  f_{lk}^+ g_{mlk}^- 
\ln\left(\frac{F_{nmlk}}{\beta_{mlk}}\right)
+ f_{lk}^- g_{mlk}^- \ln\left(-\frac{F_{nmlk}}{\beta_{mlk}}\right)
\right.\right.\nonumber\\
&&\hspace{0.5cm}
- (f_{lk}^+ g_{mlk}^- + f_{lk}^+ g_{mlk}^+) 
\ln\left(\frac{z+F_{nmlk}}{\beta_{mlk}}\right) 
- f_{lk}^+ g_{mlk}^- 
\ln\left(\frac{ S(\sigma_{mlk}, z)}{ P_{mlk}z+ Q_{mlk} }\right)    
\nonumber\\
&&\hspace{0.5cm} 
- f_{lk}^- g_{mlk}^- 
\ln\left(-\frac{ S(\sigma_{mlk}, z)}{ P_{mlk}z+ Q_{mlk} }\right)    
+ \left.
(f_{lk}^+ g_{mlk}^+ + f_{lk}^+ g_{mlk}^- ) 
\ln\left(\frac{ S(\sigma_{mlk}=0, z)}
{ P_{mlk}z+ Q_{mlk} }\right) \;\right\} 
\nonumber\\
&& \hspace{0.4cm}
+\left. \int\limits^{0}_{-\infty} dz \; G(z)
\left\{  
-(f_{lk}^- g_{mlk}^+ + f_{lk}^+ g_{mlk}^+ )
\ln\left(\frac{F_{nmlk}}{\beta_{mlk}}\right)
\right.\right.\nonumber\\
&&\hspace{0.4cm}
+ f_{lk}^- g_{mlk}^+ \ln\left(\frac{z+F_{nmlk}}{\beta_{mlk}}\right)
+ f_{lk}^- g_{mlk}^- \ln\left(-\frac{z+F_{nmlk}}{\beta_{mlk}}\right)
\nonumber\\
&&\hspace{0.4cm}
-
f_{lk}^- g_{mlk}^+ \ln\left(\frac{ S(\sigma_{mlk}=0, z)}
{ P_{mlk}z+ Q_{mlk} }\right)
- 
f_{lk}^- g_{mlk}^- \ln\left(-\frac{ S(\sigma_{mlk}=0, z)}
{ P_{mlk}z+ Q_{mlk} }\right) 
\nonumber\\
&&\hspace{0.4cm}
\left. \left.
+(f_{lk}^+ g_{mlk}^+ + f_{lk}^- g_{mlk}^+ ) 
\ln\left(\frac{ S(\sigma_{mlk}, z)}{ P_{mlk}z+ Q_{mlk} }\right)
\hspace{1cm}
\right\}
\hspace{0.2cm}
\right. 
\Bigg].
\nonumber
\end{eqnarray}
\begin{figure}[h]
\begin{center}
\begin{pspicture}(-7, -2.5)(7, 2.5)
\psline{->}(-4,-2.5)(-4,2.5)
\psline{->}(4,-2.5)(4,2.5)
\pspolygon[fillstyle=hlines,
linestyle=none, linecolor=lightgray, hatchcolor=lightgray, hatchwidth=1.2pt,
hatchsep=1.8pt](6.5,-1.9)(4, 0)(6.5,0)
\pspolygon[fillstyle=hlines,
linestyle=none, linecolor=lightgray, hatchcolor=lightgray, hatchwidth=1.2pt,
hatchsep=1.8pt,hatchangle=135](1.5,0)(4, 0)(1.5,1.9)
\psline{->}(-7,0)(-1,0)
\psline{->}(1,0)(7,0)
\psline(1.5,1.9)(6.5,-1.9)
\rput(-3.8, 2.6){$t$}
\rput(4.2, 2.6){$t$}
\rput(-1.2,-0.2){$z$}
\rput(7,-0.2){$z$}
\rput(1.5, 2.2){$\sigma_{mlk} z$}
\rput(6.,-0.5){$J_4^{+-}$}
\rput(3.5,-1.2){$J_4^{--}$}
\rput(4.5,1.2 ){$J_4^{++}$}
\rput(2.,0.5)   {$J_4^{-+}$}
\rput(-2.5, 1.5){$J_4^{++}$}
\rput(-2.5,-1.5){$J_4^{+-}$}
\rput(-5.2,1.5 ){$J_4^{-+}$}
\rput(-5.2, -1.5){$J_4^{--}$}
\psline{->}(-0.3,0)(0.3,0)
\end{pspicture}
\caption{\label{beta2} The integration region.}
\end{center}
\end{figure}
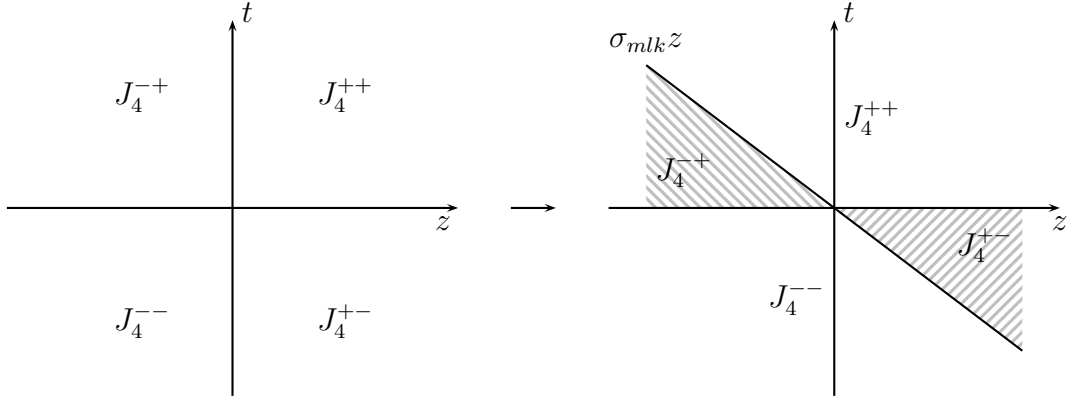
Using the formulas~(\ref{decomposea}, \ref{decomposeb}), 
$J_4$ is now presented as
\begin{eqnarray}
\label{zint-ln2-decom}
&&\hspace{-0.5cm}\dfrac{J_4}{i\pi^2} =
\sum_{k=1}^{4}\sum_{\substack{l=1\\k\neq l}}^{4}
\sum_{\substack{m=1\\m\neq l\\m\neq k}}^{4}
\frac{\Big(1-\delta(AC_{lk})\Big)\Big(1-\delta(B_{mlk})\Big)}
{AC_{lk}(B_{mlk}A_{nlk}-B_{nlk}A_{mlk})}  \times \\
&&\hspace{-0.5cm} 
\times \left[ \hspace{0.4cm}\int\limits_{0}^{\infty} dz \; G(z)
\Bigg\{
\gamma_{nmlk}^+ 
-(f_{lk}^+g_{mlk}^- + f_{lk}^+g_{mlk}^+)
 \ln\left(\dfrac{z+ F_{nmlk}}{\beta_{mlk}}\right) 
\right.
\nonumber\\
&& - f_{lk}^+g_{mlk}^-
\ln\left( S^{(\sigma)}_{mlk} z- 
S^{(\sigma)}_{mlk} Z^{(1\sigma)}_{mlk} \right)
- f_{lk}^-g_{mlk}^-
\ln\left(-S^{(\sigma)}_{mlk} z +
S^{(\sigma)}_{mlk} Z^{(1\sigma)}_{mlk} \right) \nonumber\\
&& + (f_{lk}^+g_{mlk}^+ + f_{lk}^+g_{mlk}^- )
\ln\left( S^{(0)}_{mlk} z- S^{(0)}_{mlk} Z^{(10)}_{mlk} \right)
\nonumber\\
&& 
+ (f_{lk}^+g_{mlk}^+ + f_{lk}^+g_{mlk}^-)
\ln\left(z-Z^{(20)}_{mlk} \right) 
- (f_{lk}^+g_{mlk}^- + f_{lk}^-g_{mlk}^-)
\ln\left(z-Z^{(2\sigma)}_{mlk} \right) \nonumber\\
&&  
+ (f_{lk}^-g_{mlk}^- - f_{lk}^+g_{mlk}^+)
\ln\left(P_{mlk}z + Q_{mlk}\right) 
\hspace{0.3cm}
\Bigg\}
\nonumber\\
&& \int\limits^{0}_{-\infty} dz \; G(z)
\Bigg\{
\gamma_{nmlk}^- 
+ f_{lk}^-g_{mlk}^+ \ln\left(\dfrac{z+ F_{nmlk}}{\beta_{mlk}}\right) 
+ f_{lk}^-g_{mlk}^- \ln\left(-\dfrac{z+ F_{nmlk}}{\beta_{mlk}}\right) 
\nonumber\\
&& 
- f_{lk}^-g_{mlk}^+ 
\ln\left( S^{(0)}_{mlk} z- S^{(0)}_{mlk} Z^{(10)}_{mlk} \right)
- f_{lk}^-g_{mlk}^- 
\ln\left(-S^{(0)}_{mlk} z + S^{(0)}_{mlk} Z^{(10)}_{mlk} \right)
\nonumber\\
&& + (f_{lk}^+g_{mlk}^+ + f_{lk}^-g_{mlk}^+) 
\ln\left( S^{(\sigma)}_{mlk} z - 
S^{(\sigma)}_{mlk} Z^{(1\sigma)}_{mlk} \right)
+ (f_{lk}^+g_{mlk}^+ + f_{lk}^-g_{mlk}^+) 
\ln\left(z-Z^{(2\sigma)}_{mlk} \right)
\nonumber\\
&& 
- (f_{lk}^-g_{mlk}^+ + f_{lk}^-g_{mlk}^-) 
\ln\left(z-Z^{(20)}_{mlk} \right)
+ (f_{lk}^-g_{mlk}^- - f_{lk}^+g_{mlk}^+)
\ln\left(P_{mlk}z + Q_{mlk}\right) 
\hspace{0.3cm}
\Bigg\}
\nonumber\\
&& - 2\pi\; i\;
\Big(f_{lk}^-g^-_{mlk} 
\theta[\text{Im}(Q_{mlk})]
+f^+_{lk}g_{mlk}^+\theta[-\text{Im}(Q_{mlk})]\Big)
\times \nonumber\\
&&\hspace{6cm}
\left. 
\times \int\limits^{\infty}_{-\infty} dz\; G(z)\theta 
\left(-A_{mlk}^0 z^2 -B_{mlk}^0 z- C_{mlk}^0 
\right)\;\; \right].
\nonumber
\end{eqnarray}
Where $\gamma_{nmlk}^{\pm}$ are given by
\begin{eqnarray}
\gamma_{nmlk}^+ &=& f^+_{lk}g_{mlk}^- 
\ln\left(\dfrac{F_{nmlk}}{\beta_{mlk}}\right)  
+f^-_{lk}g_{mlk}^- 
\ln\left(-\dfrac{F_{nmlk}}{\beta_{mlk}}\right)  
\\
&& \nonumber\\
&& - 2\pi \; i \; f^+_{lk}g_{mlk}^-
\theta[\text{Im}(S^{(\sigma)}_{mlk}Z^{(1\sigma)}_{mlk})]
\theta[\text{Im}(Z^{(2\sigma)}_{mlk})]
\nonumber\\
&& \nonumber\\
&& + 2\pi \; i \; f^-_{lk}g_{mlk}^-
\theta[\text{Im}(S^{(\sigma)}_{mlk}Z^{(1\sigma)}_{mlk})]
\theta[-\text{Im}(Z^{(2\sigma)}_{mlk})]
\nonumber\\
&& \nonumber\\
&& + 2\pi \; i \; (f^+_{lk}g_{mlk}^+ + f^+_{lk}g_{mlk}^-)
\theta[\text{Im}(S^{(0)}_{mlk}Z^{(10)}_{mlk})]
\theta[\text{Im}(Z^{(20)}_{mlk})]
\nonumber\\
&& \nonumber\\
\gamma_{nmlk}^- &=& -(f^-_{lk}g_{mlk}^+
+ f^+_{lk}g_{mlk}^+)
\ln\left(\dfrac{F_{nmlk}}{\beta_{mlk}}\right)  \\
&& \nonumber\\
&& - 2\pi \; i \; f^-_{lk}g_{mlk}^+
\theta[\text{Im}(S^{(0)}_{mlk}Z^{(10)}_{mlk})]
\theta[\text{Im}(Z^{(20)}_{mlk})]
\nonumber\\
&& \nonumber\\
&& + 2\pi \; i \; f^-_{lk}g_{mlk}^-
\theta[\text{Im}(S^{(0)}_{mlk}Z^{(10)}_{mlk})]
\theta[-\text{Im}(Z^{(20)}_{mlk})]
\nonumber\\
&& \nonumber\\
&&+ 2\pi \; i \; 
(f^+_{lk}g_{mlk}^+ + f^-_{lk}g_{mlk}^+) 
\theta[\text{Im}(S^{(\sigma)}_{mlk}Z^{(1\sigma)}_{mlk})]
\theta[\text{Im}(Z^{(2\sigma)}_{mlk})].
\nonumber
\end{eqnarray}

The $z$-integrals now are split into three basic 
integrals which are
\begin{eqnarray}
\label{basicsINTZ}
\mathcal{R}_1 &=& \int\limits_{0}^{\infty} \;G(z)dz, \nonumber 
\\
\mathcal{R}_2 &=& \int\limits_{0}^{\infty} 
\;\ln(M_{mlk} z + N_{mlk} )G(z)dz, \\
\mathcal{R}_3 &=& \int\limits_{-\infty}^{\infty} 
\theta\left(-A_{mlk}^0 z^2 -B_{mlk}^0 z - C_{mlk}^0\right) G(z)dz.
\nonumber
\end{eqnarray}
Where
\begin{eqnarray}
M_{mlk} &=& \pm 1, \pm \frac{1}{\beta_{mlk}}, 
\pm S_{mlk}^{\sigma}, \pm D_{mlk}, P_{mlk}, \quad 
\text{with} \quad M_{mlk} \in \mathbb{R}.
\end{eqnarray}
and 
\begin{eqnarray}
N_{mlk} &=& \pm\frac{F_{nmlk}}{\beta_{mlk}}, 
\pm S_{mlk}^{\sigma} Z_{mlk}^{1\sigma},\pm Z_{mlk}^{2\sigma}, 
\pm D_{mlk}Z_{mlk}^{10},\pm Z_{mlk}^{20}, 
Q_{mlk} \quad \text{with} \quad N_{mlk} \in \mathbb{C}.
\end{eqnarray}
The analytic results for the basic integrals can be 
found in appendix C. Since 
$\mathcal{R}_2$ in (\ref{basicsINTZ}) 
will contain two dilogarithm functions 
(see Appendix D), 
the $z$-integrations in $[0, \infty]$ contains the same 
dilogarithm functions of these in $[-\infty,0]$. 
Therefore, we only count the number of dilogarithm functions 
for the $z$-integrations from $0$ to $\infty$. We then find 
that $J_4$ contains $240$ dilogarithm functions instead of 
$108$ dilogarithm functions presented in~\cite{'tHooft:1978xw}. 
For the case of $D_{mlk}> 
\left(\frac{A_{mlk}}{B_{mlk}}-\alpha_{lk}\right)^2$, we refer 
\cite{khiemD0} for more detail.

For future prospects, we will consider the
evaluations for the tensor one-loop integrals by following this 
method. The tensor one-loop $N$-point integral with rank $M$ 
in the parallel and orthogonal space can be decomposed as
~\cite{Kreimer:1991wj,Kreimer:1992ps,Bauer:2001ig}
\begin{eqnarray}
 T^{N}_{\mu_1\mu_2...\mu_M}=(-1)^{\frac{p_{\perp}}{2}}
\dfrac{\Big(g_{\mu_1\mu_2}...g_{\mu_{p_{\perp}}-1}
g_{\mu_{p_{\perp}}}\Big)_{\text{sym}}}
{\mathcal{K}}\; T^{(p_0,p_1,...,p_{\perp})},
\end{eqnarray}
with 
\begin{eqnarray}
\mathcal{K} =
 \begin{cases}
          \prod\limits_{i=0}^{(p_{\perp}-2)/2}(D-J+2i),    
          & \text{if}   \;\; p_{\perp} \neq 0 ,\\
          1,     & \text{if}   \;\;p_{\perp}=0.
 \end{cases} 
\end{eqnarray}
Where $J$ is the number of parallel 
dimension (spanned by the external momenta). 
The tensor coefficients (form factors) are given
\begin{eqnarray}
T^{p_0,p_1...p_{\bot}}_N&=&
\frac{2\pi^{\frac{D-J}{2}}}{\Gamma(\frac{D-J}{2})}
\int\limits_{-\infty}^{\infty} d l_{0}dl_{1}... dl_{J-1}
\int\limits_{0}^{\infty} l_{\bot}^{D-J-1}d l_{\bot} 
\frac{l_0^{p_0}l_{1}^{p_1}...l_{J-1}^{p_{J-1}}
l_{\bot}^{p_{\bot}}}{P_1P_2\cdots P_N}.
\end{eqnarray}
Following tensor reduction for one-loop 
integrals developed in Refs.~\cite{Passarino:1978jh, Denner:2005nn}, 
the form factors will be obtained by contracting the Minkowski metric 
($g_{\mu\nu}$) and external momenta into the tensor integrals.
Solving a system of linear equations written in terms of the form factors, 
the Gram determinants appear in the denominators. When the determinants 
become very small, the reduction method will spoil numerical 
stability. The framework in this paper can be extended to calculate 
the form factors (or tensor one-loop integrals) directly. 
This will be devoted to future publication~\cite{khiemtensor}. 
It, therefore, opens a new approach to solve the problem 
analytically.
\section{Numerical checks and applications} 
This section is devoted to numerical checks and applications.
\subsection{Numerical checks} 
XLOOPS-GiNaC written in {\tt $C^{++}$} using the {\tt GiNaC} 
library~\cite{Bauer:2001ig} can handle scalar and tensor 
one-loop two-, three-point functions with real internal masses. 
In our previous work~\cite{khiemD0}, part of the program for 
evaluating scalar one-loop four-point functions has been implemented 
into {\tt $C^{++}$}, {\tt ONELOOP4PT.CPP}. It can evaluate numerically 
$J_4$ with real/complex masses in the case of at least one time-like 
external momentum.  Here, we implement
$J_1, J_2, J_3$ and new analytic results for $J_4$ 
in this report into {\tt Mathematica} (version $9$) 
package and {\tt FORTRAN} program. The program now can evaluate 
numerically scalar one-loop integrals with real/complex masses at 
general configurations of external momenta.

The syntax of these functions is as follows
\begin{eqnarray}
&&\text{ONELOOP1PT}(m^2, \rho), \\
&&\text{ONELOOP2PT}(p^2, m_1^2, m_2^2, \rho), \\
&&\text{ONELOOP3PT}(p_1^2,p_2^2, p_3^2, m_1^2, 
m_2^2, m_3^2,\rho), \\
&&\text{ONELOOP4PT}(p_1^2, p_2^2, p_3^2, p_4^2,
 s,t, m_1^2, m_2^2, m_3^2, m_4^2, \rho),
\end{eqnarray}
with $s = (p_1+p_2)^2$ and $t = (p_2+p_3)^2$.

In this section, we compared numerically the finite parts 
of $J_1, J_2, J_3, J_4$ 
in this report with {\tt LoopTools} (version $2.14$) 
in both real and complex internal masses.

In Table~\ref{j1table1}, the finite 
parts of $J_1$ are cross-checked with {\tt LoopTools}. One finds a 
perfect agreement between this work and {\tt LoopTools} in all cases. 
\begin{table}[H]
\begin{center}
\begin{tabular}{|c|l|} \hline\hline
$m^2$	
& {\tt This work}                      \\
& {\tt LoopTools}                      \\ \hline \hline
$100$  
&$-360.51701859880916$                 \\
&$-360.51701859880916$                 \\ \hline
$100 - 5\; i$  
&$-360.39207063012879+23.027932702631205\; i$ \\
&$-360.39207063012879+23.027932702631201\; i$ \\ \hline\hline
\end{tabular}
\caption{\label{j1table1} 
Comparing $J_1$ in this work with {\tt LoopTools}.
We set $\rho=10^{-15}.$}
\end{center}
\end{table}
In Table~\ref{j2Table1} (Table~\ref{j2Table2}), 
we compare the finite parts of $J_2$ in this paper
with {\tt LoopTools} in real (complex) internal masses 
respectively. We find that the results from this work and {\tt LoopTools} 
are in good agreement in both cases. In Table~\ref{j2Table1}, 
when $q^2$ is above threshold, or $q^2 > (m_1+m_2)^2$, the 
imaginary part of $J_2$ is non-zero.  While in the region 
(below threshold) $q^2< (m_1+m_2)^2$, the imaginary part of
$J_2$ is zero. It is understandable and the data demonstrates 
these points clearly in Table~\ref{j2Table1}.
\begin{table}[H]
\begin{center}
\begin{tabular}{|c|l|} \hline\hline
$q^2$	& {\tt This work}   \\
& {\tt LoopTools}                \\ \hline \hline
$100$  
&$-1.5232812522570536 + 1.5390597961942369\; i$ \\
&$-1.5232812522570538 + 1.5390597961942367\; i $\\ \hline
$5$  
&$-2.9052747784384294$ \\
&$-2.9052747784384296$\\ \hline
$-100$ 
&$-3.5554506187875863$         \\
&$-3.5554506187875869$\\ \hline\hline
\end{tabular}
\caption{\label{j2Table1} In case of 
$(m_1^2,m_2^2)=(10, 30)$, and $ \rho=10^{-15}.$}
\end{center}
\end{table}
\begin{table}[H]
\begin{center}
\begin{tabular}{|c|l|} \hline\hline
$q^2$	
& {\tt This work}  \\  & {\tt LoopTools}\\ \hline \hline
$100$  
&$-1.8207774022530800 + 1.9494059717871977\; i$ \\
&$-1.8207774022530803 + 1.9494059717871979\; i$\\ \hline
$-100$ 
&$-3.4154066334121885 + 0.0503812303761450\; i$         \\
&$-3.4154066334121893 + 0.0503812303761446\; i$\\ \hline\hline
\end{tabular}
\caption{\label{j2Table2} In case of
$(m_1^2,m_2^2)=(10-i, 20-2i)$ and $ \rho=10^{-15}.$}
\end{center}
\end{table}
In Tables~\ref{j3Table1}, \ref{j3Table2}, \ref{j3Table3}, 
we check the finite parts of $J_3$ with {\tt LoopTools} 
in real/complex internal masses. In Table~\ref{j3Table1}, 
the external momentum configurations are changed. 
While the internal masses are varied in Table~\ref{j3Table2}. 
The numerical checks for $J_3$ 
with complex internal masses are presented in 
Table~\ref{j3Table3}. We find a good agreement between 
the results from our work and {\tt LoopTools} in all cases. 
\begin{table}[H]
\begin{center}
\begin{tabular}{|c|l|} \hline\hline
$(p_1^2, p_2^2, p_3^2)$	
& {\tt This work}  \\
& {\tt LoopTools}                          
\\ \hline \hline
$(10, 200, 30)$  
&$-0.0030107310119513389 $ \\
&$-0.0030107310119513362 $\\ \hline
$(10, 200, -30)$  
&$-0.0029430832998831477$ \\
&$-0.0029430832998831465 $\\ \hline
$(10, -200, -30)$  
&$-0.0023628639374042822$ \\
&$-0.0023628639374042820 $\\ \hline
$(-10, -200, -30)$  
&$-0.0023432664930439692$ \\
&$-0.0023432664930439890 $\\ \hline\hline
\end{tabular}
\caption{\label{j3Table1} In case of 
$(m_1^2,m_2^2, m_3^2)=(100,200,300)$, 
and $\rho=10^{-15}.$}
\end{center}
\end{table}
In Table~\ref{j3Table2}, we observe the contribution of 
the imaginary part of $J_3$ in the region of $p_2^2 > (m_2+m_3)^2$. 
While in the region of $p_i^2 <(m_j+m_k)^2$ for $i \neq j \neq k$, 
the imaginary part of $J_3$ is zero. We observe these points clearly 
in this Table.    
\begin{table}[H]
\begin{center}
\begin{tabular}{|c|l|} \hline\hline
$m_1^2,m_2^2,m_3^2$	
& {\tt This work}  \\
& {\tt LoopTools}               \\ \hline \hline
$(0,0,0)$  
&$0.047432337875041175 - 0.055989586160261080\; i$ \\
&$0.047432337875041190 - 0.055989586160261083\; i$\\ \hline
$(10, 20, 30)$  
&$-0.016727876585043306 - 0.030266162066846561\;i$ \\
&$-0.016727876585043308 - 0.030266162066846562\;i$\\ \hline
$(50, 50 ,50)$  
&$-0.014058959690622575$ \\
&$-0.014058959690622567$\\ \hline\hline
\end{tabular}
\caption{\label{j3Table2} In case of 
$(p_1^2, p_2^2, p_3^2)=(10, 150, -30)$
and $\rho=10^{-15}.$}
\end{center}
\end{table}
In the Table~\ref{j3Table3}, cross-checking $J_3$ with complex 
internal masses in our work and {\tt LoopTools} are presented. 
We find three programs give perfect agreement results. 
\begin{table}[H]
\begin{center}
\begin{tabular}{|c|l|} \hline\hline
$(p_1^2, p_2^2, p_3^2)$	
& {\tt This work}                          \\
& {\tt LoopTools}                          \\  \hline \hline
$(100, 200, -300)$  
&$0.000302117943631926 - 0.022175834817012830\;i $ \\
&$0.000302117943631917 - 0.022175834817012831\;i $ \\ \hline
$(100, -200, -300)$  
&$-0.012274730929707654  - 0.005253630073729939\;i$ \\
&$-0.012274730929707652  - 0.005253630073729934\;i$ \\ \hline
$(-100, -2000, -300)$  
&$-0.003332905358821172 - 0.000146109020218081\;i$ \\
&$-0.003332905358821172 - 0.000146109020218078\;i$ \\ \hline\hline
\end{tabular}
\caption{\label{j3Table3} In case of 
$(m_1^2,m_2^2, m_3^2)=(10-3\;i,20 -4\; i,30-5\;i)$, 
and $\rho=10^{-15}.$}
\end{center}
\end{table}
Lastly, we compare $J_4$ in this work with {\tt LoopTools}.
One focuses on the case of all space-like of the kinematic 
invariants. In Table~\ref{j4Table1}, real internal masses 
are considered with varying the external momentum 
configurations. 
\begin{table}[H]
\begin{center}
\begin{tabular}{|c|l|} \hline\hline
$(p_1^2, p_2^2, p_3^2, p_4^2, s,t)$	
& {\tt (This work)}$\times 10^{-4}$             \\
& {\tt (LoopTools $)\times 10^{-4}$ }           \\ \hline \hline
$(-10, 60, 10, 90, 200, 5 )$  
&$-11.268763555152268 + 14.171199111711949\;i$  \\
&$-11.268763555152266 + 14.171199111711946\;i$  \\ \hline
$(-10, -60, -10, -90, 200, 5 )$  
&$2.1724718865815314 + 2.1645195914946021 \;i$  \\
&$2.1724718865815296 + 2.1645195914946018\;i$   \\ \hline
$(-10, -60, -10, -90, 200, -5 )$  
&$2.0640147463938176 + 2.1335567055149013\;i$   \\
&$2.0640147463938226 + 2.1335567055149037\;i$   \\ \hline
$(-10, -60, -25, -90, -20, -5 )$  
&$1.5152693508494305 + \mathcal{O}(10^{-21})\;i$ \\
&$1.5152693508494312 + \mathcal{O}(10^{-19}) \;i$ \\ \hline
$(17, 1, -1, -1, -17, -1 )$
&$3.1193150913091765 + \mathcal{O}(10^{-16})\;i$ \\
&$3.1193150913092383 + \mathcal{O}(10^{-18}) \;i$ \\ \hline\hline
\end{tabular}
\caption{\label{j4Table1} In case of 
$(m_1^2,m_2^2, m_3^2, m_4^2)=(10,20,30,40)$, 
and $\rho=10^{-15}$. There exists 
$AC_{lk} = \mathcal{O}(10^{-15})$ in 
the configuration of kinematic invariants in
the last line of this Table.}
\end{center}
\end{table}
In Table~\ref{j4Table2}, one examines all 
$p_1^2, p_2^2, p_3^2, p_4^2, s,t <0$ with the complex-mass 
cases. We change the value of $m_4^2$ in Table~\ref{j4Table2}. 
We find a good agreement between the results in this work 
and {\tt LoopsTools} in all cases. 
\begin{table}[H]
\begin{center}
\begin{tabular}{|c|l|} \hline\hline
$(m_1^2, m_2^2, m_3^2, m_4^2)$	
& {\tt (This work)}$\times 10^{-4}$             \\
& {\tt (LoopTools)$\times 10^{-4}$ }            \\  \hline \hline 
$(10-2\;i, 20, 30 -3\;i, 40  )$  
&$1.4577467887809371 + 0.13004659190213070\;i$ \\
&$1.4577467887809479 + 0.13004659190214078\;i$\\ \hline 
$(10-2\;i, 20, 30 -3\;i, 80  )$  
&$1.0235403166014101 + 0.0874193853007884\;i$ \\
&$1.0235403166014069 + 0.0874193853007809\;i$ \\ \hline 
$(10-2\;i, 20, 30 -3\;i, 120 -10\;i )$  
&$0.79634677966095624 + 0.1085714569206661\;i$ \\
&$0.79634677966095526 + 0.1085714569206717\;i$ \\ \hline \hline
\end{tabular}
\caption{\label{j4Table2} In case of 
$(p_1^2,p_2^2, p_3^2, p_4^2, s,t)
=(-10, -70, -20, -100, -15, -5)$, 
and $\rho=10^{-15}.$}
\end{center}
\end{table}
\subsection{Applications} 
In this subsection, we discuss on a typical example 
in which scalar box integrals may develop a leading 
Landau singularity~\cite{Boudjema:2008zn,Ninh:2008cz}. 
The process mentioned in this study is one-loop 
electroweak corrections to $gg \rightarrow b\bar{b}H$ at 
the LHC. We are not going to discuss the 
analytic structure of Landau singularities for one-loop 
integrals. One refers to Refs.~\cite{Boudjema:2008zn,Ninh:2008cz} 
for more detail of this topic. In this paper, we only evaluate
again triangle and box diagrams appear in this process 
which their numerical results have reported in 
Refs.~\cite{Boudjema:2008zn,Ninh:2008cz}. 

\begin{figure}[!ht]
\begin{center}$
\begin{array}{cc}
\includegraphics[width=3.3in,height=10cm,angle=0]
{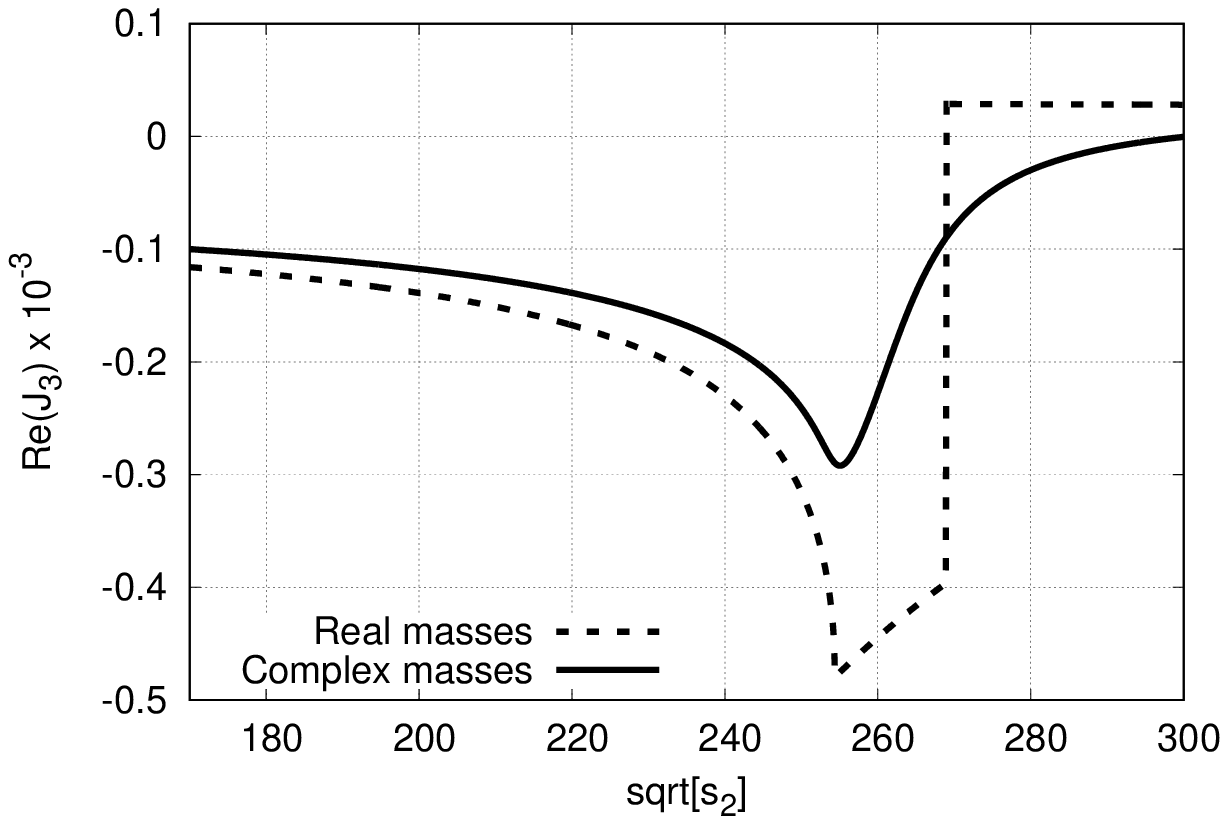} &
\includegraphics[width=3.3in,height=10cm,angle=0]
{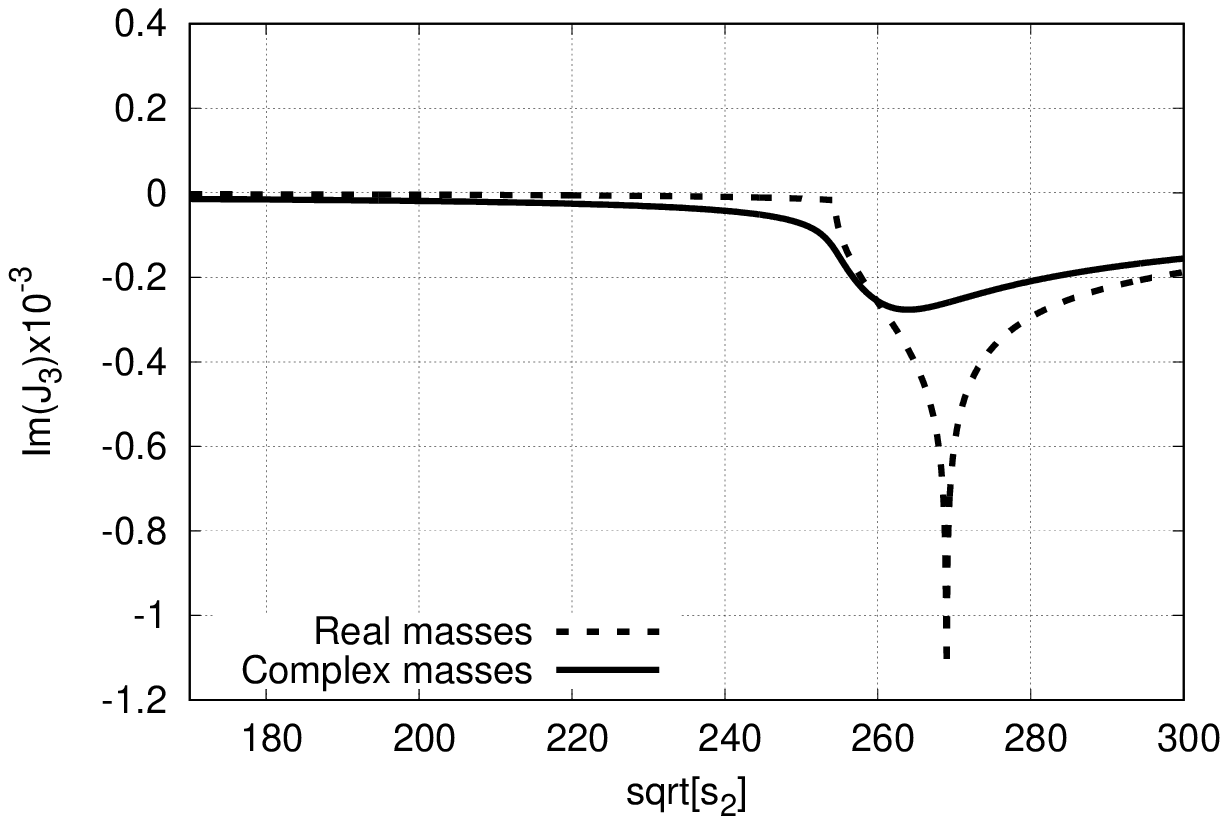}
\end{array}$
\end{center}
\caption{\label{triLANDAU} 
Impact of the width of the internal line particles 
(W boson, top) on the real and imaginary part of 
scalar one-loop three-point function in 
$gg \rightarrow b\bar{b}H$. }
\end{figure}

To regularize the Landau singularities, Feynman 
one-loop diagrams with complex internal masses
are interested. In this study, one sets the internal 
masses as
\begin{eqnarray}
\label{complexmasses1}
M_W^2 &=& M_{0W}^2 -i\; M_{0W}\Gamma_W, \\
\label{complexmasses2}
M_t^2 &=& M_{0t}^2 -i\; M_{0t}\Gamma_t. 
\end{eqnarray}
\begin{figure}[H]
\begin{center}
\includegraphics[width=4in,height=5cm,angle=0]{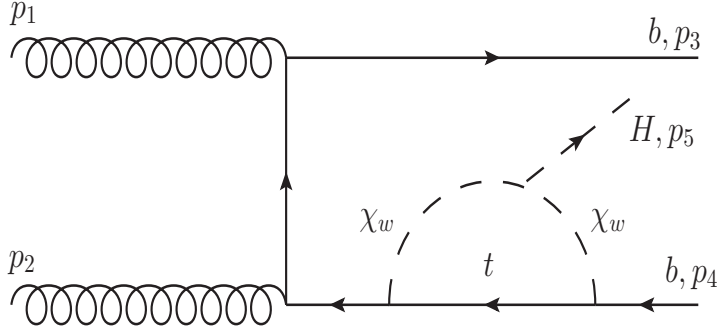}
\end{center}
\caption{\label{triLANDAUfig} Triangle diagram 
appears in the computation for one-loop corrections 
to the process $gg \rightarrow b\bar{b}H$.}
\end{figure}
The first diagram is mentioned as shown in Fig.~(\ref{triLANDAUfig}). 
We take the same input parameters in \cite{Ninh:2008cz}. In detail,
mass of Higgs boson is $160.08$ GeV\footnote{For checking
our program, we set the Higgs mass as taken 
in Refs.~\cite{Boudjema:2008zn,Ninh:2008cz}.}, mass of bottom quark 
is $M_b=4.18$ GeV. The internal masses are taken $M_{0W} = 80.3766$ GeV
with $\Gamma_W =2.1$ GeV and $M_{0t} =174.0$ GeV, $\Gamma_t =1.5$ GeV. 
In our program, scalar one-loop three-point function of this diagram 
is called 
\begin{eqnarray}
&&\text{ONELOOP3PT}(M_b^2,M_H^2,s_2=(p_4+p_5)^2, M_W^2, 
M_W^2, M_t^2,\rho), 
\end{eqnarray}
with $\rho =10^{-15}$. In Fig.~(\ref{triLANDAU}), $J_3$ 
is presented as a function of $\sqrt{s_2} \in [160.0,300.0]$ GeV. 
The case of real masses (complex masses) mean
that $\Gamma_W=\Gamma_t =0$ ($\Gamma_W=2.1$ GeV, $\Gamma_t =1.5$ GeV)
in Eqs.~(\ref{complexmasses1}, \ref{complexmasses2}) respectively.
At the threshold $M_H =2 M_{0W}$, one finds the peaks at 
$\sqrt{s_2}\sim M_{0t} + M_{0W}$ for both real (left panel) 
and imaginary part (right panel) of $J_3$, see solid lines 
in Fig.~(\ref{triLANDAU}). The dashed lines present
the impact of the width
of internal line particles (W boson, top)
on real and imaginary part of $J_3$. 

\begin{figure}[!ht]
\begin{center}
\includegraphics[width=4in,height=5.5cm,angle=0]
{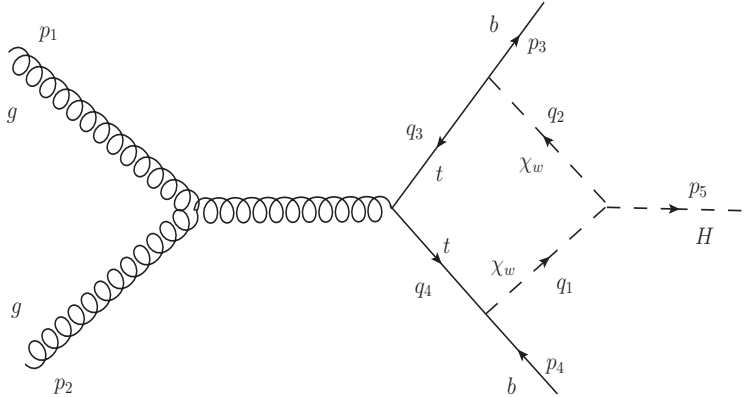}
\end{center}
\caption{\label{box-landau} 
A box diagram appear in one-loop corrections to 
process $gg \rightarrow b\bar{b}H$ which may contain 
a Landau singularity.}
\end{figure}

\begin{figure}[!ht]
\begin{center}$
\begin{array}{cc}
\includegraphics[width=3.3in,height=10cm,angle=0]
{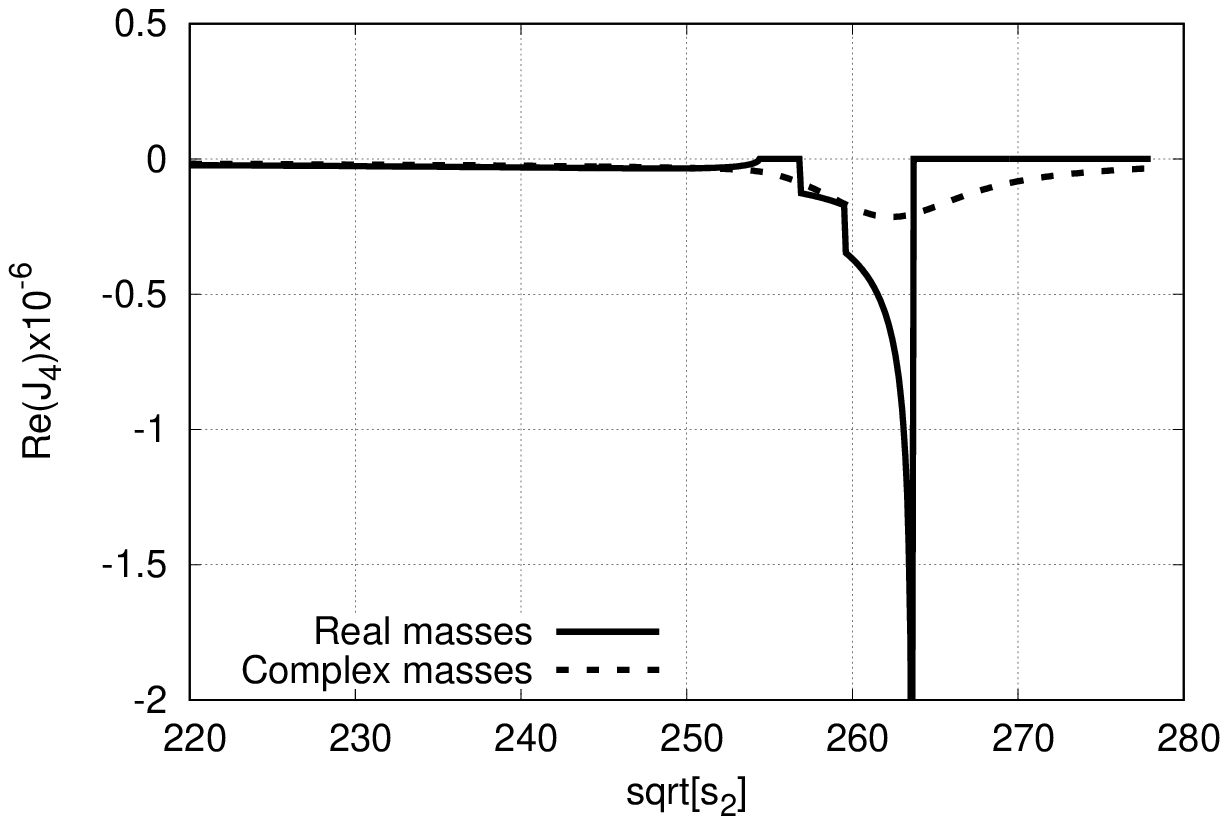} &
\includegraphics[width=3.3in,height=10cm,angle=0]
{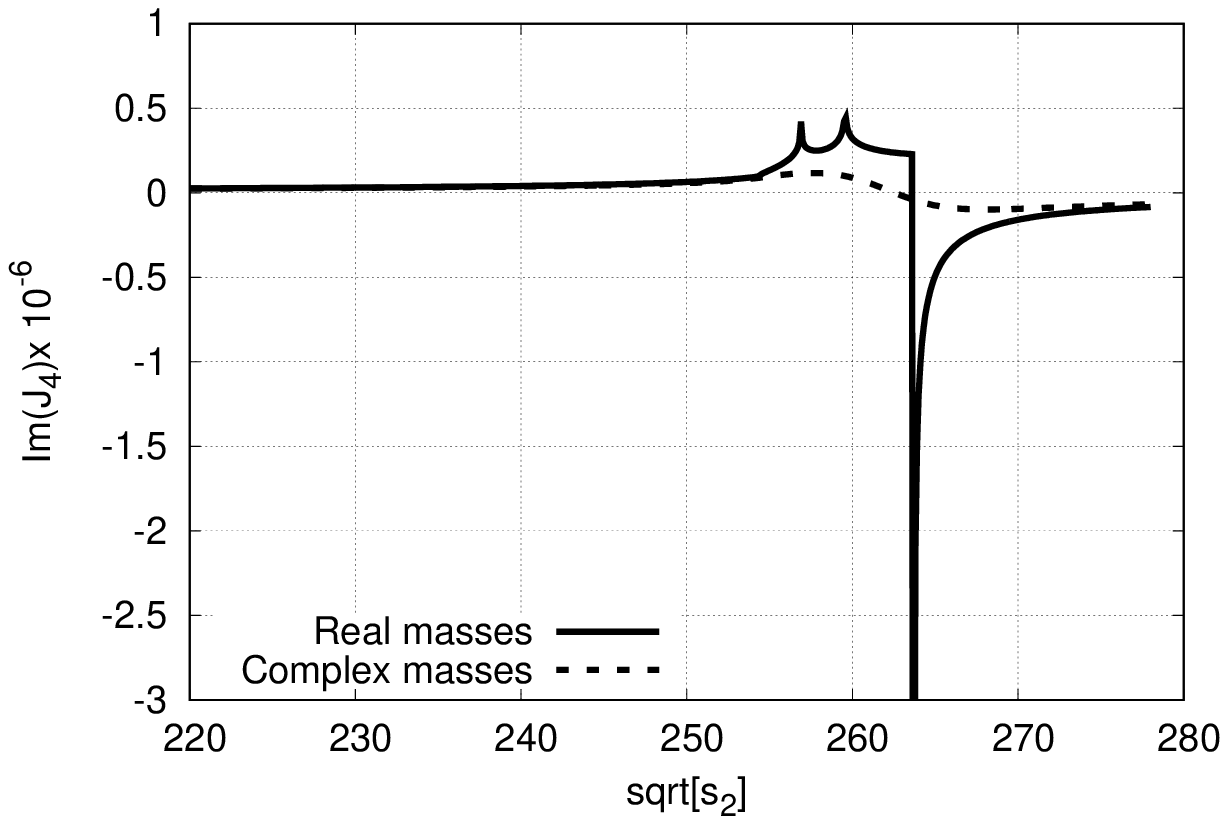}
\end{array}$
\end{center}
\caption{\label{res-boxlandau} Impact of the width 
of the internal line particles (W boson, top) 
on the real and imaginary part of scalar one-loop 
four-point function in $gg \rightarrow b\bar{b}H$.}
\end{figure}

We next consider scalar one-loop four-point function
which develops a leading Landau singularity as indicated
in Fig.~(\ref{box-landau}). For this diagram, our input 
parameters are as follows
\begin{eqnarray}
M_H &=& 165.2\;\text{GeV}, 
\quad M_b = 4.18\;\text{GeV}, 
\quad \sqrt{s} = 353.0\; \text{GeV},
\quad \sqrt{s_1} = 271.06\; \text{GeV}, \nonumber \\
M_{0W} &=& 80.3766\;\text{GeV}, \quad \Gamma_W = 2.1\; \text{GeV}\quad 
\text{and} \quad
M_{0t} = 174.0\;\text{GeV}, \quad \Gamma_t = 1.5\; \text{GeV}.
\nonumber
\end{eqnarray} 
In our program scalar box integral is evaluated by calling
\begin{eqnarray}
&&\text{ONELOOP4PT}(M_H^2, m_b^2, s, m_{\bar{b}}^2,
 s_1, s_2, M_W^2, M_t^2, M_t^2, M_W^2, \rho),
\end{eqnarray}
with $s_1 = (p_3+p_5)^2$ and $s_2 = (p_4+p_5)^2$ 
as shown in Fig.~(\ref{box-landau}). In this study, 
we set $\rho=10^{-15}$. 
In Figs.~(\ref{res-boxlandau}), $J_4$ is a function of
$\sqrt{s_2}$ which it varies from $220.0$ GeV to $280.0$ GeV. 
One find a leading Landau singularity at $\sqrt{s_2}
\sim 264.0$ GeV as indicated as solid lines in these 
figures. Again, the dashed lines in these figures 
indicate the effect of the width
of internal line particles (W boson, top)
on real and imaginary part of $J_4$ in
$gg \rightarrow b\bar{b}H$. 
All above data shown in this section confirms again 
the results in \cite{Boudjema:2008zn,Ninh:2008cz}. 
\section{Conclusions}
\noindent
In this paper, we studied systematically scalar 
one-loop two-, three-, four-point integrals with real/complex 
internal masses. The calculations are considered at general case 
of external momentum assignments. In the 
numerical checks, we compared this work with {\tt LoopTools}. 
We found a good agreement between the results generated 
from this work and those from {\tt LoopTools}. 
We also used this work to evaluate several Feynman diagrams
which appear in $gg\rightarrow b\bar{b}H$ at the LHC. It 
is shown that our program are applied well in computing 
one-loop corrections (in both Real- and Complex-Mass Schemes) 
in real processes at the colliders. 

Moreover, this work provides a framework which can be extended 
to calculate directly tensor one-loop integrals. 
It may provide a analytic solution for 
the inverse Gram determinant problem. In future work, we will 
proceed to the evaluation for tensor one-loop 
integrals~\cite{khiemtensor}.
\\

\noindent
{\bf Acknowledgment:}~
This work is funded by Vietnam's National
Foundation for Science and  Technology  Development 
(NAFOSTED) under the grant number $103.01$-$2016.33$.
The authors are grateful to  Dr. D. H. Son,
N. D. Dieu for fruitful discussions and
their contributions to this work. We also would like 
to thank P. Q. Trung, N. L. D. Thinh and C. T. Nhan 
for help reading the manuscript. 
\section*{Appendix A: One-loop one-point functions}
We consider the simplest case which is scalar one-loop
one-point functions. Their Feynman integrals
are defined as
\begin{eqnarray}
\label{j1feynappendix}
J_1(m^2) &=& \int d^Dl\;\dfrac{1}{l^2 - m^2 + i\rho}.
\end{eqnarray}
Here, $l$ is loop-momentum in space-time dimension $D$
and $l^2 = l_0^2 -l_1^2-\cdots -l^2_{D-1}$.
In the Complex-Mass Scheme, the square of internal mass 
has the form
\begin{eqnarray}
m^2 = m_0^2 - i\; m_0 \Gamma,
\end{eqnarray}
which $\Gamma$ is the decay width of the unstable particle. 
The Feynman prescription is $i\rho$. In this paper, 
dimensional regularization is performed within space-time 
dimension $D = 4-2\varepsilon$.

Performing Wick rotation 
as $l_0 \rightarrow il_0$, one can convert $J_1$ 
from Minkowski into Euclid space. The result reads
\begin{eqnarray}
J_1 &=&-i\int d^Dl_E \;\dfrac{1}{l^2_E + m^2 - i\rho}
= -\dfrac{2\pi^{\frac{D}{2} }\; i}{\Gamma(\frac{D}{2})} 
\int\limits_0^{\infty} 
dl_E\;\dfrac{l_E^{D-1} }{l^2_E + m^2 - i\rho}
\end{eqnarray}
with $l_E^2 = l_0^2 +l_1^2+\cdots+l^2_{D-1}$.
Setting $z = l_E^2$, we presented $J_1$ in terms of 
$\mathcal{R}$-function as follows
\begin{eqnarray}
J_1 &=& -\dfrac{\pi^{\frac{D}{2} }\; i}
{\Gamma\left(\frac{D}{2}\right)} 
\int\limits_0^{\infty}\; dz 
\; \dfrac{z^{\frac{D}{2}-1}}{z+m^2 -i\rho}
= -i \pi^{\frac{D}{2} }\Gamma\left(\frac{2-D}{2}\right) 
\mathcal{R}_{\frac{D-2}{2}} (1; m^2-i\rho). 
\end{eqnarray}

Applying the formula (\ref{Rz1}) in appendix B, 
we obtained
\begin{eqnarray}
\label{j1res}
J_1(m^2) &=& -i\pi^{\frac{D}{2} }
\Gamma\left(\frac{2-D}{2}\right)(m^2-i\rho)^{\frac{D-2}{2}}.
\end{eqnarray}
This result has been presented in many papers, such as
\cite{'tHooft:1978xw,FranzPHD}, etc. As
$m^2 \in \mathbb{R^+}$ or $m^2 \in \mathbb{C}$, 
one can take $\rho \rightarrow 0$ in Eq.~(\ref{j1res}).
In space-time dimension $D=4-2\varepsilon$, the 
$J_1(m^2)$ was expanded in terms of $\varepsilon$ up to 
$\varepsilon^0$. The expansion gives
\begin{eqnarray}
\label{j1expansionqm0}
\dfrac{J_1}{i\pi^2}  = m^2\left(
\frac{1}{\varepsilon} -\gamma_E - \ln(\pi) 
+1-\ln m^2\right).  
\end{eqnarray}
This formula is valid not only for real masses 
but also for complex masses.
\section*{Appendix B} 
In Appendix B, we present here several useful formulas. 
The first ones that we mention are 
\begin{eqnarray}
\label{logdecompose}
\ln(a b)&=&\ln (a)+\ln (b)+\eta(a,b),\\
\ln\left( \frac{a}{b} \right)&=&
\ln (a)-\ln (b)+\eta\left( a,\frac{1}{b} \right).
\end{eqnarray}
The $\eta$-function is defined as
\begin{eqnarray}
\eta(a,  b)&=&
2\pi i\Bigg\{\theta\Big[-\text{Im}(a)\Big]
\theta\Big[-\text{Im}(b)\Big]\theta\Big[\text{Im}(ab)\Big]
-\theta\Big[\text{Im}(a)\Big]\theta
\Big[\text{Im}(b)\Big]\theta\Big[-\text{Im}(ab)\Big]\Bigg\}. 
\nonumber\\
\end{eqnarray}
We consider several special cases
\begin{itemize}
\item If $\text{Im}(a) \text{Im}(b)<0$, one confirms that $ \eta(a,  b) =0$.
\item If $\text{Im}(a) \text{Im}(b)>0$, one has $\eta\left(a,1/b\right) =0$.
\end{itemize}

Furthermore, one can derive the following relation
\begin{eqnarray}
\label{abrhogeneral}
 (ab)^{\alpha} = e^{\alpha \ln(ab)} = e^{\alpha \eta(a,b)} a^{\alpha}b^{\alpha}.
\end{eqnarray}
In the case of $a, b \in \mathbb{R}$, one has the following relation
\begin{eqnarray}
\label{abrho}
(ab \pm i\rho)^{\alpha} = (a \pm i\rho')^{\alpha}
\left( b \pm \frac{i\rho}{a} \right)^{\alpha}
\quad \text{for} \quad a,b \in \mathbb{R}.
\end{eqnarray}
Here, $\rho'$ has the same sign with $\rho$.

The Beta function is also given by
\begin{eqnarray}
\label{beta}
 \mathcal{B}(x,y) = \dfrac{\Gamma(x)\Gamma(y)}{\Gamma(x+y)}.
\end{eqnarray}
In this paper, we also use this basic integral
\begin{eqnarray}
\label{lbot}
\int\limits_0^{\infty} \dfrac{z^{\alpha -1}}{Z_k +z} dz
&=& \mathcal{B}(1-\alpha, \alpha) (Z_k)^{\alpha -1},
\end{eqnarray}
provided that Arg$(Z_k)<\pi$.
\section*{Appendix C} 
The $\mathcal{R}$-function~\cite{B.C.carlson} is defined as
\begin{eqnarray}
\label{rintegral}
 &&\int\limits_r^{\infty} (x-r)^{\alpha-1}
 \prod\limits_{i=1}^k(z_i+w_ix)^{-b_i}dx \nonumber\\
 &&\hspace{1cm}=
 \mathcal{B}(\beta -\alpha, \alpha)\mathcal{R}_{\alpha -\beta}\left(
 b_1, \cdots, b_k, r+ \frac{z_1}{w_1}, \cdots, r+ \frac{z_k}{w_k}
 \right)\prod\limits_{i=1}^k\; w_i^{-b_i},
\end{eqnarray}
with $\beta = \sum\limits_{i=1}^kb_i$.
We also used several useful relations for 
$\mathcal{R}$-function. The following formulas 
are collected from \cite{carlsonFD}.
\begin{eqnarray}
\label{Rzz}
\mathcal{R}_{-a} (b_1, b_2, \cdots, b_N; z, z, \cdots, z) &=& z^{-a}, \\
\label{Rz1}
\mathcal{R}_{-a} (b;z)  &=& z^{-a}, 
\end{eqnarray}
and
\begin{eqnarray}
\label{Rz0}
&& \mathcal{B}(a,a') 
\mathcal{R}_{-a} (b_1, b_2, \cdots, b_N; z_1, z_2, \cdots, z_k, \vec{0}_{N-k}) 
\nonumber\\
&&
\hspace{0.5cm}
= \mathcal{B}(a,a'-\sum\limits_{j=k+1}^{N}b_j) 
\mathcal{R}_{-a} (b_1, b_2, \cdots, b_k; z_1, z_2, \cdots, z_k),    
\quad a -a' = \sum\limits_{j=1}^{N}b_j, \\
\label{R2F}
&&\mathcal{R}_{-a} (b_1, b_2, \cdots, b_N; z_1, z_2, \cdots, z_N) = \nonumber\\
&& 
\hspace{0.5cm}
= F_D\left(a; b_1, b_2, \cdots, b_N; c = \sum\limits_{j=1}^{N}b_j; 1-z_1, 1-z_2,\cdots, 
1-z_N\right),  
\end{eqnarray}
provided that $|1-z_i|<1$, Arg$(z_i)<\pi$ for $i=1,2,\cdots, N$. Where $F_D$
is Lauricella hypergeometric function. Additionally, reduction formula 
for $F_D$ is presented
\begin{eqnarray}
\label{FN2FN1}
&& F_D\left(a; b_1, b_2, \cdots, b_N;c; z_1, z_2,\cdots, z_N\right) =\nonumber\\
&& \hspace{0.5cm}
= (1-z_N)^{-a} F_D\left(a; b_1, b_2, \cdots, b_{N-1}, c; 
\frac{z_1-z_N}{1-z_N}, \frac{z_2-z_N}{1-z_N},\cdots, 
\frac{z_{N-1}-z_N}{1-z_N}\right).
\end{eqnarray}
Expansion for R-functions in terms of 
$\varepsilon$ are presented 
in the following paragraphs.
\begin{enumerate}
\item \underline{$\mathcal{R}_{-\varepsilon}
\left( -\frac{1}{2}+\varepsilon, 1, z_1, z_2\right)$:}
Expansion for this function was  presented in 
Refs.~\cite{Kreimer:1991wj, FranzPHD}. 
\begin{eqnarray}
\label{expR1}
\mathcal{R}_{-\varepsilon}
\left( -\frac{1}{2}+\varepsilon, 1, z_1, z_2\right)
&=& \frac{1}{2} (z_1)^{-\varepsilon} 
\left(\dfrac{z_1}{z_2}\right)^{-\varepsilon} 
\left(u_2 u_1^{2\varepsilon} + u_1 u_2^{2\varepsilon}\right), 
\end{eqnarray}
with $u_{1,2} = 1\pm \sqrt{1-\dfrac{z_1}{z_2}}$.
\item \underline{$\mathcal{R}_{-2\varepsilon}\left(1, \varepsilon, 
\varepsilon,x, y, z\right)$:}
Expansion for this function was presented in 
Refs.~\cite{Kreimer:1992ps, FranzPHD}. 
\begin{eqnarray}
\label{expR2}
&& \hspace{-0.5cm}
\mathcal{R}_{-2\varepsilon}\left(\varepsilon, \varepsilon,1;
x, y, z\right) =
1- 2\varepsilon \ln z + 2\varepsilon^2\Bigg\{
\text{Li}_2\left(1-\frac{x}{z}\right)
+\ln\left(1-\frac{x}{z}\right)\eta\left(x, \frac{1}{z}\right) \nonumber\\
&&\hspace{0.5cm}
+ \ln(z)\Big[\eta\left(x-z, \frac{1}{1-z}\right) 
-\eta\left(x-z, -\frac{1}{z}\right) \Big] + (\ln z)^2 
+ \text{Li}_2\left(1-\frac{y}{z}\right)   \\
&& \hspace{0.5cm}
+\ln\left(1-\frac{y}{z}\right)\eta\left(y, \frac{1}{z}\right) 
+ \ln(z)\Big[\eta\left(y-z, \frac{1}{1-z}\right) 
-\eta\left(y-z, -\frac{1}{z}\right) \Big]  \Bigg\}. \nonumber
\end{eqnarray}
\item \underline{ $\mathcal{R}_{1-2\varepsilon}\left(\varepsilon, 
\varepsilon,x, y\right)$:}
Expansion for this function was presented in 
Refs.~\cite{FranzPHD}.
\begin{eqnarray}
\label{expR3}
\mathcal{R}_{1-2\varepsilon}\left( \varepsilon, \varepsilon; 
x, y\right) &=&
\frac{1}{2}(x+y) - [y\ln(y)+x\ln(x)]\; \varepsilon.
\end{eqnarray}
\item \underline{$\mathcal{R}_{-\varepsilon}\left(1, \varepsilon;
 y, z\right)$:}
Expansion for this function was presented in 
Refs.~\cite{FranzPHD}.
\begin{eqnarray}
\label{expR4}
\mathcal{R}_{-\varepsilon}\left(\varepsilon, \varepsilon; y, z\right) &=&
1- \varepsilon \ln y + \varepsilon^2\Bigg\{
\text{Li}_2\left(1-\frac{z}{y}\right) + \frac{1}{2} (\ln y)^2 
\\
&& \hspace{-1.5cm}
+ \ln y \Big[\eta\left(z-y, \frac{1}{1-y}\right) 
-\eta\left(z-y, -\frac{1}{y}\right) \Big]  
+ \ln\left(1-\frac{z}{y}\right)\; \eta\left(z, \frac{1}{y}\right) 
 \Bigg\}. \nonumber
\end{eqnarray}
\end{enumerate}

Useful relations for $\mathcal{R}$-functions are also 
listed in this appendix. The formulas shown here are collected
from Ref.~\cite{B.C.carlson}. We denote that $b$, $z$ and $e_i$ 
are $k$-tuple
\begin{eqnarray}
 b &=& (b_1, b_2, \cdots, b_k), \nonumber\\
 z &=& (z_1, z_2, \cdots, z_k), \nonumber\\
 e_i &=&(0,0,\cdots, 1,0, \cdots, 0) \quad
 \text{where the $1$ is located at the $i$th entry}. 
 \nonumber
\end{eqnarray}
They are presented as follows
\begin{eqnarray}
\mathcal{R}_t(b, z) 
&=& \sum\limits_{i=1}^k 
\frac{b_i}{\beta} \mathcal{R}_t(b+e_i,z), \label{relation1} 
\\
\mathcal{R}_{t+1}(b, z) 
&=& \sum\limits_{i=1}^k 
\frac{b_i}{\beta}\; z_i\;\mathcal{R}_t(b+e_i,z), 
\label{relation2} \\
\beta \mathcal{R}_{t}(b, z) 
&=& (\beta+t)\mathcal{R}_{t}(b+e_i, z)  
- t z_i\mathcal{R}_{t-1}(b+e_i, z), 
\label{relation3} \\
\partial_{z_i} \mathcal{R}_t(b, z)
&=& \frac{b_i}{\beta}t \mathcal{R}_{t-1}(b+e_i,z), 
\label{relation4}\\
\mathcal{R}_t(b, z) 
&=& \prod\limits_{i=1}^k z_i^{-b_i}
\mathcal{R}_{-\beta-t}(b+e_i,z^{-1}), 
\quad \text{Euler's transformation} 
\label{relation5} \\
\mathcal{R}_t(b, \lambda z) 
&=& \lambda^t \mathcal{R}_t(b, z) 
\quad \text{scaling law}. 
\label{relation6} 
\end{eqnarray}
\section*{Appendix D} 
We consider three basic integrals in the 
following paragraphs. Here, all the shown formulas 
are taken from \cite{khiemD0}.
\begin{enumerate}
\item \underline{Basic integral $I$:}
 The basics integral $I$ is defined as
\begin{eqnarray}
\label{R1master}
\mathcal{R}_1(x,y) 
= \int\limits_{0}^{\infty}\frac{1}{(z+x)(z+y)}dz 
= \frac{\ln(x)-\ln(y)}{x-y},
\end{eqnarray}
with $x,y\in \mathbb{C}$.\\
\item \underline{Basic integral $II$:} 
The basic integral $II$ is
\begin{eqnarray}
\mathcal{R}_2(r,x,y) &=&  
\int\limits_{0}^{\infty}\frac{\ln(1+rz)}{(z+x)(z+y)}dz  
=-\frac{1}{x-y}\Big[ \text{Li}_2(1-rx)-\text{Li}_2(1-ry) \Big]
\nonumber\\
&&-\frac{1}{x-y} 
\Big[  \eta(x,r)\ln(1-rx)  -\eta(y,r)\ln(1-ry)     \Big],
\end{eqnarray}
with $r , x,y\in \mathbb{C}$.
\item \underline{Basic integral $III$:} 
The basic integral $III$ has the form of
\begin{eqnarray}
 \mathcal{R}_3 
 &=& 
\int\limits_{-\infty}^{\infty}
\theta\left(-A_{mlk}^0 z^2 -B_{mlk}^0 z - C_{mlk}^0\right) G(z)dz,
\end{eqnarray}

If $-A_{mlk}^0 z^2 -B_{mlk}^0 z - C_{mlk}^0 \geq 0$
in the region $\Omega \subset \mathbb{R}$, one then has
\begin{eqnarray}
\mathcal{R}_3  
&=& \int\limits_{\Omega} dz\; G(z).\nonumber
\end{eqnarray}
The integral in right-hand side of this equation can 
be reduced to basic integral $I$.
\end{enumerate}

\section*{Appendix E} 
The Gauss hypergeometric series are given (see Eq.~($1.1.1.4$) 
in Ref.~\cite{Slater})
\begin{eqnarray}
\label{gauss-series}
\Fh21\Fz{a,b}{c}{z} 
= \sum\limits_{n=0}^{\infty} \dfrac{(a)_n (b)_n}{(c)_n} \frac{z^n}{n!},
\end{eqnarray}
provided that $|z|<1$. Here, the Pochhammer symbol, 
\begin{eqnarray} 
(a)_n =\dfrac{\Gamma(a+n)}{\Gamma(a)},
\end{eqnarray} 
is taken into account.

The integral representation for Gauss hypergeometric functions is
(see Eq.~(1.6.6) in Ref.~\cite{Slater})
\begin{eqnarray}
\label{gauss-int}
 \Fh21\Fz{a,b}{c}{z} 
= \dfrac{\Gamma(c)}{\Gamma(b)\Gamma(c-b)} \int\limits_0^1 du
  \; u^{b-1} (1-u)^{c-b-1} (1-zu)^{-a},
\end{eqnarray}
provided that  $|z|<1$ and Re$(c)>$Re$(b)>0$.

The series of Appell $F_1$ functions are  given (see Eq.~(8.13) 
in Ref.~\cite{Slater})
\begin{eqnarray}
\label{appell-series}
F_1(a; b, b'; c; x, y) 
= \sum\limits_{m=0}^{\infty}\sum\limits_{n=0}^{\infty}
  \dfrac{(a)_{m+n} (b)_m (b')_n}{(c)_{m+n}\; m! n!} x^m y^n,
\end{eqnarray}
provided that $|x|<1$ and $|y|<1$.

The single integral representation for $F_1$ is 
(see Eq.~(8.25) in Ref.~\cite{Slater})
\begin{eqnarray}
\label{appell-int}
F_1(a; b, b'; c; x, y) 
=  \dfrac{\Gamma(c)}{\Gamma(c-a)\Gamma(a)} \int\limits_0^1 du\;
   u^{a-1} (1-u)^{c-a-1} (1-xu)^{-b}(1-yu)^{-b'},
\end{eqnarray}
provided that Re$(c)$ $>$ Re$(a)>0$ and $|x|<1$, $|y|<1$.
 
Analytic continuation for Gauss hypergeometric functions are 
given \cite{Slater}
\begin{eqnarray}
\Fh21\Fz{a,b}{c}{z} 
&=& \frac{\Gamma(c) \Gamma(b-a)}{\Gamma(b) \Gamma(c-a)} (-z)^{-a} 
\Fh21\Fz{a,1-c+a}{1-b+a}{\frac{1}{z} } \nonumber \\
&&+ \frac{\Gamma(c) \Gamma(a-b)}{\Gamma(a) \Gamma(c-b)} (-z)^{-b} 
\Fh21\Fz{b,1-c+b}{1-a+b}{\frac{1}{z} }\label{Fz21z}.
\end{eqnarray}

We have reduction formula for Appell $F_1$ 
function~\cite{Slater}
\begin{eqnarray}
\label{F12GAUSS}
F_1(a; b_1, b_2; c=b_1+b_2; x, y) 
&=& (1-x)^{-a} \Fh21\Fz{a, b_2}{c}{\frac{y-x}{1-x}} \\
&=& (1-y)^{-a} \Fh21\Fz{a, b_1}{c}{\frac{x-y}{1-y}}.
\end{eqnarray}
Moreover,
\begin{eqnarray}
F_1\Big(a;b,b';c;x,y\Big)=(1-x)^{-b}(1-y)^{-b'}
F_1\Big(c-a;b,b';c;\frac x{x-1},\frac y{y-1}\Big).
\end{eqnarray}
\section*{Appendix F} 
Scalar one-loop $N$-point Feynman integrals 
are defined
\begin{eqnarray}
\label{npoint}
J_{N}(D; \{p_ip_j\}, \{m_i^2\}) =
\int d^Dl \dfrac{1}
{\mathcal{P}_1\mathcal{P}_2\dots\mathcal{P}_N}.
\end{eqnarray}
Where the inverse Feynman propagators are given
\begin{eqnarray}
\label{FeynProp}
\mathcal{P}_i &=& (l+q_i)^2-m_i^2+i\rho, \quad 
\text{for $i=1,2,\cdots, N$.}
\end{eqnarray}
Here $p_i$ and $m_i$ for $i=1,2,\cdots, N$ are the external momenta 
and the internal masses respectively. We have used the same convention 
with~\cite{Denner:2005nn}. In particular, the momenta of internal lines 
are $q_1 =p_1, q_2 =p_1+p_2, \cdots, q_i = \sum\limits_{j=1}^{i}p_j,$ 
and $q_N=\sum\limits_{i=1}^{N}p_i=0$ thanks to momentum conservation. 

The determinants of Cayley and Gram matrices
are defined as follows
\begin{eqnarray}
\hspace{0cm} S_{N} =  \left|
\begin{array}{cccc}
S_{11}  & S_{12}  &\ldots & S_{1N} \\
S_{12}  & S_{22}  &\ldots & S_{2N} \\
\vdots  & \vdots  &\ddots & \vdots \\
S_{1N}  & S_{2N}  &\ldots & S_{NN}
\end{array}
         \right|,~~~~~~
\label{caylay}
\end{eqnarray}
\begin{eqnarray}
\hspace{0.0cm}
G_{N} &=& ~-2^N \left|
\begin{array}{cccc}
\! p_1^2 & p_1p_2   &\ldots & p_1p_{N-1} \\
\! p_1p_2  &  p_2^2 &\ldots & p_2p_{N-1} \\
\vdots  & \vdots  &\ddots   & \vdots \\
\!p_1p_{N-1}  & p_2p_{N-1}  &\ldots & p_{N-1}^2
\end{array}
\right|, 
\label{gram} \\
&& \nonumber\\
&& \nonumber\\
\hspace{-0.0cm}
S_{ij}&=&-(q_i-q_j)^2+m_i^2+m_j^2, 
\quad \text{for} \quad i,j =1,2,\cdots, N.
\end{eqnarray}

The Feynman parameter integral for $J_N$
is obtained
\begin{eqnarray}
\label{JNfeyn}
\hspace{-0.7cm}
J_{N}\left(D; \{p_ip_j\}, \{m_i^2\} \right) &=&
(-1)^N\; \Gamma\left(N -\frac{D}{2} \right)
\left(\prod\limits_{i=1}^N \int\limits_{0}^{\infty} dx_i \right)
\dfrac{\delta(1-\sum\limits_{i=1}^N x_i) }
{\left( {\bf X}^T \cdot S_N \cdot {\bf X}/2 - i\rho \right)^{N -D/2} },
\nonumber\\
\end{eqnarray}
with ${\bf X}^T = (x_1, x_2, \cdots, x_N)$.

We discuss briefly on scalar one-loop five- and six-point 
functions. The reduction for scalar one-loop $N$-point 
integrals ($N\geq 5$) has been derived in
~\cite{Tarasov:1996br,Binoth:1999sp,
Binoth:2002xh, Melrose:1965kb,vanOldenborgh:1989wn,Denner:1991kt}
\begin{eqnarray}
\label{recursionJN}
\hspace{-0.5cm}
(D-N +1)J_{N}(D+2; \{p_ip_j\}, \{m_i^2\}) 
&=& \left[ \dfrac{S_{N}}{G_{N}} 
+\sum\limits_{k=1}^N \left(
\dfrac{\partial_{k} S_{N} }{ G_{N} } \right) \;
{\bf k^-} \right] J_N (D; \{p_ip_j\}, \{m_i^2\}).
\nonumber\\
\end{eqnarray}
Where the operator ${\bf k^-}$ is understood that 
it reduces $J_N$ to $J_{N-1}$  
by shrinking an $k$-th propagator in the integrand of 
one-loop $N$-point integrals, see Ref.~\cite{Tarasov:1996br}
 for more detail. The $S_N$ and $G_N$ are 
the determinants of Cayley and Gram matrices which are 
defined as in Eqs.~(\ref{caylay}, \ref{gram}).

In the space-time $D=4-2\varepsilon$, the 
$J_N(6-2\varepsilon; \{p_ip_j\}, \{m_i^2\})$ for $N\geq 5$ 
are the finite integrals. Thus, the term in the right hand side of 
(\ref{recursionJN}) corresponding to $N=5$ 
will vanish in the limit of $\varepsilon \rightarrow 0$. 
The resulting equation for $N=5$ reads 
\begin{eqnarray}
\label{J5J4}
J_5 (4; \{p_ip_j\}, \{m_i^2\}) = 
-\sum\limits_{k=1}^5 \left(
\dfrac{\partial_{k} S_{5} }{ S_{5} }
\right)
\;
{\bf k^-}\; J_5 (4; \{p_ip_j\}, \{m_i^2\}).
\end{eqnarray}
We also arrive at the same relation for $J_6$, or
\begin{eqnarray}
\label{J6J5}
J_6 (4; \{p_ip_j\}, \{m_i^2\}) = 
-\sum\limits_{k=1}^6 
\left( \dfrac{\partial_{k} S_{6} }{ S_{6} } 
\right) \;
{\bf k^-}\; J_6 (4; \{p_ip_j\}, \{m_i^2\}).
\end{eqnarray}
It means that scalar one-loop pentagon and 
hexagon integrals are also reduced to scalar 
one-loop box integrals (\ref{J5J4}, \ref{J6J5}) 
at the final stage. 

\end{document}